\documentclass[traditabstract]{aa} 
\usepackage{graphicx}
\usepackage{marginnote}
\usepackage{txfonts}
\usepackage{subfigure}
\usepackage{natbib}
\usepackage{amstext}
\usepackage[verbose]{placeins}
\usepackage{tikz}
\usepackage{array}
\usepackage{amssymb}
\usepackage{xcolor}
\usepackage{multirow}
\usepackage{longtable}
\usepackage{url}

\usepackage{pdflscape}
\usepackage{color, colortbl}
\definecolor{Gray}{gray}{1} 

\usepackage{multicol}
\newcommand{\bd}{\begin{displaymath}}
\newcommand{\ed}{\end{displaymath}}
\newcommand{\be}{\begin{equation}}
\newcommand{\ee}{\end{equation}}
\newcommand{\beaa}{\begin{eqnarray*}}
\newcommand{\eeaa}{\end{eqnarray*}}
\newcommand{\bea}{\begin{eqnarray}}
\newcommand{\eea}{\end{eqnarray}}

\newcommand{\tabspace}{  \rule{0pt}{2.5ex}}

\def\GLEE{\textsc{Glee}}
\def\GLaD{\textsc{Glad}}
\def\emcee{\textsc{Emcee}}
\def\tensorflow{\textsc{TensorFlow}}
\def\JAX{\textsc{Jax}}
\def\gigalens{\textsc{Giga-Lens}}
\def\GG{\GLEE \, \& \GLaD}
\def\gleeautopy{\textit{glee}$\_ $\textit{auto.py}}
\def\gleetoolspy{\textit{glee}$ \_ $\textit{tools.py}}

\newcommand{\sersic}{S\'{e}rsic}

\defcitealias{schuldt21a}{S21b}
\defcitealias{schuldt23a}{S23}
\usepackage{hyperref}

\begin{document}

   \title{HOLISMOKES - X. Comparison between neural network and semi-automated traditional modeling of strong lenses}

   \titlerunning{HOLISMOKES - X. Comparison between neural network and semi-automated traditional modeling}

   \author{S.~Schuldt\inst{1}\inst{,2}
     \and
     S.~H.~Suyu\inst{1}\inst{,2}\inst{,3}
     \and
     R.~Ca\~{n}ameras\inst{1}
     \and
     Y.~Shu\inst{1}\inst{,4}
     \and
     S.~Taubenberger\inst{1}
     \and
     S.~Ertl\inst{1}\inst{,2}
     \and
     A.~Halkola\inst{5}
          }

   \institute{Max-Planck-Institut f\"ur Astrophysik, Karl-Schwarzschild Str.~1, 85748 Garching, Germany \\
     \email{schuldt@mpa-garching.mpg.de}
     \and  
     Technical University of Munich, TUM School of Natural Sciences, Department of Physics, James-Franck-Str.~1, 85748 Garching, Germany
     \and
     Academia Sinica Institute of Astronomy and Astrophysics (ASIAA), 11F of ASMAB, No.1, Section 4, Roosevelt Road, Taipei 10617, Taiwan
     \and
     Ruhr University Bochum, Faculty of Physics and Astronomy, Astronomical Institute (AIRUB), German Centre for Cosmological Lensing, 44780 Bochum, Germany
     \and
     Py\"orrekuja 5 A, FI-04300 Tuusula, Finland
             }

   \date{Received --; accepted --}


  \abstract
      {Modeling of strongly gravitationally lensed galaxies is often required in order to use them as astrophysical or cosmological probes. With current and upcoming wide-field imaging surveys, the number of detected lenses is increasing significantly such that automated and fast modeling procedures for ground-based data are urgently needed. This is especially pertinent to short-lived lensed transients in order to plan follow-up observations. Therefore, we present in a companion paper a neural network predicting the parameter values with corresponding uncertainties of a singular isothermal ellipsoid (SIE) mass profile with external shear. In this work, we also present a newly developed pipeline \gleeautopy \, that can be used to model any galaxy-scale lensing system consistently. In contrast to previous automated modeling pipelines that require high-resolution space-based images, \gleeautopy \, is optimized to work well on ground-based images such as those from the Hyper-Suprime-Cam (HSC) Subaru Strategic Program or the upcoming Rubin Observatory Legacy Survey of Space and Time. We further present \gleetoolspy, a flexible automation code for individual modeling that has no direct decisions and assumptions implemented on the lens system setup or image resolution. Both pipelines, in addition to our modeling network, minimize the user input time drastically and thus are important for future modeling efforts. We applied the network to 31 real galaxy-scale lenses of HSC and compare the results to traditional, Markov-Chain Monte-Carlo sampling-based models obtained from our semi-autonomous pipelines. In the direct comparison, we find a very good match for the Einstein radius. The lens mass center and ellipticity show reasonable agreement. The main discrepancies pretrain to the external shear, as is expected from our tests on mock systems where the neural network always predicts values close to zero for the complex components of the shear. In general, our study demonstrates that neural networks are a viable and ultra fast approach for measuring the lens-galaxy masses from ground-based data in the upcoming era with $\sim10^5$ lenses expected.}


   \maketitle
%
\section{Introduction}
\label{sec:introduction}

Gravitational lensing, which means that the light of a background object is gravitationally deflected by a massive object in the foreground, gives us the opportunity to probe the Universe in various aspects. This includs the study of high-redshift systems \citep[e.g.,][]{dye18, lemon18, mcgreer18, rubin18, salmon18, shu18}, the study of the nature and distribution of dark matter \citep[e.g.,][]{schuldt19, baes21, basak22, gilman21, shajib21, wang22}, and cosmological parameter measurements \citep[e.g.,][]{refsdal64, chen19, birrer20, millon20a, shajib20, shajib22,  wong20}.

Therefore, huge effort is spent on large strong lens detection surveys. So far only hundreds of lenses have been confirmed and thousands of candidates have been identified \citep[e.g.,][]{bolton06, cabanac07, treu11, brownstein12, sonnenfeld15, sonnenfeld18a, shu16b, shu17, cornachione18, wong18, chan20, jaelani20a, jaelani21a}. Many lenses have been detected through deep learning networks scanning large image sets, from example, from the Panoramic Survey Telescope and Rapid Response System \citep[Pan-STARRS, e.g., ][]{canameras20}, the Hyper Suprime-Cam Subaru Strategic Program \citep[HSC-SSP,][]{canameras21b, shu22, jaelani22}, the Dark Energy Survey \citep[DES,][]{jacobs19, rojas22}, and the Canada-France Imaging Survey \citep[CFIS,][]{savary22}. Besides the deep learning technique that has become popular over the last years thanks to the fast application to very large data sets, multiple other techniques have been used to identify strong lenses, including pattern-based searches \citep[e.g.,][]{cabanac07, gavazzi12, gavazzi14}, spectroscopic searches \citep[e.g.,][]{bolton06, brownstein12, shu16a, talbot18, talbot21}, and modeling searches \citep[e.g.,][]{sonnenfeld18a, chan20}. In the near future, the  Legacy Survey of Space and Time (LSST) at the Vera C. Rubin Observatory \citep{ivezic08}, which will cover around 18,000 deg$^2$ in six filters ($u,g,r,i,z,y$), and the Euclid telescope \citep{laureijs11}, covering around 15,000 deg$^2$ throughout its six-year-long mission, will provide billions of galaxy images including on the order of 100,000 additional lenses \citep{collett15}.

After a lens system is detected, a mass model is necessary for nearly all further studies. Lens models are often described through parameterized profiles such as a \sersic \, profile \citep{DeVaucouleurs48, sersic63} for the lens light. There are different profiles assumed for the mass distribution, where the choice of profile depends on the image resolution and scientific goal. For ground-based images of galaxy-scale lensing systems, one typically adopts a singular isothermal sphere or singular isothermal ellipsoid (SIE) profile, possibly together with an external-shear $\gamma_\text{ext}$ component, while for high-resolution images, for example from the Hubble Space Telescope (HST), more complex profiles can be adopted. In that case, the dark matter component can be described independently through an NFW \citep*{navarro97} profile and the baryonic matter through a mass-follows-light profile or power-law profile \citep[e.g., SPEMD profile,][]{barkana98}. The best fitting parameter values are often obtained with Markov chain Monte Carlo (MCMC) sampling \citep[e.g.,][]{jullo07, suyu10a, suyu12b, fowlie20, sciortino20} as it is able to sample a high-dimensional parameter space and yields the posterior distribution for the uncertainties and degeneracies. However, this method is very time and resource consuming due to its computational time and due to the required user inputs, such that the modeling of a single lens system can take weeks. Therefore the current techniques are already insufficient for the known lens candidates as well as for upcoming surveys such as LSST and Euclid.

One possibility is to automate the modeling procedure while still relying on Bayesian inference such as MCMC sampling \citep[e.g.,][]{nightingale18, rojas22, savary22, ertl22, etherington22, gu22, schmidt23}, which reduces the user input dramatically, resulting in an overall runtime on the order of days. A further speed-up can be achieved by using Graphical Processing Units \citep[GPUs, e.g.,][]{gu22}. Another option is to use machine learning \citep[hereafter \citetalias{schuldt21a} and \citetalias{schuldt23a}, respectively]{hezaveh17, levasseur17, morningstar18, morningstar19, pearson19, pearson21, schuldt21a, schuldt23a}. Convolutional neural networks (CNNs), including CNN-based residual networks (ResNets) have become one of the major tools in (astronomical) image processing on very large data sets \citep[e.g.,][]{paillassa20, tohill21, wu20, tanoglidis21, cavanagh21, grover21, schuldt21b, vegaferrero21} and thus also in recent lens detections \citep{jacobs17, jacobs19, petrillo17, lanusse18, schaefer18, davies19, metcalf19, canameras20, canameras21b, canameras22a, he20, huang20, li20, rojas22, savary22, jaelani22, shu22}. The main requirement of neural networks (NNs) is a large enough training set on the order of a hundred thousand images. Since there are not that many known lenses so far, the training data need to be simulated. While the training data were previously completely simulated \citep{jacobs17, jacobs19, petrillo17, schaefer18, davies19, metcalf19}, recent efforts favor the use of real galaxy images and simulate only the lensing effect for the background object, that is paint the lensed arcs on top of a real galaxy image, which is then the lens galaxy \citep[e.g.,][]{canameras20, canameras21b, canameras22a, savary22, schuldt21a, shu22}. Here the galaxies used as lens galaxies are typically limited to luminous red galaxies (LRGs) given their higher lensing cross section.

The main advantage of machine learning is the fully autonomous procedure and the huge speed-up compared to MCMC sampling methods since a trained network is able to predict the mass parameters within fraction of a second. The main challenges of this method are the requirement of a training, validation, and test set, which need to be, at least partly, simulated, and the difficulty in translating the model parameters into e.g., a $\chi^2$ for checks of their accuracy. \citet{pearson21} performed a detailed comparison between the network predictions and conventionally obtained models of a variety of complex mock lensing systems. This, however, still allows the question about the network performance on real observed lenses. Therefore, we apply the network described in \citetalias{schuldt23a} to a sample of grade A (i.e., secure) galaxy-scale lenses from the Survey of Gravitationally-lensed Objects in HSC Imaging (SuGOHI) program \citep{sonnenfeld18a, wong18, sonnenfeld19, chan20, jaelani20a, sonnenfeld20, jaelani21a}. This work, which is part of our ongoing Highly Optimized Lensing Investigations of Supernovae, Microlensing Objects, and Kinematics of Ellipticals and Spirals \citep[HOLISMOKES, ][]{suyu20} program, is the first time a trained modeling network is applied to real ground-based images instead of mock images. Further, we compare each model predicted by the network to a model that we obtained with traditional MCMC sampling methods. For the traditional modeling, we have developed \gleeautopy, an automated code that is optimized for HSC-like ground-based images and thus will also be very helpful beyond this comparison project. Given the expected similarity in data quality, \gleeautopy \, can also be used to model lenses observed in the near future with LSST. For more specific and detailed follow-up modeling, we have developed \gleetoolspy, a flexible code to automate optimization steps selected by the user without assuming anything on the lens system setup. Thanks to its flexibility, \gleetoolspy \, will be very useful for many forthcoming projects including lens observations from LSST, but also from Euclid or the James Webb Space Telescope \citep{rigby22}.

The outline of the paper is as follows. In Sect.~\ref{sec:dataset} we present the SuGOHI lenses used in this work. In Sect.~\ref{sec:mcmcmodels} we describe \gleeautopy, our automated MCMC based modeling code for ground-based images, our flexible automation code \gleetoolspy, and the resulting models. We then introduce our modeling network from \citetalias{schuldt23a} and present the output models from the network in Sect.~\ref{sec:networkmodels}. A detailed comparison of the mass models is given in Sect.~\ref{sec:comparison}, and of the predicted image positions and time delays in Sect.~\ref{sec:ImPoscomparison}. We summarize our results in Sect.~\ref{sec:conclusion}.

Throughout this work we assume a flat $\Lambda$CDM cosmology with a Hubble constant $H_0 = 72\, \text{km}\, \text{s}^{-1}\, \text{Mpc}^{-1}$ \citep{bonvin17} and $\Omega_\text{M} =1 -\Omega_\Lambda = 0.32 $ \citep{planck20} for consistency with corresponding work presented in \citetalias{schuldt21a} and \citetalias{schuldt23a}. Unless specified otherwise, each quoted parameter estimate is the median of its 1D marginalized posterior probability density function, and the quoted uncertainties show the 16$\text{th}$ and 84$\text{th}$ percentiles (i.e., the bounds of a 68\% credible interval).

\FloatBarrier
\section{Comparison data set}
\label{sec:dataset}

\begin{table*}[ht!]
\begin{center}
\caption{Overview of all 31 SuGOHI lenses modeled with \GG \, for a direct comparison to the predictions of our ResNet.}
\label{tab:overviewLenssample}
\begin{tabular}{c|cc|cc|c}
  Name    & RA       & DEC         & $z_\text{d}$ & $z_\text{s}$ & References\\
        &          &             &             &              &           \\
\hline
  HSCJ015618$-$010747   &  29.0755 &	$-1.1298  $ & 0.542 & 1.167       & (a), (b)\\
  HSCJ020141$-$030946   &  30.4249 &	$-3.1628  $ & 0.362 & $-$         & (c) \\ 
  HSCJ020241$-$064611   &  30.6725 &	$-6.7698  $ & 0.502 & 2.748       & (b), (c)\\
  HSCJ020955$-$024442   &  32.4809 &	$-2.7450  $ & (0.56)& $-$         & (d), (e)\\ 
  HSCJ021737$-$051329   &  34.4049 &	$-5.2248  $ & 0.646 & 1.847       & (b), (d), (f), (g), (h), (i)\\ 
\hline
  HSCJ022346$-$053418   &  35.9423 &	$-5.5718  $ & 0.499 & 1.444       & (b), (d), (f), (g)\\ 
  HSCJ022610$-$042011   &  36.5444 &	$-4.3366  $ & 0.496 & $-$         & (b), (c), (g) \\ 
  HSCJ023217$-$021703   &  38.0724 &	$-2.2844  $ & 0.508 & $-$         & (c)\\ 
  HSCJ023322$-$020530   &  38.3443 &	$-2.0918  $ & (0.49)& $-$         & (d), (e)\\ 
  HSCJ085046$+$003905   & 132.6942 &	$ 0.6515  $ & (0.84)& $-$         & (e)\\ 
\hline
  HSCJ085855$-$010208   & 134.7333 &	$-1.0357  $ & 0.468 & 1.421       & (b), (c)\\
  HSCJ090429$-$010228   & 136.1239 &	$-1.0411  $ & 0.957 & 3.403       & (d), (j)\\
  HSCJ094427$-$014742   & 146.1145 &	$-1.7951  $ & 0.539 & 1.179       & (b), (k)\\
  HSCJ120623$+$001507   & 181.5994 &	$ 0.2520  $ & 0.563 & 3.120       & (b), (c)\\
  HSCJ121052$-$011905   & 182.7187 &	$-1.3181  $ & 0.700 & 2.295       & (b), (c)\\
\hline
  HSCJ121504$+$004726   & 183.7685 &	$ 0.7906  $ & 0.642 & 1.297       & (b), (k)\\
  HSCJ124320$-$004517   & 190.8365 &	$-0.7550  $ & 0.654 & $-$         & (a)\\ 
  HSCJ125254$+$004356   & 193.2275 &	$ 0.7323  $ & 0.649 & $-$         & (a)\\ 
  HSCJ135138$+$002839   & 207.9122 &	$ 0.4778  $ & 0.461 & $-$         & (a)\\ 
  HSCJ141136$-$010215   & 212.9022 &	$-1.0377  $ & 0.949 & 3.021       & (e)\\
\hline
  HSCJ141815$+$015832   & 214.5656 &	$ 1.9756  $ & 0.556 & 2.139       & (b), (c)\\
  HSCJ142720$+$001916   & 216.8356 &	$ 0.3211  $ & 0.551 & 1.266       & (b), (c)\\
  HSCJ144320$-$012537   & 220.8359 &	$-1.4270  $ & (1.16)& $-$         & (e), (l)\\ 
  HSCJ145242$+$425731   & 223.1789 &	$42.9589  $ & 0.718 & $-$         & (a)\\ 
  HSCJ150021$-$004936   & 225.0876 &	$-0.8269  $ & (0.41)& $-$         & (e)\\ 
\hline
  HSCJ150112$+$422113   & 225.3007 &	$42.3537  $ & (0.27)& $-$         & (d)\\ 
  HSCJ223733$+$005015   & 339.3897 &	$ 0.8377  $ & 0.604 & 2.143       & (b), (c)\\
  HSCJ230335$+$003703   & 345.8965 &	$ 0.6176  $ & 0.458 & 0.936       & (a), (b), (k)\\
  HSCJ230521$-$000211   & 346.3403 &	$-0.0366  $ & 0.492 & $-$         & (a), (h)\\ 
  HSCJ233130$+$003733   & 352.8770 &	$ 0.6259  $ & 0.552 & $-$         & (a), (h)\\ 
\hline
  HSCJ233146$+$013845   & 352.9434 &	$ 1.6460  $ & 0.476 & $-$         & (a)\\ 

\end{tabular}
\end{center}
Note. From left to right we give the name used to reference each lens, right ascension (J2000), declination (J2000), spectroscopic (photometric) lens redshift $z_\text{d}$ and source redshift $z_\text{s}$. The last column gives the references: (a) \citet{wong18}, (b) \citet{sonnenfeld19}, (c) \citet{sonnenfeld18a}, (d) \citet{jaelani20a}, (e) \citet{sonnenfeld20}, (f) \citet{gavazzi14}, (g) \citet{sonnenfeld13}, (h) \citet{jacobs19}, (i) \citet{cabanac07}, (j) \citet{jaelani20b}, (k) \citet{brownstein12}, (l) \citet{chan20}. 
\end{table*}

For our comparison we use HSC images observed with the 8.2m Subaru Telescope in Hawaii. The HSC survey covers around 1400deg$^2$ in the second public data release wide layer in at least one filter. It provides images with very good quality and with a pixel size of $0.168\arcsec$ in different filters, including $griz$, which our network is trained on \citepalias{schuldt21a, schuldt23a}. The quality is expected to match that from LSST such that our results should also hold for those images.

All lenses of our sample were detected as part of the SuGOHI program\footnote{Webpage: \url{http://www-utap.phys.s.u-tokyo.ac.jp/~oguri/sugohi/} .}, a large and extensive lens search in HSC data using various methods \citep{sonnenfeld18a, sonnenfeld19, sonnenfeld20, wong18, chan20, jaelani20a, jaelani21a}. For our comparison, we select only the grade A candidates detected by the SuGOHI program to have a very reliable, partly spectroscopically confirmed, sample without false-positive lens candidates. We further select galaxy-scale lenses as the network is trained for such systems. From the resulting sample, we reject systems HSCJ023307$-$043838 \citep{more12, more16b, sonnenfeld13, gavazzi14, jacobs19, sonnenfeld19, chan20}, HSCJ144132$-$005358 \citep{sonnenfeld20}, and HSCJ135138+002839 \citep{wong18} as those look more like cluster- or group-scale lenses based on their environment and image separation although they are listed as galaxy-scale systems on the webpage. This results in a sample of 31 lenses which we summarize in Tab.~\ref{tab:overviewLenssample}. 

\begin{figure*}
  \includegraphics[trim=0 0 0 0, clip, width=0.92\textwidth]{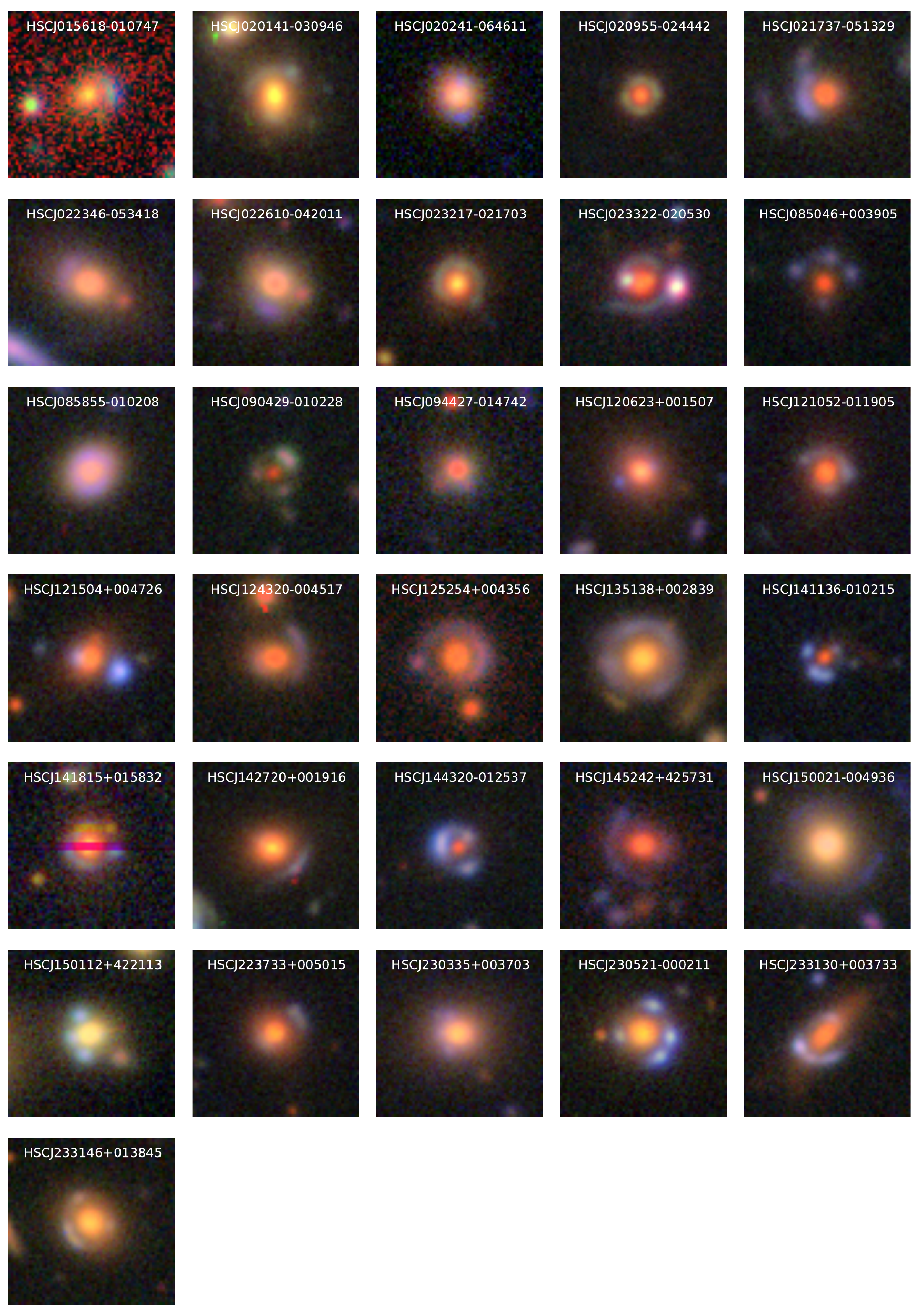}
  \caption{Color images based on the $gri$ filters of the 31 HSC SuGOHI lenses used for direct comparison. All images have the size of $64\times 64$ pixels (i.e. $\sim 10\arcsec \times \sim 10\arcsec$) and are oriented such that north is up and east is left. The name of each lens is given on the top of each image.\label{fig:colorimage}}
\end{figure*}

In the table, we also quote the spectroscopic redshifts of the lenses and -- if available -- of the sources. In case there is no spectroscopic lens redshift available, we report the photometric redshift in brackets. Since the SIE+$\gamma_\text{ext}$ parameters are independent of the redshifts, we can model all systems even if not all redshifts are measured yet. For our comparison we use images of the lenses in the four filters $g,r,i,$ and $z$ as our network was trained on these filters \citepalias{schuldt23a}. A mosaic of $gri$ color images of our comparison sample is shown in Fig.~\ref{fig:colorimage}.

\section{Mass models through automated \GG \, software}
\label{sec:mcmcmodels}

For our comparison, we need a mass model of the 31 lens systems described in Sect.~\ref{sec:dataset} that is obtained with a traditional, MCMC sampling method. A subset was already modeled by \citet{sonnenfeld13, sonnenfeld19} for studies on the stellar initial mass function, but they adopted only an SIE profile parameterized by the lens-mass center $x_\text{l}$ and $y_\text{l}$, the axis ratio $q_\text{lm}$ with position angle $\phi_\text{lm}$, and an Einstein radius $\theta_\text{E}$. Here, we also include an external-shear component parameterized by a shear strength $\gamma_\text{ext}$ and orientation $\phi_\text{ext}$ to account for perturbations due to the environment of the lens system and to match the adopted parameterization of our neural network (see \citetalias{schuldt23a} and Sect.~\ref{sec:networkmodels}). Therefore, we model all 31 lenses using \GLEE \, \citep[Gravitational Lens Efficient Explorer,][]{suyu10a, suyu12b} and its extension \GLaD \, \citep[Gravitational Lensing and Dynamics,][]{chirivi20}, which are both well tested modeling codes that provide several different parameterized profiles and optimization algorithms. These codes support e.g., MCMC sampling using the Metropolis Hastings algorithm \citep[hereafter MCMC refers to Metropolis Hastings MCMC unless specified otherwise]{hastings70, robert04}, a highly parallelized ensemble sampler, \emcee \, \citep{foreman-mackey13}, simulated annealing, a generalized version called dual annealing\footnote{Python Package available here: \url{https://docs.scipy.org/doc/scipy/reference/generated/scipy.optimize.dual_annealing.html}} \citep{tsallis88, tsallis96, xiang97, xiang00, xiang13, mullen14}, and basin hopping\footnote{Python Package available here: \url{https://docs.scipy.org/doc/scipy/reference/generated/scipy.optimize.basinhopping.html}} \citep{wales97}.

Such traditional modeling of lens systems is very time and resource consuming. Especially it requires a lot of input from a user with specific modeling expertise, e.g., to create the required input files, including a configuration file specifying the adopted light and mass profiles with the initial starting values, and the optimization details such as the chain length, step size, and range for the different sampling methods like MCMC and simulated annealing, the prior ranges on each parameter, and several other details. Each optimization run will lead to an updated configuration file with the newest best set of parameter values. After a possible update with e.g., the selection of varying parameters which get typically iteratively optimized, a new optimization run is started. This will be repeated until the sampled parameter values stabilize and represent the observation to an acceptable level. Thus, this procedure is a completely iterative process and thus the user input time is relatively high.

We develop \gleeautopy, a code to automate the modeling procedure and thus to minimize the user input time where we adapt partly the code and procedure presented in \citet{ertl22} who model high-resolution HST images of lensed quasars. The implemented procedure and decision criteria were extensively tested on the presented sample, such that the code is able to model a broad range of typical galaxy-scale lenses from ground-based surveys, where most of the lenses are detected. The final procedure and criteria are presented in Sect.~\ref{sec:mcmcmodels:automatedcode}.

Since each lensing system is special in its own way and thus requires specific treatment in the modeling, the presented automated code will not obtain for every lensing system a fit that perfectly represents the observation. This is expected because of the huge variety of galaxy light distribution, orientations, line-of-sight objects and similar aspects, but provides at least a very good initial model for further refinement of the parameter values with minimal user input and in an acceptable amount of computational run time. Since the main goal of this work is a direct comparison between network predictions and traditionally obtained models on real lenses, we conduct a specific follow-up modeling for several individual lenses of our sample to improve the fit. For this, we introduce  \gleetoolspy\ in Sect.~\ref{sec:mcmcmodels:gleetoolspy}, a code that automates individual optimization steps specified by the user but without its own decisions implemented. This allows full control of the modeling sequence while still reducing notable user input time.

The resulting best models are presented in Sect.~\ref{sec:mcmcmodels:results} where we also discuss details of the code limitations. We compare \gleeautopy \, to other modeling codes in Sect.~\ref{sec:mcmcmodels:comparison}.

\subsection{Automated modeling code \gleeautopy}
\label{sec:mcmcmodels:automatedcode}

\begin{figure*}[p]
\begin{tabular}{p{0.5\textwidth} p{0.5\textwidth}}
  \vspace{0pt} \includegraphics[trim=25 290 25 30, clip, width=0.75\columnwidth]{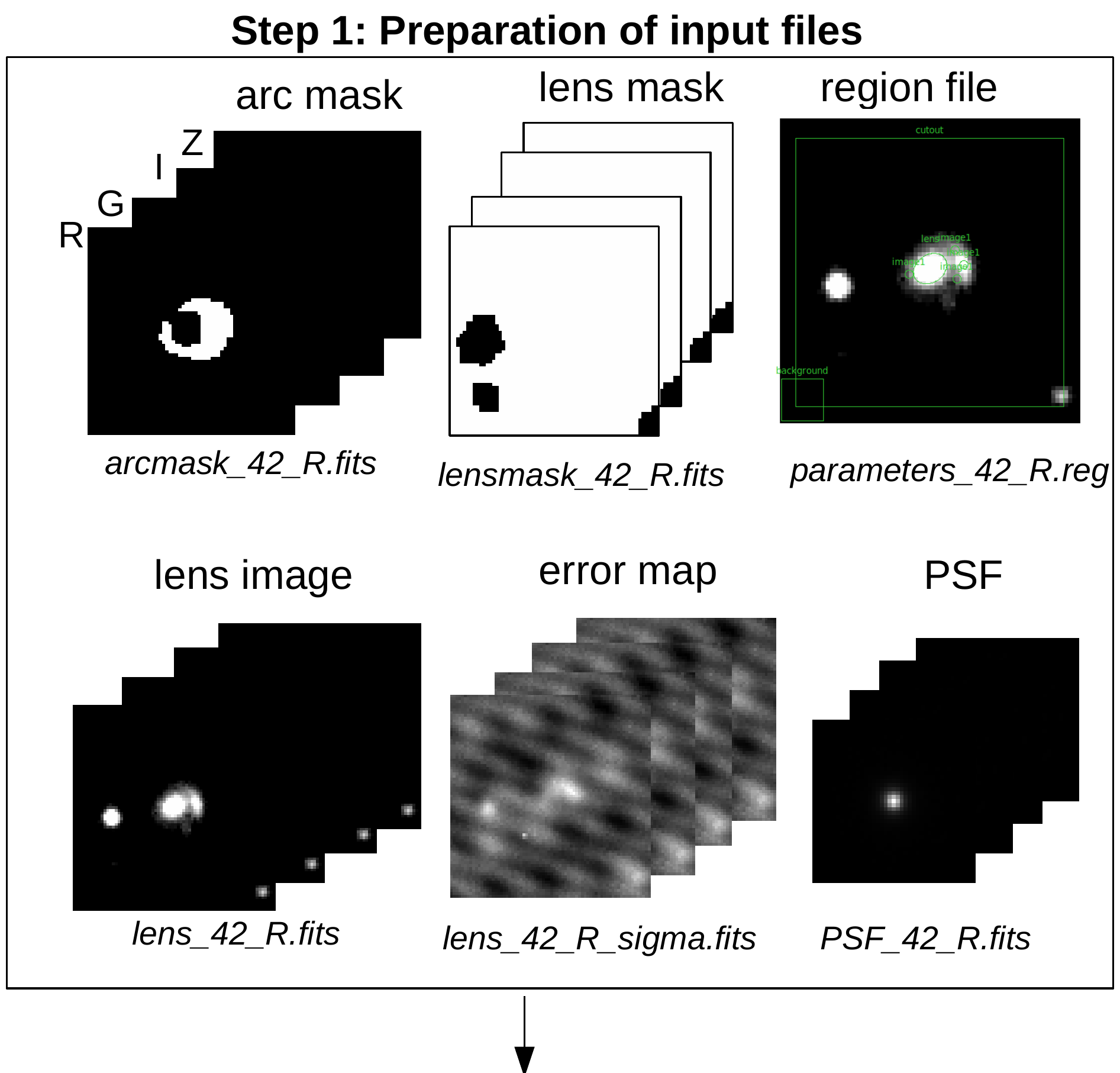}
  \vspace{0pt} \includegraphics[trim=25  50 25 30, clip, width=0.75\columnwidth]{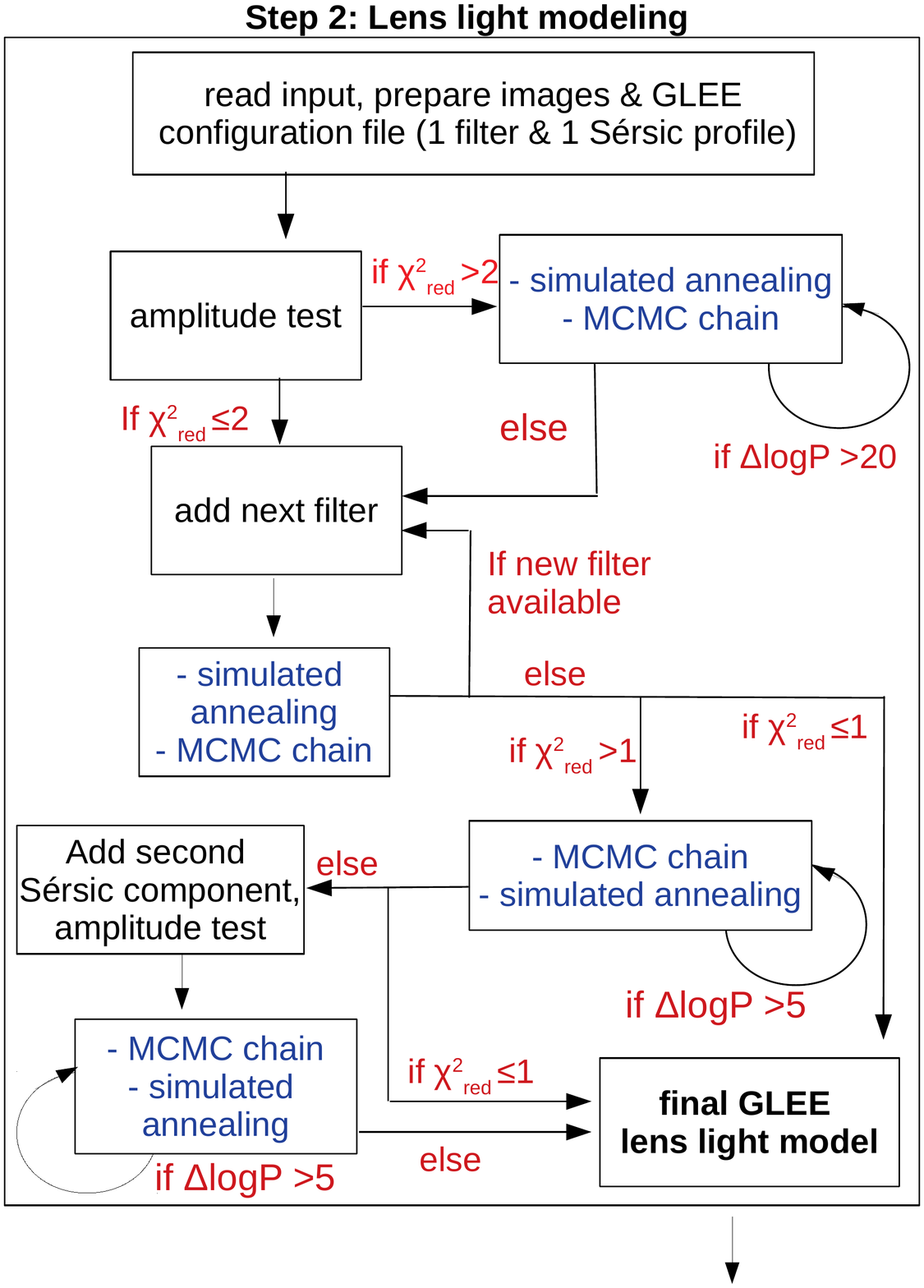}
  \vspace{0pt} \includegraphics[trim=25 320 25 30, clip, width=0.75\columnwidth]{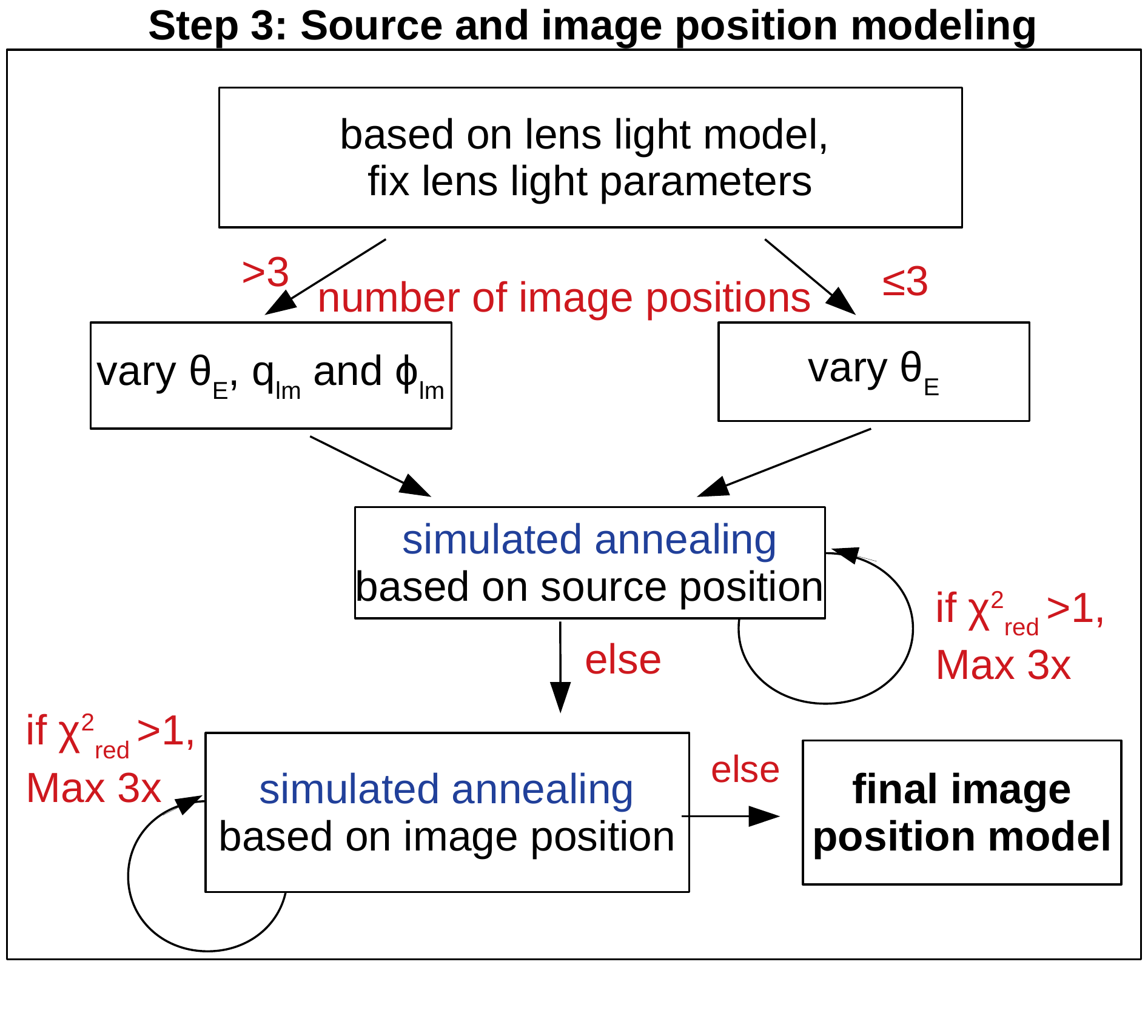}&
  \vspace{0pt} \includegraphics[trim=25  36 25 30, clip, width=0.85\columnwidth]{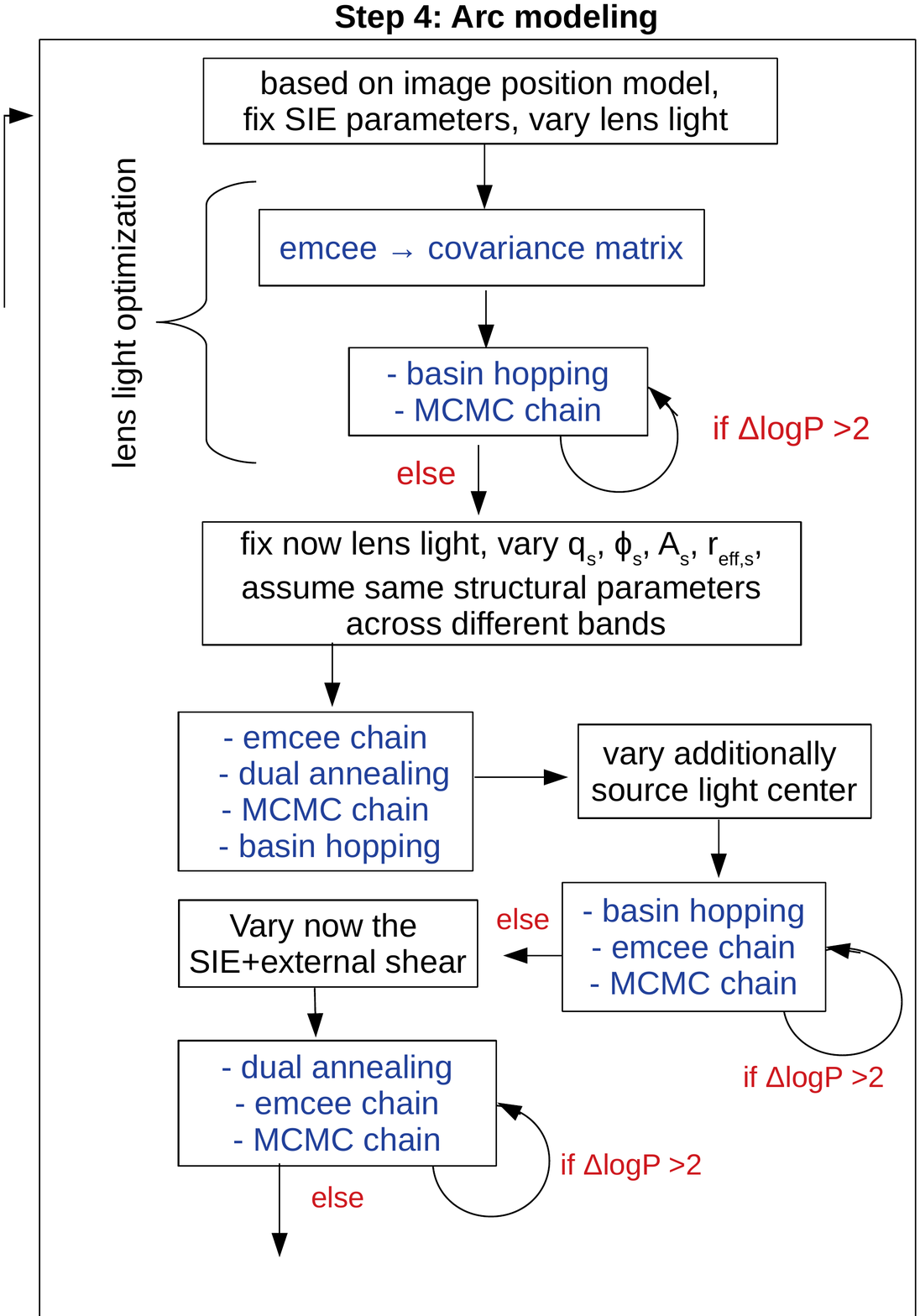}
  \vspace{0pt} \includegraphics[trim=25  50 25 60, clip, width=0.85\columnwidth]{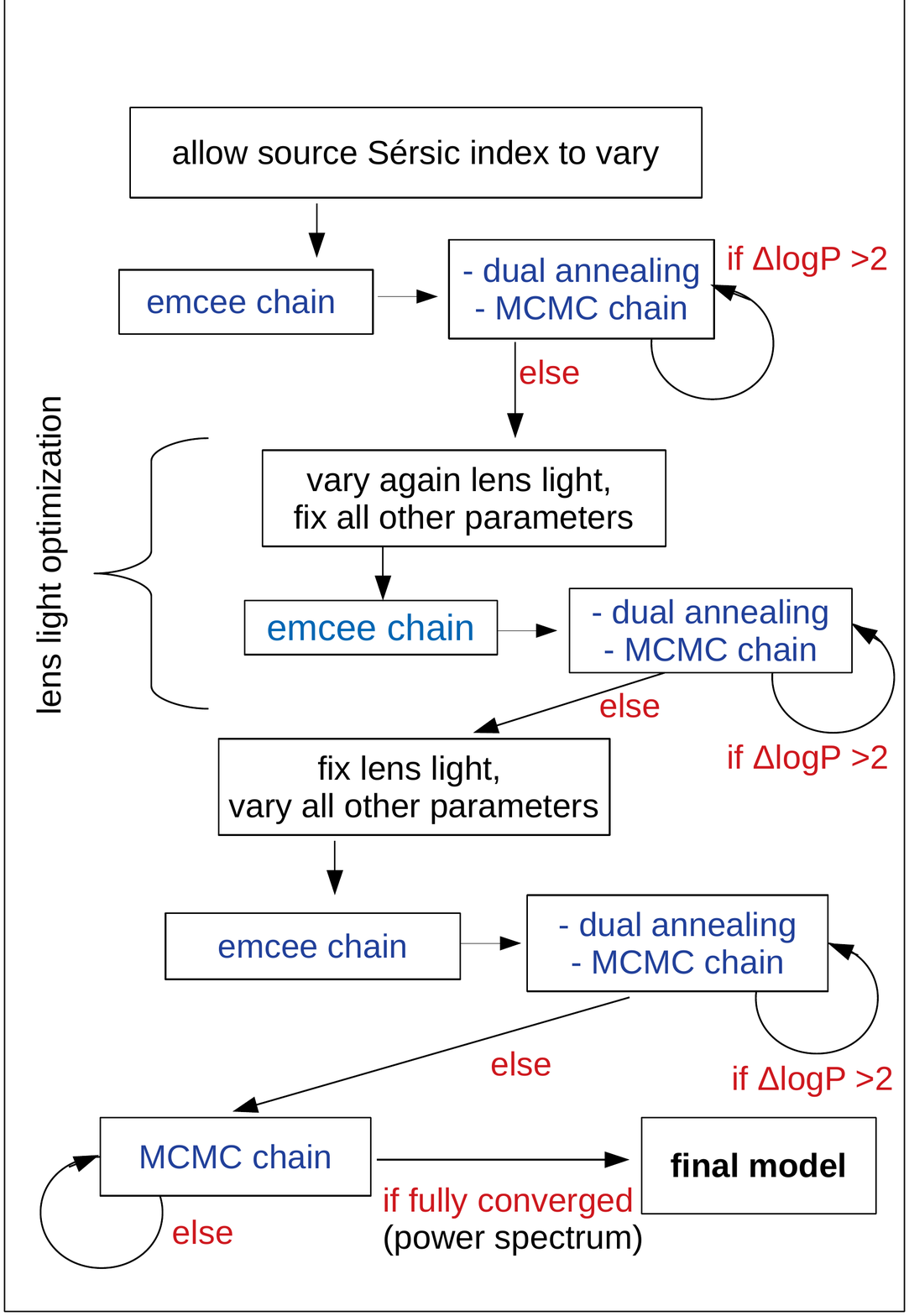}
\end{tabular}
\caption{Diagram of \gleeautopy, our automated procedure for galaxy-scale lens modeling with \GG, which is divided into four main steps indicated on the top of each panel. For the optimization we make use of simulated annealing, basin hopping, dual annealing, and Metropolis Hastings MCMC sampling. The MCMC, and also \emcee, are used to get an estimate of the sampling covariance matrix and provide the probability distribution used as decision criteria (indicated as logP). The ``amplitude test'' refers to a quick check of a newly included \sersic \, profile's amplitude to determine the best order of magnitude to use as an initial starting point. This example lensing system has an identification number of 42 and the filter is denoted as $R$. Further details on the procedure can be found in Appendix \ref{app:modelprocedure}.\label{fig:flowchart}}
\end{figure*}

Our newly developed automated modeling code \gleeautopy \, is divided into four individual parts (see also details in Appendix \ref{app:modelprocedure}). In the first step, the user simply prepares the input files which are used for the modeling. In the second step, the lens light distribution is modeled, where we adopt \sersic \, profiles parameterized as
\be
I^\text{sersic}(r) = A e^{-\zeta(n) ~\left( \left( \frac{r}{r_\text{eff}} \right) ^{1/n}-1 \right)} \ ,
\label{eq:sersicintensity}
\ee
with an elliptical radius
\be
r = \sqrt{x^2 + \frac{y^2}{q_\text{ll}^2}}. \label{eq:ellipticalradius}
\ee
Here ($x,y$) are the coordinates aligned along the semi-major and semi-minor axis of the lens light, $q_\text{ll}$ is the lens-light axis ratio, and $\phi_\text{ll}$ is the position angle. The amplitude is denoted as $A$ and the effective radius as $r_\text{eff}$. The constant $\zeta(n)$ depends on the \sersic \, index $n$ and ensures that the effective radius encloses half of the projected light \citep{ciotti99, cardone04, dutton11}. Therefore, the effective radius is also called half-light radius. For our modeling, we assume the same structural parameters ($x_\text{l}$, $y_\text{l}$, $q_\text{ll}$, $\phi_\text{ll}$, $r_\text{eff,ll}$, $n_\text{ll}$) across different filters, where ($x_\text{l}, y_\text{l}$) are the lens light center. This step is necessary to remove the light from the main lens and other line-of-sight objects (where the pixels associated with other line-of-sight objects are in the so-called lens mask and discarded in the light model fitting, see Fig.~\ref{fig:flowchart}), resulting in an image of the arcs alone.

Before modeling the arcs on the pixel level, we include as the third step (see Fig.~\ref{fig:flowchart}) a short optimization of the SIE mass parameters based on the multiple lensed image positions identified in Step 1. The convergence (also called dimensionless surface mass density) of the adopted SIE profile can be expressed as
\be
\kappa(r) = \frac{\theta_\text{E}}{(1+q_\text{lm})r}
\ee
rotated by an angle $\phi_\text{lm}$, measured counterclockwise from the positive $x$-axis in our implementation, and with an elliptical radius defined in analogy to Eq.~(\ref{eq:ellipticalradius}) with axis ratio $q_\text{lm}$. Such an optimization is typically performed before the lens light model, but in our case we adopt the lens light center as the center of the SIE profile. This step only takes around a minute, but updates the initial mass parameter values such that the multiple images map to the same source position. This is crucial for the arc light modeling in step four. Additionally, it gives us a preliminary source position which is used as a starting value for the source light center. In this part, the external shear is neglected given the low number of constraints (four data points for a doubly and eight for a quadruply lensed system).

The arc light modeling and source light reconstruction is then performed as the fourth step, where the external shear is included. Here we assume one \sersic \, profile to describe the source light distribution. A parameterized profile for the background source, which is supported only by \GLaD, is preferred over the pixelized source surface brightness (SB) reconstruction implemented in \GLEE, as we use ground-based images that only resolve the main structure of the arcs. This is the reason for making use of \GLaD, even though we do not include dynamical modeling, which is the key part of \GLaD. For the same reason, we adopt only one \sersic \, profile instead of two as done for the lens light distribution. The input data with assumed nomenclature, the individual optimization steps and implemented criteria of the modeling routine are presented as a flow diagram in Fig.~\ref{fig:flowchart} and explained in more detail in Appendix \ref{app:modelprocedure}.

This procedure was developed through extensive tests on all our 31 SuGOHI lenses and is therefore optimized for ground-based observations with a parameterized source light distribution. The individual sequences typically consists of one MCMC or \emcee \, chain to obtain a covariance matrix, and then alternate between either simulated annealing, dual annealing, or basin hopping, and an MCMC chain to obtain an updated covariance matrix but also optimized parameter values. The code is able to predict the lens-light model within a few hours, the source- and image-position model within around a minute, and the extended image model with source SB reconstruction within around a day. It runs on a single core and automatically launches 60-core parallelized jobs for the \emcee \, optimizations. This allows a uniform modeling of a larger sample of galaxy-scale lenses without much user input to provide a basic model of the observations.

\subsection{Flexible modeling code \gleetoolspy}
\label{sec:mcmcmodels:gleetoolspy}

Using the automated procedure described in Sect.~\ref{sec:mcmcmodels:automatedcode}, we modeled all 31 SuGOHI lenses uniformly. Because every lens system is peculiar in its own way, the automated procedure does not work perfectly for all of them. Since the main focus of this work is the comparison between network predictions and conventional methods rather than discussing the automation code limitations, we improved several models afterwards by hand until the residuals were acceptable and no further improvement was achieved. For this, we tested for lenses with stronger residuals the improvement when including additional profiles like a third \sersic \, profile for the lens light or a second component for the source light. The automation code at least gives a very good starting point for further individual optimization with minimal user input time.

For the individual modeling, we developed \gleetoolspy, a flexible \GLEE-based code to automate several optimization steps when modeling with \GG. This means one provides as usual a configuration file to the code that specifies the data, the number of profiles, the varying parameters, starting values, the adopted cosmology, and other required information. One can then specify a list of optimization iterations that the code shall sequentially perform without further input from the user. This helps to reduce the user input time and waiting time for the start of the next iteration of optimization while giving the flexibility to assume any setup (e.g., number of filters or profiles, kinematic data, single or multiplane lensing etc.). The list of tasks can also include saving the best set of parameters from a MCMC or \emcee \, chain as well as computing the covariance matrix and updating the configuration file, which are normally done always manually by hand.

Since the code does not include any decision criteria as the code presented in Sect.~\ref{sec:mcmcmodels:automatedcode}, \gleetoolspy \, can be used for any lens system configuration and does not rely on the assumption of galaxy-scale lensing. This means, it can be used to model ground-based images like those from HSC or soon from LSST, but also high-resolution images from space or by using the adaptive optics technique. Moreover, it is independent of the mass scale (galaxy, galaxy group, or galaxy cluster), which means it is helpful for modeling any lensing system.

In addition to the sampling opportunities, \gleetoolspy \, is equipped with several other frequently needed tools for e.g., visualization of the obtained fits (compare Figs.~\ref{fig:model_47} to \ref{fig:model_1976}) with \GG\footnote{This plotting tool is adapted from a code developed by Dr.~Giulia Chiriv{\`i} with the \GLaD \, extension \citep{chirivi20}.}, running the amplitude test used in \gleeautopy \, (see Appendix~\ref{app:modelprocedure}), updating all linked parameters within the configuration file, generating masks such as the required arc mask and lens mask\footnote{This tool was written by Dr.~Yiping Shu \citep{shu16b}.}, and converting the complex ellipticity and external shear with their uncertainties into the normal parameterization (axis ratio and shear strength with corresponding position angles) or vice versa (see also Sect.~\ref{sec:networkmodels}).

\subsection{Results and discussion of MCMC modeling}
\label{sec:mcmcmodels:results}

We model each of the presented lenses in the sample with \gleeautopy. Depending on the $\chi^2$, MCMC chain convergence, and residuals, we improve the models further manually where we make extensive use of \gleetoolspy. Since we are mainly interested in the comparison to the network predictions instead of demonstrating the power of our automated code, we only give a short quantitative summary of the performance of our automated pipeline here and  afterwards report the results of the final models in detail.

For 29 out of 31 systems, we obtain a $\chi^2_\text{red}$ over all four bands of less than 2 directly with \gleeautopy, and for 15 out of 31 systems a $\chi^2_\text{red}$ of less than 1.5. As usual, $\chi^2_\text{red}$ is defined as $\chi^2$ divided by the number of degrees of freedom (the number of modeled pixels minus the number of free parameters) with 
\be
\chi^2 = \sum_{j=1}^{N_\text{p}} \frac{ \left( I_j^\text{obs} - \text{PSF} \otimes I_j^\text{sersic} \right) ^2}{\sigma_{\text{tot},j}^2} \, .
\label{eq:chi2}
\ee
In this equation, $N_\text{p}$ denotes the number of pixels, $\sigma_{\text{tot},j}$ the total noise of pixel $j$ provided by the HSC error map, $\otimes$ represents the convolution of the point spread function (PSF) and the predicted intensity $I_j^\text{sersic}$ from Eq.~(\ref{eq:sersicintensity}) including both lens galaxy and the lensed source galaxy, and $I^\text{obs}_j$ describes the observed intensity of pixel $j$.

When visually inspecting the residuals, we identified six systems where \gleeautopy \, produced inadequate model or failed completely, while for seven systems (nearly) no further improvement was necessary to obtain our final model presented below (compare also Figs.~\ref{fig:model_47} to \ref{fig:model_1976}). However, a $\chi^2_\text{red} \sim 1$ does not necessarily mean that the model is good as the code sometimes predicted unrealistic parameter values (e.g., very low axis ratio values of $<0.1$) but low residuals which might come from known degeneracies and relatively large prior ranges. For reducing this possibility, stronger priors would help (e.g., on the \sersic \, index or mass axis ratio).

The median values with $1 \sigma$ uncertainties computed from our final MCMC chain for the SIE and the external shear parameters, i.e. after possible manual refinement with \gleetoolspy, are reported in Tab.~\ref{tab:comparison_normal} (white background). We further quote in Tab.~\ref{tab:comparison_normal} the $\chi^2$ and the $\chi^2_\text{red}$ values (compare Eq.~(\ref{eq:chi2})), which give an indication of how good the obtained \GG \, model is. From this list, we see that the $\chi^2_\text{red}$ is above 1.5 for only five of the 31 lenses ({HSCJ020141$-$030946}, {HSCJ023322$-$020530}, {HSCJ135138+002839}, {HSCJ150112+422113}, {HSCJ230521$-$000211}). Nonetheless, we included them in our comparison as the fits are overall still reasonable.


In contrast to the network, we model with the traditional method directly the lens light and source light as \sersic \, profiles, which also influence the quoted $\chi^2$ and $\chi^2_\text{red}$. The resulting parameter values are given in Appendix \ref{app:modeldetails}. The best fit values for the \sersic \, parameters of the lens light are listed in Tab.~\ref{tab:lenslightpars}, and of the source light in Tab.~\ref{tab:sourcelightpars}. We further show all 31 final models as Figs.~\ref{fig:model_47} to \ref{fig:model_1976}. Each plot shows from left to right the observed image, the model, and the normalized residuals. The four rows correspond to the four different filters in the order $g,r,i,$ and $z$.

In the course of the modeling, we have made several observations which we discuss in the following.


In general, the lens center, which is assumed to be the same for the light and mass distribution, is in all models very well constrained ($1\sigma \sim 0.001\arcsec$). The offsets with respect to the image center
are also relatively small, only seven systems 
have a difference larger than half a pixel (0.084\arcsec) and none has an offset larger than one pixel (0.168\arcsec). The source center is not as well constrained as the lens center with a typical $1\sigma$ uncertainty of $ < 0.1\arcsec$.

The estimated Einstein radius is, apart from the lens HSCJ015618$-$010747 with $\theta_\text{E}=0.99\arcsec$, always above 1$\arcsec$. The lens HSCJ150021$-$004936 has the largest Einstein radius with 3.063\arcsec, followed by system HSCJ135138+002839 with 2.216\arcsec. All other systems have an Einstein radius between 1\arcsec and 2\arcsec.

When comparing the lens axis ratio between light $q_\text{ll}$ and mass $q_\text{lm}$, we find notable differences, especially that some mass distributions seem to be very elongated. A quantitative comparison to lens models from the Strong Lensing Legacy Survey (SL2S) program \citep{sonnenfeld13} reveals similar differences, although there a simple SIE profile was adopted by default and an external shear component was included only when the SIE-only model led to strong residuals. Given that \citet{arneson12} observed no bias for the axis ratio with spectroscopically identified lenses, this might be an effect of the imaging selection process used in SL2S and SuGOHI. Also, several sources seem to be very elongated; twelve out of 31 have an axis ratio below 0.2 and 24 out of 31 below 0.5. This could be because of the lens search strategy from SuGOHI. Since the arcs must be bright, galaxies with higher surface area are more likely to be detected sources. These are then typically edge-on galaxies, i.e. they have very low $q$. This is in agreement with a relative low \sersic \, index; 16 out of 31 systems have $n_\text{s} < 1$. Additionally, it is known that some of these parameters are degenerate. Therefore, it might be necessary to reconsider whether more stringent prior ranges than our broad and flat priors would be better for images of ground-based resolution. For instance, there could be a Gaussian prior for the lens mass axis ratio $q_\text{lm}$ centered on the lens light axis ratio $q_\text{ll}$, or for the shear strength or the source and/or lens \sersic \, index. Although we generally consider these models to be more trustworthy than the network predictions, this demonstrates that also the models from \GG \, cannot be considered as true reference models and include some inaccuracies also from parameter degeneracies.

In the following we discuss aspects of individual lens systems that were not mentioned above.

\begin{itemize}
\item \textbf{HSCJ023322$-$020530}: This lens system has one very bright source, potentially an active galactic nucleus, which is doubly lensed. Since those two lensed images are extremely compact, this configuration is very hard to model with \GLaD, which is optimal for modeling extended sources but not point-like objects. Additionally, it seems that the PSF is not perfectly symmetric, leading to significant residuals of the point-like images. Since there is another fainter source lensed into extended arcs, we manually included here a second source at the same redshift. However, even with this second source included, visible residuals remain, resulting in a relatively high $\chi^2_\text{red}$ of 1.87. We tried several different options such as including additional profiles or relaxing assumptions on the structural parameters but obtained no notable improvement that would justify the increase in model complexity.
\item \textbf{HSCJ090429$-$010228}: The lens system HSCJ090429$-$010228 appears also like a point-source such as an active galactic nucleus. However, based on \citet{jaelani20b}, it is a compact Lyman alpha break galaxy. As mentioned already above, \GLaD \, works best for extended sources such that residuals are expected. Interestingly, we find only moderate residuals in the $i$ band but nearly no residuals in the other filters and obtain a good $\chi^2_\text{red}$ of 1.22. 
\item \textbf{HSCJ135138+002839}: This lens system has overall relatively low residuals, resulting in a good $\chi^2$. The somewhat higher reduced $\chi^2$ of 1.65 is related to the relatively large part of the image in the south-west (bottom-right) that is masked out due to luminous objects. This reduces the number of modeled pixels and thus the number of degrees of freedom. Although those masked pixel are not taken into account when computing the $\chi^2$, the reduction in the number of degrees of freedom effectively increases the $\chi^2_\text{red}$, which needs to be taken into account when comparing it to the $\chi^2_\text{red}$ of other lens systems. There are two additional areas in the image that have been masked, one on the south-east of the lens and the other on the north-west. Given the orientation, shape and color (compare Fig.~\ref{fig:colorimage}), this could be from a second source behind the lens. To confirm this, either a further multiplane model analysis, which is beyond of the scope of this work, or spectroscopic observations would be needed.
\item \textbf{HSCJ141815+015832}: Two images ($g$ and $r$ bands) of this lensing system are unfortunately slightly corrupted, which also leads to the wrongly colored stripes in the color image shown in Fig.~\ref{fig:colorimage}. Nonetheless, we modeled this system masking out the affected regions. We find that the remaining lensing information is still enough to constrain the parameter values and provide a reasonable fit, most likely as we model the different filters simultaneously and assume that they have the same structural parameters for the lens light. This was previously not obvious as the corrupted pixels belong to the most relevant filters and both masked areas go directly through the lens and arcs. 
\item \textbf{HSCJ150112+422113}: The best model of this system represents the observed structure in lens and arcs relatively well, but shows still notable residuals, both in the lens light as well as in the arcs, which results in a higher reduced $\chi^2$ of 1.63. To improve the model a flexible lens and source center across the different bands might help. 
\item \textbf{HSCJ230521$-$000211}: The final model of this lens system reproduces the observed structure in lens and arcs, but has slight differences especially also in the positions. A different lens light center for each band might improve the fit slightly but is incompatible with the uniform modeling needed for our comparison in Sect.~\ref{sec:networkmodels}.
\end{itemize}

\subsection{Comparison to other automated modeling codes}
\label{sec:mcmcmodels:comparison}

In this section we compare the main properties of the two modeling pipelines \gleeautopy \, (Sect.~\ref{sec:mcmcmodels:automatedcode}) and \gleetoolspy \, (Sect.~\ref{sec:mcmcmodels:gleetoolspy}) with those of other similar approaches from the literature. While there are no assumptions on the lensing system, prior ranges, or profiles in \gleetoolspy, \gleeautopy \, is dedicated to model galaxy-scale lenses, preferentially in ground-based data given our parameterized source SB. This is the main difference to previous investigations done for instance by \citet{nightingale18} and \citet{etherington22}, which both assumed HST image resolution, but focused also on galaxy-scale lenses. For instance, \citet{etherington22} modeled 59 high-resolution systems from the SLACS \citep{bolton06, auger10} and GALLERY \citep{shu16a, shu16b} sample fully autonomously and achieved a physically plausible fit for 54 of the 59 systems without further changes in e.g., the data preprocessing steps. Given the good data quality, they assumed a power-law mass distribution with external shear, i.e. they had an additional a slope parameter to constrain. In an attempt to accelerate the modeling, \citet{gu22} presented \gigalens, a modeling code tested on simulated HST-like lenses that still relies on a Bayesian framework but is much faster through the use of GPUs, high parallelization, and implementations in \tensorflow \, \citep{abadi15} and \JAX \, \citep{bradbury18}. In contrast to those codes for lensed galaxies, \citet{schmidt23} and \citet{ertl22} have developed both modeling pipelines that are dedicated to HST images of strongly lensed quasars to derive mass models more rapidly which is necessary to predict time delays and give a good initial model with minimal user input required for cosmological analysis. 

Besides these codes for high-resolution images, \citet{rojas22} and \citet{savary22}, who carried out dedicated galaxy-lens search programs in DES and CFIS, respectively, presented a modeling pipeline based on MCMC sampling and a particle swarm optimizer. This code was used to model their best candidates for further refinement. Their code is fully automated, whereas in our pipelines the initial preparation of input files such as the masks and lens/arcs identification is still manual. The advantage of a fully autonomous procedure is clearly the further speed up and also the applicability to much larger samples (hundreds to thousands of lenses if the run time permits). The drawback is the risk of further inaccuracies and miss-identifications for some models, as already pointed out by \citet{rojas22} and \citet{savary22}. Therefore, we prefer a non-fully automated procedure given our sample size of 31 lenses and the aim to obtain very good models for the comparison to the neural network modeling. But even beyond this immediate objective, \gleeautopy \, will be very helpful for future modeling as it is perfectly suited for observations from LSST. In addition, \gleetoolspy \, will be useful for further refinement, as already in this work, independent of the image quality or type of lensing system.

\subsection{Comparison to previously published models}
\label{sec:mcmcmodels:comparisonSonnenfeld}

\begin{figure}[ht!]
\begin{center}
\includegraphics[trim=0 0 0 0, clip, width=0.85\columnwidth]{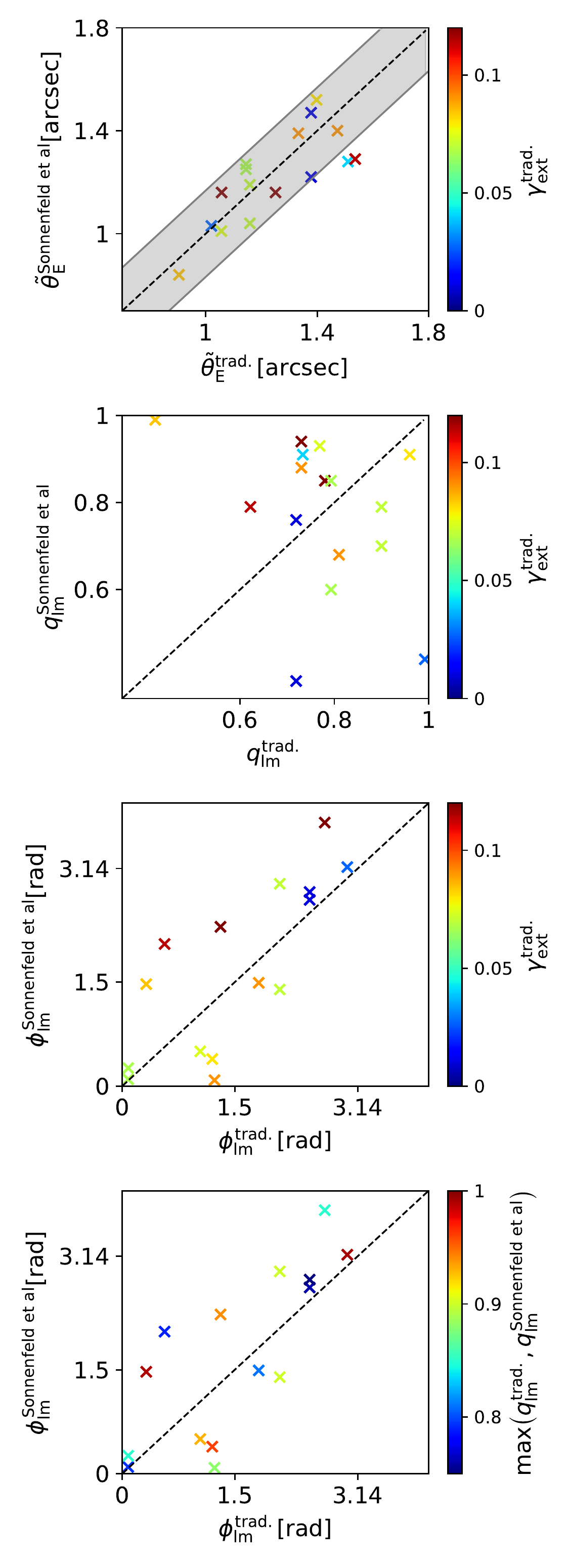}
\caption{Comparison of the SIE parameter values obtained with \GG \, using SIE+$\gamma_\text{ext}$ to those values from \citet{sonnenfeld13, sonnenfeld19} adopting mainly SIE-only. The gray shaded region in the first panel indicates the 1 pixel range. The discrepancy is mostly due to the difference in the adopted mass model in this work and in \citet{sonnenfeld13, sonnenfeld19}.\label{fig:Sonnenfeld_comparison}}
\end{center}
\end{figure}

As mentioned in the beginning of this section, 17 of our lensing systems were already modeled by \citet{sonnenfeld13, sonnenfeld19} adopting an SIE profile for describing the lens mass distribution, and a single De Vaucouleurs profile \citep{DeVaucouleurs48} or a single \sersic \, profile \citep{sersic63} for describing the lens light distribution. Despite these differences, we compare briefly the values obtained for the Einstein radius $\tilde{\theta}_\text{E}$, the lens-mass axis ratio $q_\text{lm}$, and the position angle $\phi_\text{lm}$ in Fig.~\ref{fig:Sonnenfeld_comparison}. Here we adopt the parameterization of \citet{sonnenfeld13, sonnenfeld19}, i.e. we convert our Einstein radius $\theta_\text{E}$ to
\be
\tilde{\theta}_\text{E} = \frac{2\sqrt{q_\text{lm}}}{1+q_\text{lm}} \theta_\text{E}
\ee
as this leads to an Einstein radius that should be more independent of the axis ratio. In this comparison, we exclude the lens system {HSCJ094427$-$014742} with its very faint counter image, as \citet{sonnenfeld19} adopted here the lens mass model parameter values measured by \citet{brownstein12} using HST data since it was not possible for them to obtain a robust model with HSC.


In the comparison panels in Fig.~\ref{fig:Sonnenfeld_comparison}, we color code the points by either the external shear strength or the axis ratio as possible reasons for the differences between our results and that of \citet{sonnenfeld13, sonnenfeld19}.  In particular, we color code the first three panels in Fig.~\ref{fig:Sonnenfeld_comparison} by the external shear strength $\gamma_\text{ext}$ since a higher external shear in our models could explain differences in the reconstructed parameters. Furthermore, since an axis ratio $q_\text{lm}\sim1$ leads to no constraints on the position angle, we show the comparison of the position angle, in the fourth panel, color-coded by the axis ratio $q_\text{lm}$ either from \GG \, or \citet{sonnenfeld13, sonnenfeld19}, whichever is higher. We obtain a median offset of 0.097\arcsec in the Einstein radius (corresponding to 0.58 pixels) while the largest difference is 0.28\arcsec (1.7 pixels) for a system with a high external shear of around 0.1. For better comparison to the pixel size, we show a gray band corresponding to the range of 1 pixel (i.e. $0.168\arcsec$). The axis ratio shows stronger differences (median offset 0.16, highest offset 0.57). Interestingly, the systems with the largest discrepancies do not have high external shear. For the position angle we obtain a median offset of 0.76 radians and the highest difference is 1.49 radians (corresponding to 43.5$^\circ$ and 85.1$^\circ$, respectively), but we see a stronger correlation to the external shear and axis ratio that explains the larger differences. In other words, if $\gamma_\text{ext} \leq 0.05$ and $q_\text{lm} \leq 0.85$, the position angle matches very well with a median offset of 0.25 radians (14$^\circ$) and a maximum of 0.30 radians (17.3$^\circ$).

All in all, these differences demonstrate that the lens mass model parameter values depend to some extent on the modeling assumptions, such as the mass and light profiles and relations between mass and light.  Parameter degeneracies, such as that between the external shear and axis ratio, also affect the resulting parameter constraints.  We therefore should keep in mind these scatters in the parameter values obtained with \GG \, for our model parameter comparison in Sec.~\ref{sec:comparison} between the traditional and neural network modeling results. Nonetheless, the Einstein radius is overall well recovered within $\sim$$0.1\arcsec$, irrespective of differences in modeling assumptions.

\FloatBarrier
\section{Mass models predicted by the neural network}
\label{sec:networkmodels}

In \citetalias{schuldt23a}, we present a ResNet to model galaxy-scale lens images of HSC quality. It was trained on simulated images using real HSC LRGs as lens images and galaxies from the Hubble ultra deep field (HUDF) as background sources. The lens redshift, peaking at $z \sim 0.5$, and velocity dispersion, ranging from $\sim 100 \, \text{km\,s}^{-1}$ to $\sim 500 \, \text{km\,s}^{-1}$ and peaking at $\sim 280 \, \text{km\,s}^{-1}$, are taken from SDSS, while the redshifts for the sources are directly provided by HUDF. Details on the simulation procedure and network training are in \citetalias{schuldt21a} and \citetalias{schuldt23a}. We now apply this network to our sample of 31 known real lenses. Within few seconds, the network predicts the full set of parameter values with corresponding 1$\sigma$ uncertainties for all lenses in the sample. This set of parameters includes the SIE mass parameters, namely the lens mass center $x_\text{l}$ and $y_\text{l}$, the ellipticity $e_\text{x}$ and $e_\text{y}$, and the Einstein radius $\theta_\text{E}$. The complex ellipticity of the lens mass can be converted into the axis ratio
\be
q_\text{lm} = \sqrt{ \frac{1-\sqrt{e_\text{x}^2 + e_\text{y}^2} }{ 1 + \sqrt{e_\text{x}^2 + e_\text{y}^2} } }
\label{eq:qmass}
\ee
and into the position angle
\be
\phi_\text{lm} = \left\{ \begin{array}{ccc} f & \text{if } e_\text{x} \geq 0 \text{ and } e_\text{y} \geq 0\\
  f + \pi & \text{if } e_\text{x} \geq 0 \text{ and } e_\text{y} < 0\\
  \frac{1}{2}\pi-f  & \text{if } e_\text{x} < 0 \end{array} \right.
\label{eq:PA}
\ee
with
\be
f = \frac{1}{2} \arcsin \left( e_\text{y} \frac{1+q_\text{lm}^2}{1-q_\text{lm}^2} \right)
\ee
and
\be
f \in [-\pi/4, +\pi/4].
\ee

The network further predicts the external shear $\gamma_\text{ext,1}$ and $\gamma_\text{ext,2}$, which can be translated into a shear strength
\be
\gamma_\text{ext} = \sqrt{ \gamma_\text{ext,1}^2 + \gamma_\text{ext,2}^2 }
\label{eq:gamext}
\ee
that is rotated by
\be
\phi_\text{ext} = \left\{\begin{array}{cccc} s & \text{if } \gamma_\text{ext,1} \geq 0 \text{ and } \gamma_\text{ext,2} \geq 0\\ \frac{1}{2} \pi - s & \text{if } \gamma_\text{ext,1} < 0 \text{ and } \gamma_\text{ext,2} \geq 0 \\ \frac{\pi}{2}+s & \text{if } \gamma_\text{ext,1} < 0 \text{ and } \gamma_\text{ext,2} < 0 \\ \pi-s & \text{if } \gamma_\text{ext,1} \geq 0 \text{ and } \gamma_\text{ext,2} < 0 \end{array} \right. \
\label{eq:PAext}
\ee
with
\be
s = \frac{1}{2} \arcsin \left( \frac{ |\gamma_\text{ext,2}|}{\gamma_\text{ext} } \right) .
\ee
We report in Tab.~\ref{tab:comparison_normal} the network-predicted values and uncertainties, which we converted to the parameterization of \GG \, (i.e. $\gamma_\text{ext}$, $\phi_\text{ext}$, $q_\text{lm}$, $\phi_\text{lm}$). We provide the values in complex notation, as directly obtained from the network, in  Tab.~\ref{tab:comparison_complex} as well, where we also include the converted values obtained with \GG. While the median values are directly convertible through Eq.~(\ref{eq:gamext}) and Eq.~(\ref{eq:PAext}), this is not straightforward for the uncertainties. Therefore, we implemented in \gleetoolspy \, the option to convert values with Gaussian uncertainties. To this end, a sample of values is internally generated based on the given median and $\sigma$ width, which is then converted into the complex notation. From that new, converted sample, the median and 1$\sigma$ values are computed. Given the constraint $e_\text{x}^2 + e_\text{y}^2 \leq 1$ to obtain physically possible values in Eq.~(\ref{eq:qmass}), we exclude all nonphysical values, if any, from the sample. Because of the conversion, the uncertainties are no longer symmetric about the median value and are thus reported individually. 

\FloatBarrier
\section{Comparison and discussion}
\label{sec:comparison}

\begin{figure*}[ht!]
\begin{center}
\includegraphics[trim=0 1150 0 0, clip, width=0.46\textwidth]{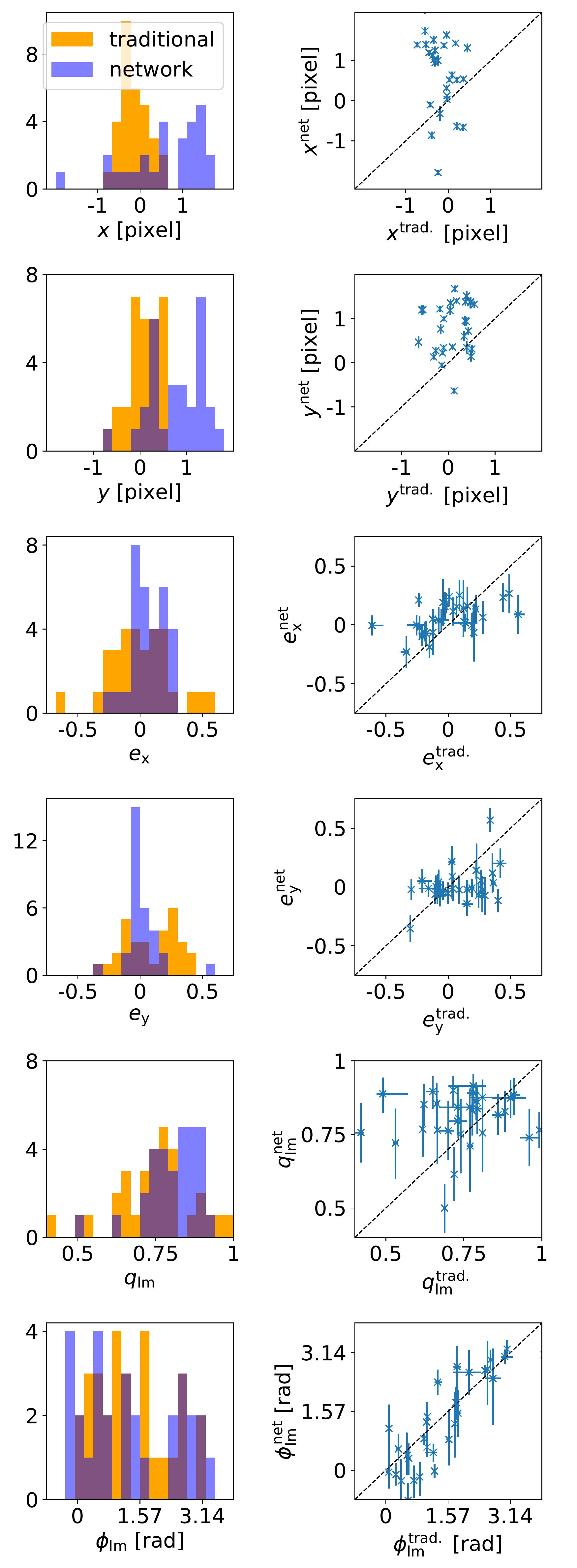}\includegraphics[trim=0 0 0 1370, clip, width=0.46\textwidth]{plots/comparison_plot_SuGOHI.pdf}
\caption{Comparison of the SIE+$\gamma_\text{ext}$ values obtained with the traditional \GG \, method (orange histogram) and our ResNet (blue histogram). We further include a comparison where the \GG \, values are plotted against the ResNet values.\label{fig:SuGOHI_comparison}}
\end{center}
\end{figure*}

\begin{figure}[ht!]
\begin{center}
\includegraphics[trim=0 1010 0 0, clip, width=0.215\textwidth]{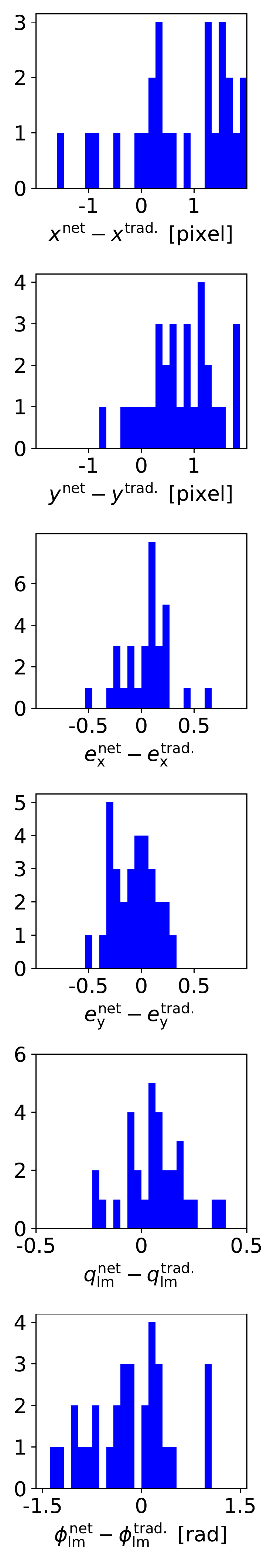}\includegraphics[trim=0 0 0 1220, clip, width=0.215\textwidth]{plots/comparison_plot_SuGOHI_difference.pdf}
\caption{Difference of the SIE+$\gamma_\text{ext}$ values obtained with the traditional method using \GG \, and our ResNet.\label{fig:SuGOHI_comparison_difference}}
\end{center}
\end{figure}

After modeling our lens sample in the traditional way with \GG \, and with our ResNet from \citetalias{schuldt23a}, we compare directly the obtained SIE+$\gamma_\text{ext}$ parameter values. Fig.~\ref{fig:SuGOHI_comparison} shows them as histograms (left) and plotted against each other (right), with the traditional obtained values on the $x$-axis and the network predictions on the $y$-axis. We further show the difference between the traditional and network-based values as histograms in Fig.~\ref{fig:SuGOHI_comparison_difference}. 

As we can see from Fig.~\ref{fig:SuGOHI_comparison}, the Einstein radius is very well determined through the traditional procedure because the image positions used at the beginning already constrained the Einstein radius very well and then further refined through the extended image modeling. Also the ResNet performs overall well on our comparison sample (see also Fig.~\ref{fig:SuGOHI_comparison_difference}). For $\theta_\text{E} \lesssim2$, we find overall a good match between both methods, although a few are not within the 1$\sigma$ range. In agreement with \citetalias{schuldt23a}, the network underpredicts the Einstein radius for system HSCJ150021$-$004936 with $\theta_\text{E}=1.8\arcsec$, notably lower than the $\theta_\text{E}=3.1\arcsec$ obtained with \GG. Since the number of systems with $\theta_\text{E} \geq 2\arcsec$ is significantly lower than the number of systems with $\theta_\text{E} \leq 2\arcsec$  in the training set, the network shows a bias towards lower separations on the test set.



The coordinates of the lens mass center $x_\text{l}$ and $y_\text{l}$ are very well constrained by both methods but we observe some differences between these two methods. The traditional modeling predicts a lens center very close to the image center, i.e. within $\pm 1$ pixel. Here we have to remember that we assume the light center to be coincident with the lens mass center. Since the lens light has a relatively large influence on the $\chi^2$ and thus on the lens center, the predicted value will be highly influenced by the lens light. A possible offset to the true mass center can be compensated through a change in the external shear. This could be a reason why the network predicts for several systems a larger offset to the image center. The largest offset is $0.484\arcsec$ for lens HSCJ021737$-$051329, corresponding to nearly three pixels. The fact that we can model nearly all lenses with \GG \, by assuming a coincident lens light and mass center, implies that we could also adopt this assumption when generating our network training data. Moreover, if we assume the traditionally obtained value to be more accurate, a lens-center offset of $\pm 1$ pixel instead of the currently used $\pm 3$ pixels would be enough when creating the mocks. This could simplify the task for the network and thus increase the performance on the lens center and also on the other parameters. On the other hand, for most of the lens systems, a slight offset with respect to the cutout center were found with the traditional procedure. Therefore, depending on the science goal, it can be important to include a variable lens center and to predict all five parameters of the SIE profile, instead of assuming that the lens light and mass center fall exactly on the cutout center and thus predicting only three parameters (ellipticity and Einstein radius) as done in other modeling networks \citep[e.g.,][]{hezaveh17, levasseur17, pearson19, pearson21}.

The ellipticity shows better agreement between traditional and network-based modeling than the lens centroids. In general, the network predicts values closer to zero than the traditional modeling, which was expected from the performance on the test data (compare \citetalias{schuldt23a}). This is most likely the result of having nearly two orders of magnitude more systems with ellipticity $\sim 0$ than $\sim \pm 0.5$ in the training sample. A further possible reason is that SuGOHI lenses tend to have more elliptical mass distributions than the training set in \citetalias{schuldt23a}, since the inner caustic covers a larger area when the mass distribution is more elliptical, leading to a higher magnification that makes the elliptical lenses easier to detect. Because the ellipticity in our training data is set by real observations of LRGs, a flatter distribution, which would most likely lead to an improvement on the currently underrepresented values, is difficult to achieve.

Finally, the external shear is very difficult to estimate. This is especially true for the ResNet, but also the traditionally obtained error bars for are relatively large, especially $\gamma_\text{ext,1}$, indicating the difficulty to constrain that parameter. Nonetheless, it is interesting to see that the shear orientation is roughly recovered, even if the network predicts relatively large uncertainties.

\begin{figure}[ht!]
\begin{center}
  \includegraphics[trim=0 0 0 0, clip, width=\columnwidth]{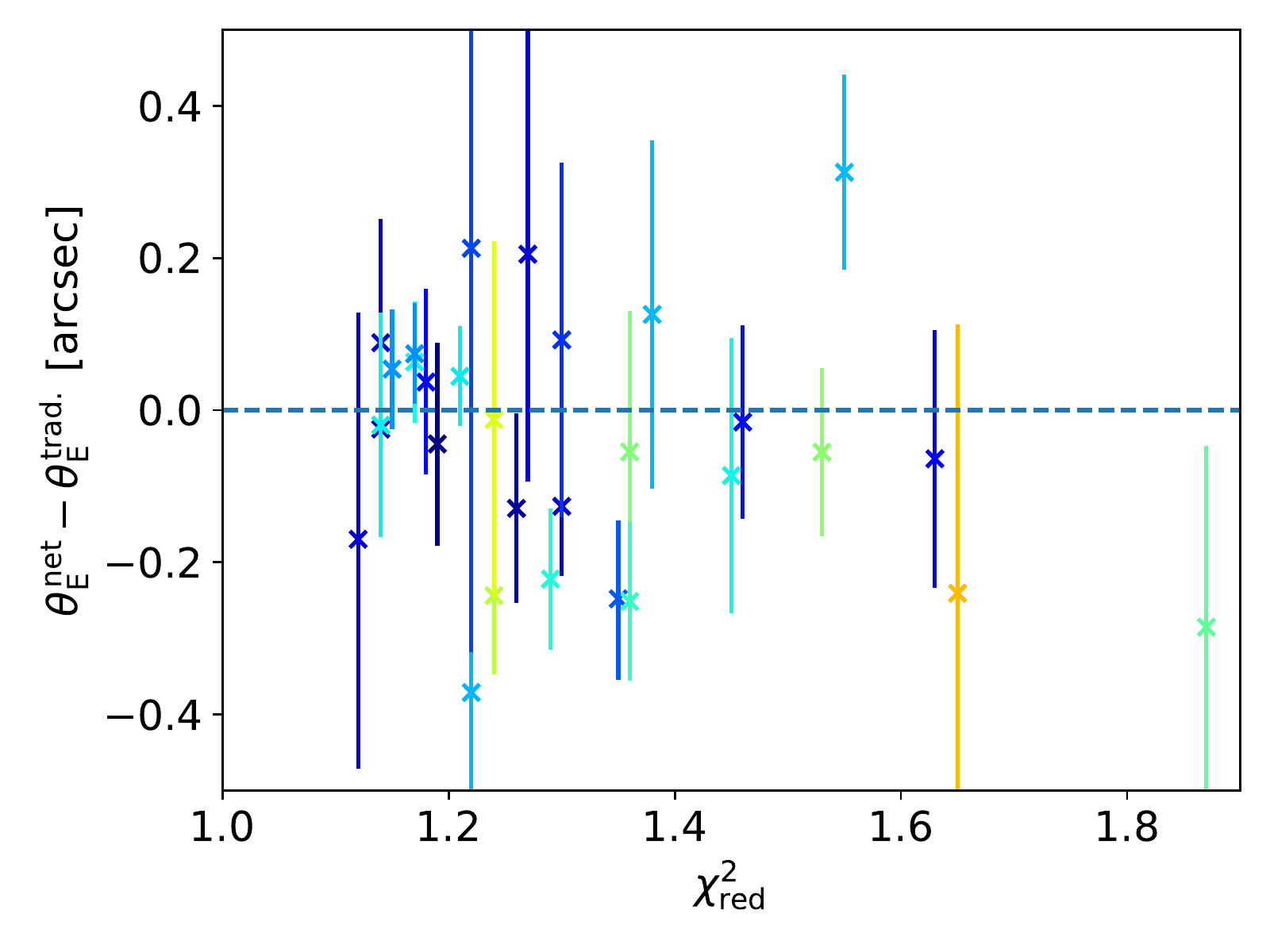}
  \includegraphics[trim=0 0 0 0, clip, width=\columnwidth]{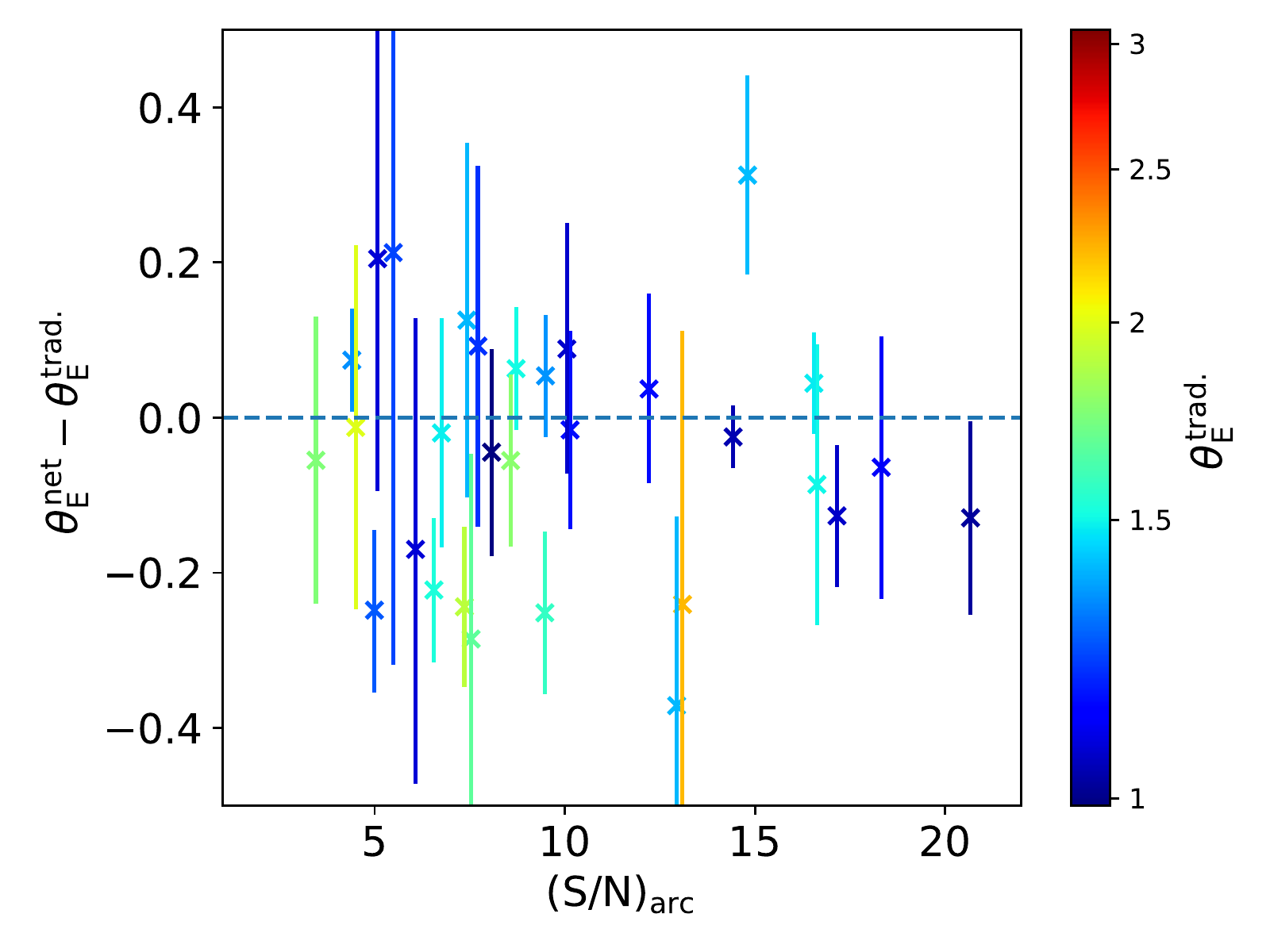}
\caption{Dependency of the Einstein radius difference on the $\chi^2_\text{red}$ obtained with \GG \, (top panel) and the S/N of the arcs (bottom panel).\label{fig:SuGOHI_comparison_chi2}}
\end{center}
\end{figure}

While the conventional method is considered to give more accurate estimations in general, some of the models show significant residuals in their fit as can be seen from Figs.~\ref{fig:model_47}~-~\ref{fig:model_1976}, resulting in a higher $\chi^2_\text{red}$ as noted in Sect.~\ref{sec:mcmcmodels:results}. We therefore compare the $\theta_\text{E}$ inference as a function of $\chi^2_\text{red}$ in the top panel of Fig.~\ref{fig:SuGOHI_comparison_chi2}. We do not see a direct correlation between the overall accuracy of the \GG \, model and the agreement of the Einstein radius between the traditional and the network-based approach. On the other hand, we find a small correlation between the signal-to-noise ratio (S/N) of the arcs and the Einstein radius (see Fig.~\ref{fig:SuGOHI_comparison_chi2}, bottom panel, especially for systems with $\theta_\text{E}^\text{trad}\sim1\arcsec$) or the S/N and the complex ellipticity. This is not too surprising, as systems with higher S/N have arcs that are more prominent and thus both methods can better constrain the parameters. This fact could be used in the future as an additional criterion to probe the trustworthiness of the predicted parameter values, e.g. with a limit of (S/N)$_\text{arc} > 10$.

Apart from these general trends, we note again that HSCJ023322$-$020530 was difficult to model in the automated, uniform way as it shows two very bright objects, for which we had to adopt a third \sersic \, profile for the underlying extended background source to describe the observation with acceptable accuracy. Moreover, the extended, relatively faint arcs seem to originate from another source as the one giving the bright images, such that we modeled those arcs with a separate \sersic \, component. Nonetheless, the modeling resulted in visible residuals (see Fig.~\ref{fig:model_359}) and a higher $\chi^2_\text{red}$ as noted above.

Special consideration is also required for HSCJ141815+015832, as it has corrupted data, which we tried to avoid in the training of the network. As noted in Sect.~\ref{sec:mcmcmodels:results}, the traditional modeling worked quite well regardless of the missing data. Surprisingly this holds also for the network, although only the position angle and the shear orientation are within 1 $\sigma$. The lens center is quite off with a difference of $\sim 2.5$ pixels in the $x$ direction, which can be due to the stripe artifact along the $x$-axis in the $g$ and $r$ bands, falling in the $r$ band directly on top of the lens center. The axis ratio and also the Einstein radius are both well recovered without larger uncertainties than in other systems. This demonstrates that our network is able to handle even such cases which it was not trained on.

\FloatBarrier
\section{Image position and time delay comparison}
\label{sec:ImPoscomparison}

Since the computational time of the network is extremely low, it would be perfectly suited to predict the next appearing image(s) and corresponding time delay(s)\footnote{Hereafter we use always the plural for the image position(s), time delay(s), and Fermat potential difference(s) for better readability of the text.} for a supernova in the background galaxy. In \citetalias{schuldt21a} and \citetalias{schuldt23a}, we therefore included a comparison based on the simulated test data set to see whether the precision of the network is sufficient. For completeness, we now also compute the image positions and time delays for our 31 SuGOHI lenses. Explicitly, we use the mass model and source position $\vec{\beta}^\text{trad.} = (x_\text{s}^\text{trad.}, y_\text{s}^\text{trad.})$ obtained with \GG\ to predict the image positions $\vec{\theta}^\text{trad.}$, the corresponding Fermat potential differences $\Delta \tau^\text{trad.}$ and time delays $\Delta t^\text{trad.}$ defined as
\be
\Delta t_{jk} = \frac{D_{\Delta t}}{c} \Delta \tau_{jk}
\ee
with the Fermat potential
\be
\tau = \frac{(\vec{\theta} - \vec{\beta})^2}{2} - \Psi(\vec{\theta}) \, .
\ee
Here $c$ is the speed of light, $\Psi$ the lens potential, and $D_{\Delta t}$ the time-delay distance which depends on the lens redshift $z_\text{d}$ and source redshift $z_\text{s}$, where we assume the redshifts noted in Tab.~\ref{tab:overviewLenssample} or $z_\text{s} \equiv 1$ if the source redshift is unknown. For simplicity, we assume the supernova event is located directly in the center of the source galaxy when calculating their corresponding image positions based on the \GG\ model. For the time-delay and image-position predictions based on the ResNet, we replace the mass model of \GG \, by that of the neural network and predict the source position $\vec{\beta}^\text{net}$ using the position of the first appearing image predicted by \GG. This means we obtain a coincident first appearing image, i.e. $ (\theta_\text{x,A}^\text{trad.}, \theta_\text{y,A}^\text{trad.}) \equiv (\theta_\text{x,A}^\text{net}, \theta_\text{y,A}^\text{net})$, which would be the observed image position. From the obtained source position $\vec{\beta}^\text{net}$, we can then predict with the ResNet mass model the other image positions $\vec{\theta}^\text{net}$, Fermat potential differences $\Delta \tau ^\text{net}$ and time delays $\Delta t^\text{net}$. Since the image multiplicity depends on the source position and mass model, the number of images can differ between the network and \GG \, prediction. When the predicted number of image positions $N$ matches, the obtained image positions are sorted in order to minimize 
\be
d = \Sigma_{i=1}^N \frac{\sqrt{\left( \theta_{\text{x},i}^\text{trad.}-\theta_{\text{x},i}^\text{net} \right) ^2 + \left( \theta_{\text{y},i}^\text{trad.}-\theta_{\text{y},i}^\text{net} \right) ^2 } }{N}
\label{eq:d}
\ee
to facilitate a direct comparison.

\begin{figure}[ht!]
\begin{center}
\includegraphics[trim=0 0 0 0, clip, width=0.5\textwidth]{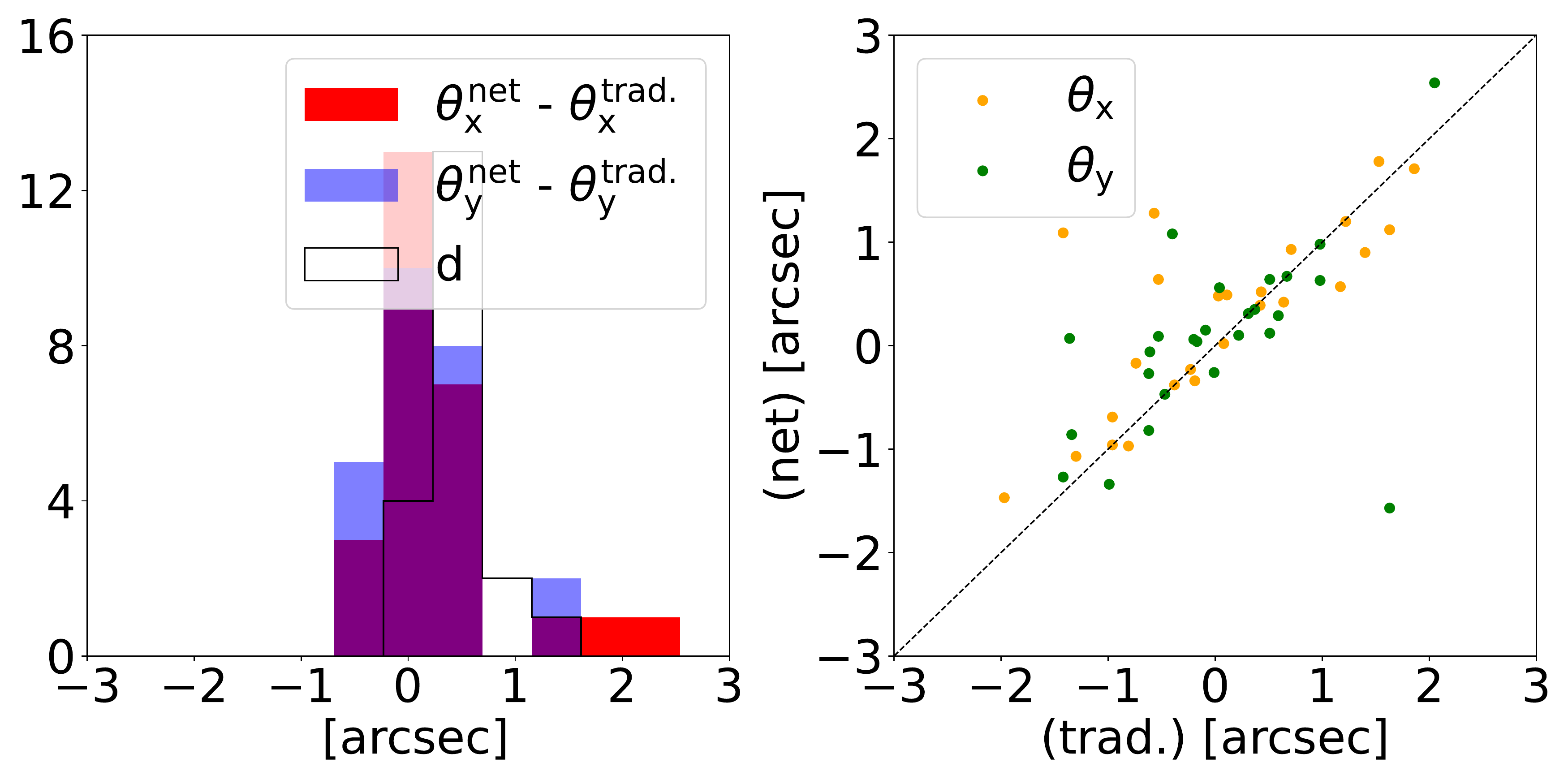}
\caption{Comparison of the next appearing image positions $(\theta_\text{x}, \theta_\text{y})$ predicted by adopting the traditional \GG \, model and the mass model of the ResNet in case of matching image mutliplicity. The first image A of each lensing system is excluded because of our assumption of a coincident first image.\label{fig:SuGOHI_comparison_ImPos}}
\end{center}
\end{figure}

The obtained image positions and source positions are reported in Tab.~\ref{tab:comparison_ImPos}. The Fermat potential differences and time delays are sorted accordingly and reported in Tab.~\ref{tab:comparison_Fermat}. When comparing the different quantities for our sample, we unfortunately find that the scatter is slightly larger than on the test set in \citetalias{schuldt21a} and \citetalias{schuldt23a}. Fig.~\ref{fig:SuGOHI_comparison_ImPos} shows the differences of the obtained image positions and the obtained values $d$ for visualization. For most of the lens systems, we find significant differences in the predicted time delays between the traditional and ResNet models. Therefore, the network prediction on the basis of ground-based data is unfortunately not good enough for accurate time-delay predictions. However, it could give a good starting model for further optimization with \GG, e.g. in a fully automated way, as it could replace the creation of the region file marking lens position and ellipticity (compare Sect.~\ref{sec:mcmcmodels:automatedcode}, step 1 in \gleeautopy). We note that the values obtained with \GG \, are more accurate and precise, but not necessarily the true values (like in test set) and also have inaccuracies given the obtained model parameter value uncertainties. Nonetheless, we consider the image positions and time delays inferred from our \GG \, model as more trustworthy. 

\FloatBarrier
\section{Summary and conclusion}
\label{sec:conclusion}

In this paper, we compared the predictions of the residual neural network presented in \citetalias{schuldt23a} to the SIE+$\gamma_\text{ext}$ parameter values obtained through modeling with \GG. For this comparison, we selected known galaxy-scale lenses detected in HSC, as the network is trained for these kind of lenses and for this image quality. This resulted in a sample of 31 grade-A lenses, which we presented in Sect.~\ref{sec:dataset}.

We modeled the full sample of 31 lenses with \GG, a software based on Bayesian optimization algorithms such as simulated annealing and MCMC sampling and thus referred to as traditional, non-machine learning technique. Because of the iterative sampling, this procedure is very time-and resource-consuming. To minimize the user input, we automated most of the modeling steps and developed \gleeautopy, a dedicated procedure to model galaxy-scale strong-lensing systems in optionally multiple filters simultaneously. The code autonomously fits the lens light with \sersic\ profiles, before including an SIE+$\gamma_\text{ext}$ profile to describe the lens mass distribution. After a quick optimization based on the user-identified image positions, the code performs a source SB reconstruction by fitting to the full image cutout. Because \gleeautopy \, is specifically optimized for ground-based observations like those in our comparison sample, we adopted a parameterized source SB reconstruction rather than a pixelated reconstruction. This means we adopt one \sersic \, profile to describe the light distribution of the background source.

Since each lens with its environment is unique, the presented uniform modeling sequence did not produce a good fit for all lenses. We therefore further refined some of the models manually. To this end, we developed \gleetoolspy, a software package that accepts any configuration file for \GG \, and a list of optimization algorithms. The specified optimization steps are then performed one after the other without any further input of the user. Since \gleetoolspy \, has no criteria on the modeling procedure incorporated, it has a very broad applicability. With the two codes, we were able to model all 31 lens systems with satisfactory quality and in an acceptable amount of time.

For the comparison, we applied our trained network to the same sample of lenses. We find very good agreement with the traditional models for the Einstein radiusalthough differences appear for systems with larger Einstein radii ($\theta_\text{E}\gtrsim 2\arcsec$, i.e. HSCJ150021$-$004936. This is expected given the performance on the test set (compare \citetalias{schuldt23a}) because of the under-representation of these systems in the training set. 
The predicted Einstein radius from the traditional modeling is comparatively well constrained, which comes at least partly from using our visually identified image positions as constraints to get a first estimate.

For the lens center, all values predicted through the traditional modeling procedure are within $\pm1$ pixel with respect to the cutout center, while the network predicts larger offsets for some systems. This can be explained by our assumption of a coincidence between lens light and mass center for the traditional modeling on the one hand and the rather generous $\pm 3$ pixel shift adopted when generating the training data for our network on the other hand. The ellipticity is relatively well constrained by both techniques, but the network tends to predict a more spherical mass distribution (i.e., values closer to zero in complex parameterization) compared to \GG. This is in agreement with the network performance on the test set in \citetalias{schuldt23a}, and a result of a realistic, but non-uniform distribution in the training sample. For future networks, it might thus make sense to enforce a more evenly distributed ellipticity in the training set, which could be achieved through data augmentation of more elliptical lenses and/or limiting the number of rounder systems. Finally, as expected, the external shear is not well predicted by the network, resulting in the prediction of large uncertainties.

All in all, the performance of the network is very good, especially when taking into account the minimal user input and extremely low computational time. We were able to predict all seven SIE+$\gamma_\text{ext}$ values for the full sample within a fraction of a minute, while the traditional modeling, even with our automated code, requires a few days per lens in addition to possible follow-up modeling. We confirm with our comparison that the network performs similarly well on real lenses as it does on the test set. This demonstrates that the mock images in \citetalias{schuldt23a} are indeed realistic and that we can expect a similar performance on a large sample of hundreds to thousands of systems, which can be modeled easily with our network. This would allow a detailed statistical analysis of lens mass properties, especially for systems with $\theta_\text{E} \gtrsim 1.5\arcsec$ or with S/N in the arcs above $\sim$$10$. In contrast to that, we are able to model a sample of dozens of lenses with our automated traditional pipeline to better accuracy and we can also evaluate the quality of the fit in terms of a $\chi^2$, which is not possible for the network output. The \gleetoolspy \, code enables us to further refine the models obtained with our fully automated procedure or also other dedicated automated modeling codes \citep[e.g.,][]{hezaveh17, levasseur17, nightingale18, nightingale21a, nightingale21b, pearson19, pearson21, adam22, ertl22, etherington22,  schmidt23}. The combination of all three codes enables us to handle different sample sizes of lenses, and thus takes us a huge step forward in handling the newly detected lenses in current 
and upcoming wide-field imaging surveys such as LSST and Euclid.

\FloatBarrier
\begin{acknowledgements}
We thank J. H. H. Chan for good discussions and helpful comments. We further thank the referee A.~Sonnenfeld for the constructive and helpful comments. SS, SHS, RC and SE thank the Max Planck Society for support through the Max Planck Research Group for SHS. This project has received funding from the European Research Council (ERC) under the European Unions Horizon 2020 research and innovation programme (LENSNOVA: grant agreement No 771776).
This research is supported in part by the Excellence Cluster ORIGINS which is funded by the Deutsche Forschungsgemeinschaft (DFG, German Research Foundation) under Germany's Excellence Strategy -- EXC-2094 -- 390783311. YS acknowledges support from the Alexander von Humboldt Foundation in the framework of the Max Planck-Humboldt Research Award endowed by the Federal Ministry of Education and Research.
\\
The Hyper Suprime-Cam (HSC) collaboration includes the astronomical communities of Japan and Taiwan, and Princeton University. The HSC instrumentation and software were developed by the National Astronomical Observatory of Japan (NAOJ), the Kavli Institute for the Physics and Mathematics of the Universe (Kavli IPMU), the University of Tokyo, the High Energy Accelerator Research Organization (KEK), the Academia Sinica Institute for Astronomy and Astrophysics in Taiwan (ASIAA), and Princeton University. Funding was contributed by the FIRST program from Japanese Cabinet Office, the Ministry of Education, Culture, Sports, Science and Technology (MEXT), the Japan Society for the Promotion of Science (JSPS), Japan Science and Technology Agency (JST), the Toray Science Foundation, NAOJ, Kavli IPMU, KEK, ASIAA, and Princeton University. This paper makes use of software developed for the Rubin Observatory Legacy Survey in Space and Time (LSST). We thank the LSST Project for making their code available as free software at http://dm.lsst.org This paper is based in part on data collected at the Subaru Telescope and retrieved from the HSC data archive system, which is operated by Subaru Telescope and Astronomy Data Center (ADC) at National Astronomical Observatory of Japan. Data analysis was in part carried out with the cooperation of Center for Computational Astrophysics (CfCA), National Astronomical Observatory of Japan. We make partly use of the data collected at the Subaru Telescope and retrieved from the HSC data archive system, which is operated by Subaru Telescope and Astronomy Data Center at National Astronomical Observatory of Japan.\\
Software Citations:
This work uses the following software packages:
\href{https://github.com/astropy/astropy}{\texttt{Astropy}}
\citep{astropy1, astropy2},
\href{https://github.com/matplotlib/matplotlib}{\texttt{matplotlib}}
\citep{matplotlib},
\href{https://github.com/numpy/numpy}{\texttt{NumPy}}
\citep{numpy1, numpy2},
\href{https://www.python.org/}{\texttt{Python}}
\citep{python},
\href{https://github.com/scipy/scipy}{\texttt{Scipy}}
\citep{scipy},
\href{https://pytorch.org}{\texttt{torch}}
\citep{torch}.

\end{acknowledgements}

\bibliographystyle{aa}
\bibliography{ModelComparison}

\FloatBarrier
\appendix

\section{Detailed description of \gleeautopy}
\label{app:modelprocedure}

In the following we describe in detail the model sequence of \gleeautopy, our automated modeling pipeline for galaxy-scale lenses introduced in Sect.~\ref{sec:mcmcmodels}. 
Because of the relatively long run-time of the last part, \gleeautopy \, can be re-started from several different steps indicated with $m$, for instance if the code gets aborted or the model needs to be refined. 

\begin{itemize}
\item[1.] \textbf{Preparation of input files}
  \begin{itemize}
  \item Creation of lens and arc masks, shown in Fig.~\ref{fig:flowchart} as the top-left and top-middle insert of the corresponding box. These masks specify the region to be modeled and can be different for each filter.
  \item Creation of a region file with the ds9 software\footnote{https://sites.google.com/cfa.harvard.edu/saoimageds9} \citep{joye03} as shown in the top-right insert of the corresponding box in Fig.~\ref{fig:flowchart}. Here one specifies the cutouts size, the lens center and the lens ellipticity, the image positions, and if needed a region to subtract the image background $\sigma_\text{bkgr}$.
  \item Renaming of all files according to the assumptions of the modeling code displayed on the bottom of each insert. Both, the alphanumerical ID specifying the lensing system, which is 42 in our example, and a name to distinguish between the filters, which is $R$ in our example, are chosen by the user.
  \end{itemize}
\item[2.] \textbf{Lens light modeling} with \GLEE
  \begin{itemize}
  \item Read in the provided files, crop the lens image and error map, subtract the background if specified and save the new image and error map cutout to disk.
  \item The code now creates automatically the initial \GLEE \, configuration file for the first filter. The starting values for the lens center and ellipticity are determined from the region file provided. At this stage one \sersic \, profile with the following parameters and prior ranges is adopted: lens light center coordinates $x_\text{l} \in [x_{\text{l},i} -2\arcsec, x_{\text{l},i} +2\arcsec]$ and $y_\text{l} \in [y_{\text{l},i} -2\arcsec, y_{\text{l},i} +2\arcsec]$, axis ratio $q_\text{ll} \in [0.3,1] $, position angle $\phi_\text{ll} \in [-\pi,+\pi]$, the amplitude $A_\text{ll} \in [0,100]$, the effective radius $r_\text{eff,ll} \in [0.01\arcsec,10\arcsec]$, and the \sersic \, index $n_\text{ll} \in [1,5]$. The initial starting values ($x_{\text{l},i}$, $y_{\text{l},i}$, $q_{\text{ll},i}$, $\phi_{\text{ll},i}$, $A_{\text{ll},i}$, $r_{\text{eff},i}$, $n_{\text{ll},i}$) are extracted from the region file created by the user in step one.
  \item Since the \sersic \, amplitude $A_\text{ll}$ is not known a priori, the code evaluates automatically each order of magnitude between $10^{-5}$ and $10^5$ and selects the correct order of magnitude defined through the minimal $\chi^2$. The code then updates the upper limit of the prior range to 100 times the estimated amplitude. If the new upper parameter range limit is lower than 10, it is set to 10. In this work, we refer to that procedure as amplitude test.
  \item If the reduced $\chi^2$, $\chi^2_\text{red}$, is above 2, optimize the model by running
    \begin{itemize}
    \item[$-$] a simulated annealing optimization and 
    \item[$-$] then an MCMC chain to use the best model of the chain as new model parameters and to include a covariance matrix in the next optimization sequence. 
    \end{itemize}
  Redo both optimization steps until the MCMC chain passes the criterion of $\Delta \log P \leq 20$ where $P$ is the likelihood probability of the corresponding MCMC chain.
  \item Add now iteratively the other filters in the order specified by the user. Assume the same structural parameters across the different filters, which means only the amplitude is added as a new parameter for each filter. 
    \begin{itemize}
    \item[$-$] For each new filter run first an amplitude test as described above, and 
    \item[$-$] then a simulated annealing optimization,
    \item[$-$] followed by an MCMC chain to continue with the best set of parameter values of that chain.
    \end{itemize}
  \item After adding all filters and if $\chi^2_\text{red}>1$, optimize all filters simultaneously further by alternating between 
    \begin{itemize}
    \item[$-$] an MCMC run and 
    \item[$-$] a simulated annealing run 
    \end{itemize}
  until $\Delta \log P \leq 5$.
  \item If still $\chi^2_\text{red}>1$, which is normally the case, the code adds a second but concentric \sersic \, profile for each filter, assuming again the same structural parameters across the different filters. 
  \item Determine again the order of magnitude of each new amplitude and set the prior range as specified above.
  \item Optimize the model further by alternating between 
    \begin{itemize}
    \item[$-$] an MCMC run and 
    \item[$-$] a simulated annealing optimization 
    \end{itemize}
  until $\Delta \log P \leq 5$.
  \item Accept this as final lens light model obtained with \GLEE.
  \end{itemize}
\item[3.] \textbf{Source and image position modeling} with \GLEE:
  \begin{itemize}
  \item This optimization step of the SIE profile parameters is based on the multiple image positions identified by the user during the preparation stage. The mass parameters are now optimized to reproduce these image positions.
  \item As starting values for the SIE central coordinates, use the obtained lens light center as lens mass center and keep it fixed for now to reduce the number of free parameters. Adopt the axis ratio and position angle from the lens light fit as well, and vary them only if three or more image positions are specified to not under-constrain the model. The Einstein radius is estimated from the identified image positions and always allowed to vary. Assume no shear for now to minimize the number of free parameters.
  \item optimize the model based on the source position with simulated annealing. Perform up to three optimizations, stopping earlier if $\chi^2_\text{red} \leq 1$.
  \item optimize the model based on the image positions with simulated annealing. Perform up to three optimizations, stopping earlier if $\chi^2_\text{red} \leq 1$.
  \end{itemize}
\item[4.] \textbf{Arc light modeling} with \GG:
  \begin{itemize}
  \item Transfer the best fit values to a \GLaD \, configuration file. Assume the source profile to be located at the predicted weighted source position ($x_{\text{s},i}, y_{\text{s},i})$ obtained from the image position model.
  \item ($m=1$) Perform again a quick lens-light-only optimization, to reduce the minimal differences in the model arising though differences between \GLEE \, and \GLaD \, in subsampling the PSF and the usage of the masks\footnote{\GLEE \, excludes directly all pixels that are specified in the mask when summing up the $\chi^2$, while in \GLaD \, the masked regions are incorporated implicitly in the error map through significantly boosting of their uncertainty values such that they contribute effectively nothing to the $\chi^2$.}.
    \begin{itemize}
    \item[$-$] To this end, first run an \emcee \, chain to obtain a new covariance matrix.
    \end{itemize}
    Alternate then between 
    \begin{itemize}
    \item[$-$] a basin-hopping iteration and 
    \item[$-$] an MCMC chain to obtain a new covariance matrix and also to update the parameter values to the new best set from the chain.
    \end{itemize}
    until $\Delta \log P \leq 2$ is achieved in the MCMC chain.
  \item ($m=2$) Fix now all lens-light parameters to the best values obtained in the previous modeling sequence. Instead, allow now the source-light axis ratio $q_\text{s} \in [0.5,1]$, the position angle $\phi_\text{s} \in [-\pi,+\pi]$, the amplitude $A_\text{s} \in [0,50]$, and the effective radius $r_\text{eff,s} \in [0.01,10]$ to vary in the specified prior ranges, but assume again the same structural parameters across the different bands. Since the source \sersic \, parameters cannot be easily estimated by the user, the initial starting values are $q_{\text{s},i} \equiv 0.9$, $\phi_{\text{s},i} \equiv 0$, $r_{\text{eff},i} \equiv 0.5$, and $n_{\text{s},i} \equiv 3$. The amplitudes, which are different for each filter, are again determined from an amplitude test. Include from now on also the regions specified in the arc mask in the optimization which were previously excluded to fit only to the light from the lens.
    \begin{itemize}
    \item[$-$] Run an \emcee \, chain to update the covariance matrix.
    \item[$-$] Perform a dual annealing optimization, followed by
    \item[$-$] an MCMC chain and
    \item[$-$] a basin-hopping optimization.
    \end{itemize}
  \item ($m=3$) Allow the coordinates of the source light center $x_\text{s} \in [x_{\text{s},i}-1\arcsec, x_{\text{s},i}+1\arcsec]$ and $y_\text{s} \in [y_{\text{s},i}-1\arcsec, y_{\text{s},i}+1\arcsec]$ to vary. Increase additionally the prior range of the source axis ratio to $q_\text{s} \in [0.01,1]$. Optimize now the model until reaching $\Delta \log P \leq 2$ by a repeated sequence of
    \begin{itemize}
    \item[$-$] a basin-hopping optimization,
    \item[$-$] an \emcee \, chain to update the covariance matrix (max 10 times in total) and
    \item[$-$] an MCMC chain to save the new best set of parameter values and update the covariance matrix.
    \end{itemize}
  \item ($m=4$) After optimizing the lens light and source light, allow now, in addition to the source light, the lens-mass axis ratio $q_\text{lm} \in [0.3,1]$, the position angle $\phi_\text{lm} \in [-\pi,+\pi]$, and the Einstein radius $\theta_\text{E} \in [0.5\arcsec, 10\arcsec]$ to vary. Moreover, include from now on also an external shear component with $\gamma_\text{ext} \in [0,0.2]$ and $\phi_\text{ext} \in [-\pi,+\pi]$. Optimize until $\Delta \log P \leq 2$ through a repeated sequence of
    \begin{itemize}
    \item[$-$] a dual annealing iteration,
    \item[$-$] an \emcee \, chain to update the covariance matrix (max 15 times in total), and
    \item[$-$] an MCMC chain to save the new best set of parameters and update the covariance matrix.
    \end{itemize}
  \item ($m=5$) Vary now additionally the source \sersic \, index $n_\text{s} \in [0.5,6]$, which was previously fixed to 3.
    \begin{itemize}
    \item[$-$] Run an \emcee \, chain to obtain a covariance matrix for the new set of varying parameters.
    \end{itemize}
    Optimize then all varying parameters through a repeated sequence consisting of 
    \begin{itemize}
    \item[$-$] a dual annealing iteration, and
    \item[$-$] an MCMC chain to save the best set of parameter values and update the covariance matrix
    \end{itemize}
    until $\Delta \log P \leq 2$ is reached.
  \item ($m=6$) After all parameters were optimized at least once, refine once more the lens light parameters which were fixed during the last optimization steps. For this, allow the lens light to vary again, but fix all other parameters, i.e the lens mass, external shear, and the source light components.
    \begin{itemize}
    \item[$-$] Run one \emcee \, chain to obtain a first covariance matrix.
    \end{itemize}
    Optimize then until $\Delta \log P \leq 2$ through a repeated sequence of 
    \begin{itemize}
    \item[$-$] a dual annealing iteration and 
    \item[$-$] an MCMC chain to save the best values and update the covariance matrix.
    \end{itemize}
  \item ($m=7$) Fix again all lens light parameters and vary again the source light, lens mass and external shear by using the same prior ranges as before, but update the parameter range for the source light center to be again $x_\text{s} \in [x_{\text{s},i}'-1\arcsec, x_{\text{s},i}'+1\arcsec]$ and $y_\text{s} \in [y_{\text{s},i}'-1\arcsec,y_{\text{s},i}'+1\arcsec]$ with ($x_{\text{s},i}', y_{\text{s},i}'$) being the previously best source position.
    \begin{itemize}
    \item[$-$] Run first one \emcee \, chain to obtain a new covariance matrix.
    \end{itemize}
    Optimize until $\Delta \log P \leq 2$ through a repeated sequence of 
    \begin{itemize}
    \item[$-$] a dual annealing iteration and 
    \item[$-$] an MCMC chain to save the best parameter set and update the covariance matrix.
    \end{itemize}
  \item ($m=8$) Double the length of the MCMC chains to 400,000 and run them until they are fully converged based on the power spectrum \citep{dunkley05}. Always take the best set of parameter values of the chain and update the covariance matrix. In case the tenth MCMC chain of this optimization sequence did not converge, the number of sampling steps is increased to 600,000.
  \end{itemize}
\end{itemize}

\FloatBarrier
\section{Details of the 31 HSC models}
\label{app:modeldetails}

\begin{figure}[ht!]
  \begin{center}
    \includegraphics[trim=0 0 0 0, clip, width=\textwidth]{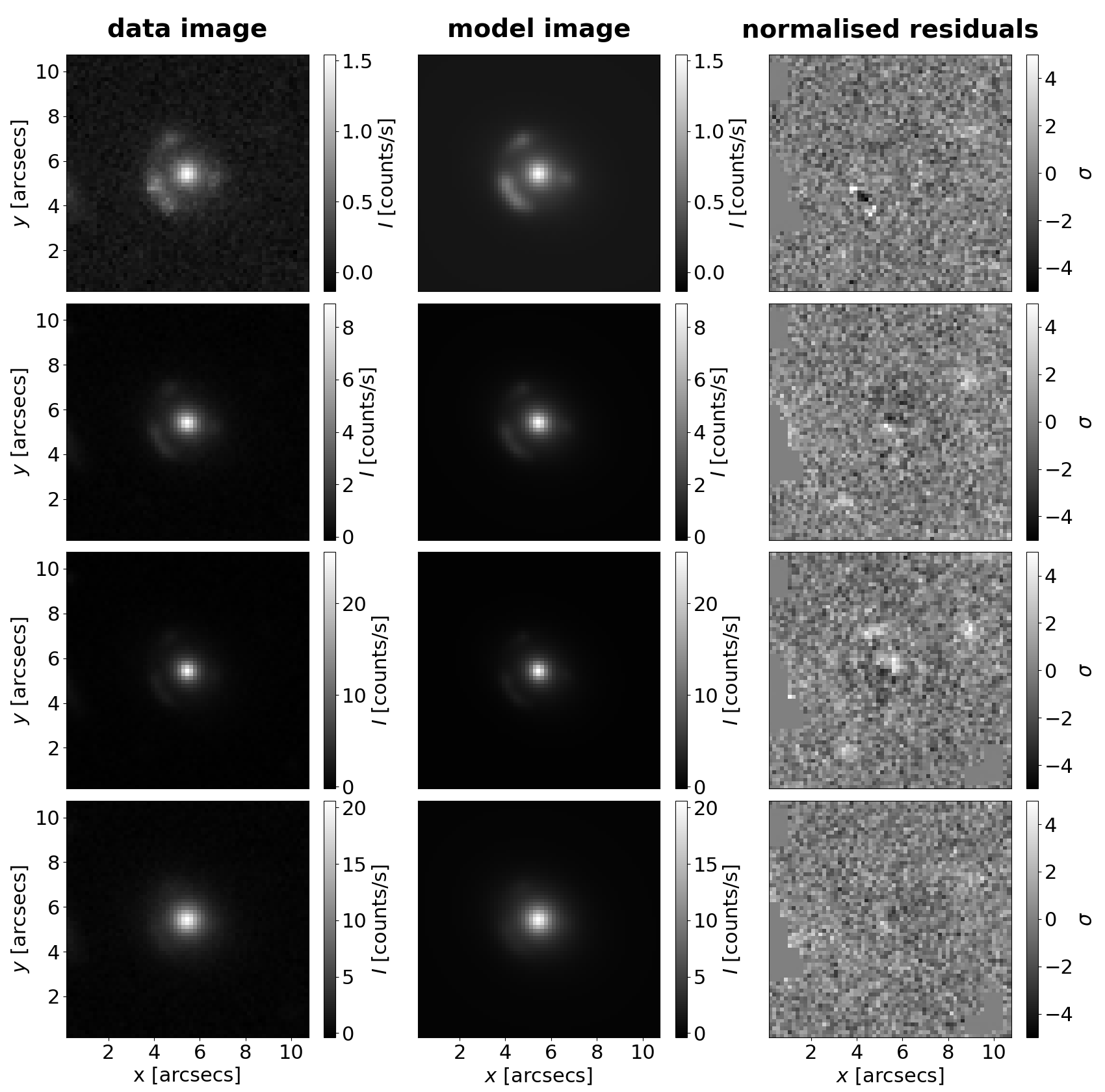}
    \caption{Fit of lens HSCJ233146$+$013845. Top to bottom: $griz$ filters.\label{fig:model_1976}}
  \end{center}
\end{figure}

In this section we provide details of the individual lens models. Tab.~\ref{tab:comparison_normal} lists the values of the SIE profile and external shear as obtained with \GG \, and the converted values from the network. In analogy, Tab.~\ref{tab:comparison_complex} lists these values in complex notation as predicted directly by the network and the converted values from \GG \, models. We further report the values from \GG \, for the lens light (Tab.~\ref{tab:lenslightpars}) and source light (Tab.~\ref{tab:sourcelightpars}) parameters, which are not obtained with the network. Furthermore, we show images of the 31 \GG \, models in all four filters in separate rows (Figs.~\ref{fig:model_47}~-~\ref{fig:model_1976}), with the observed image (left column), predicted image (middle), and normalized residuals in the range $[-5\sigma, +5\sigma]$ (right).

\onecolumn
\LTcapwidth=\textwidth
\begin{landscape}
  \small{
\begin{longtable}{c|c|cc|ccccc|cc}
\caption{Shear and mass parameter values from our ResNet and \GG \, for the 31 HSC SuGOHI lenses. Columns 3 and 4 report, respectively, the $\chi^2$ and $\chi^2_\text{red}$ for the \GG-based model. This is followed by the lens (light and mass) center $x_\text{l}$ and $y_\text{l}$ (with respect to the image cutout center), the lens mass axis ratio $q_\text{lm}$, its orientation $\phi_\text{lm}$, the Einstein radius $\theta_\text{E}$, the external shear strength $\gamma_\text{ext}$, and its orientation $\phi_\text{ext}$. The angles $\phi_\text{lm}$ and $\phi_\text{ext}$ are measured counterclockwise from the positive $x$-axis. This means $\phi_\text{lm}=0$ corresponds to a lens mass distribution with major-axis along $x$ and $\phi_\text{ext}=0$ corresponds to a shear of the lensed images along the $x$-axis.\label{tab:comparison_normal}}\\

  Name   & method & $\chi^2$ & $\chi^2_\text{red}$ & \multicolumn{5}{c|}{SIE lens mass distribution} & \multicolumn{2}{c}{external shear} \\
         &        &        &       & $x_\text{l} [\arcsec]$ & $y_\text{l} [\arcsec]$ & $q_\text{lm}$ & $\phi_\text{lm}$ [rad] & $\theta_\text{E} [\arcsec]$ & $\gamma_\text{ext}$ & $\phi_\text{ext}$ [rad] \\
  \hline
  \endfirsthead
  \caption*{\textbf{Table \ref{tab:comparison_normal} Continued:} Obtained shear and mass parameter values by using ResNet or \GG \, for the 31 HSC SuGOHI lenses.}\\
  Name  & method & $\chi^2$ & $\chi^2_\text{red}$ & \multicolumn{5}{c|}{SIE lens mass distribution} & \multicolumn{2}{c}{external shear} \\
        &        &          &       & $x_\text{l} [\arcsec]$ & $y_\text{l} [\arcsec]$ & $q_\text{lm}$ & $\phi_\text{lm}$ [rad] & $\theta_\text{E} [\arcsec]$ & $\gamma_\text{ext}$ & $\phi_\text{ext}$ [rad] \\  
  \hline
  \endhead
  \rowcolor{Gray}
  \multirow{2}{*}{HSCJ015618$-$010747} & ResNet & & & $ 0.201 \pm  0.073 $  & $ 0.38 \pm  0.11 $  & $ 0.76_{- 0.11}^{+ 0.10} $  & $ 0.57_{- 0.32}^{+ 0.40} $  & $ 0.95 \pm  0.14 $  & $ 0.050_{- 0.025}^{+ 0.032} $  & $ 2.21_{- 1.22}^{+ 0.90} $\\
  & \GG & $1.68\times 10^4$ & 1.19 & $ -0.076 \pm 0.002 $ &$ 0.070 \pm 0.002 $ & $ 0.418_{-0.013}^{+0.020} $ &$ 0.318_{-0.042}^{+0.054} $ &$ 0.994_{-0.014}^{+0.015} $ &$ 0.084 \pm 0.005 $ &$ 2.677_{-0.081}^{+0.093} $\\
  \hline
  \rowcolor{Gray}
  \tabspace
  \multirow{2}{*}{HSCJ020141$-$030946} & ResNet & & & $ 0.02 \pm  0.05 $  & $ 0.02 \pm  0.06 $  & $ 0.864_{- 0.070}^{+ 0.063} $  & $ 1.81_{- 0.50}^{+ 0.42} $  & $ 1.72 \pm  0.13 $  & $ 0.044_{- 0.022}^{+ 0.030} $  & $ 1.76_{- 1.1}^{+ 1.0} $\\
  & \GG & $1.97\times 10^4$ & 1.55  & $ -0.005 \pm 0.001 $ &$ -0.052 \pm 0.001 $ & $ 0.792 \pm 0.014 $ &$ 1.769_{-0.042}^{+0.043} $ &$ 1.406 \pm 0.007 $ &$ 0.073 \pm 0.003 $ &$ 0.764_{-0.020}^{+0.021} $\\
  \hline
  \rowcolor{Gray}
  \tabspace
  \multirow{2}{*}{HSCJ020241$-$064611} & ResNet & & & $ 0.052 \pm  0.071 $  & $ 0.16 \pm  0.11 $  & $ 0.915_{- 0.051}^{+ 0.041} $  & $ 2.45_{- 1.25}^{+ 0.80} $  & $ 1.01 \pm  0.11 $  & $ 0.047_{- 0.024}^{+ 0.030} $  & $ 2.15_{- 0.92}^{+ 0.83} $\\
\end{longtable}
  }
\end{landscape}

\begin{landscape}
\begin{longtable}{c|c|ccccc|cc}
\caption{SIE mass and external shear parameter values obtained by modeling with \GG \, (converted to complex quantities) or with our neural network (column 2, method) for the 31 HSC SuGOHI lenses used for the direct comparison. The quoted SIE mass parameters are the lens mass center $x_\text{l}$ and $y_\text{l}$, the complex lens mass ellipticity $e_\text{x}$ and $e_\text{y}$, and the Einstein radius $\theta_\text{E}$. We further give values for $\gamma_\text{ext,1}$ and $\gamma_\text{ext,2}$.\label{tab:comparison_complex}}\\
  Name  & method & $x_\text{l} [\arcsec]$ & $y_\text{l} [\arcsec]$ & $e_\text{x}$ & $e_\text{y}$ & $\theta_\text{E} [\arcsec]$ & $\gamma_\text{ext,1}$ & $\gamma_\text{ext,2}$ \\
  \hline
  \endfirsthead
  \caption*{\textbf{Table \ref{tab:comparison_complex} Continued:} SIE mass and external shear parameter values obtained by modeling with \GG \, (converted to complex quantities) or with our neural network (column 2, method) for the 31 HSC SuGOHI lenses used for the direct comparison.}\\
  Name  & method & $x_\text{l} [\arcsec]$ & $y_\text{l} [\arcsec]$ & $e_\text{x}$ & $e_\text{y}$ & $\theta_\text{E} [\arcsec]$ & $\gamma_\text{ext,1}$ & $\gamma_\text{ext,2}$ \\
  \hline
  \endhead
  \rowcolor{Gray} 
  \multirow{2}{*}{HSCJ015618$-$010747}    & ResNet & $ 0.201 \pm  0.073 $ & $ 0.38 \pm  0.11 $ & $ 0.1 \pm  0.2 $ & $ 0.20 \pm  0.13 $ & $ 0.95 \pm  0.14 $ & $ 0.008 \pm  0.044 $ & $ -0.01 \pm  0.04$\\
  & \GG & ${-0.076}\pm 0.002 $ & ${0.07}\pm 0.002 $ & ${0.562}^{+0.03}_{-0.05} $ & ${0.417}^{+0.06}_{-0.05} $ & ${0.994}^{+0.015}_{-0.014} $ & ${0.05} \pm 0.01 $ & ${-0.066}^{+0.01}_{-0.008} $ \\
  \hline
  \rowcolor{Gray}
  \tabspace 
  \multirow{2}{*}{HSCJ020141$-$030946}   & ResNet & $ 0.02 \pm  0.05 $ & $ 0.02 \pm  0.06 $ & $ -0.095 \pm  0.085 $ & $ -0.050 \pm  0.095 $ & $ 1.72 \pm  0.13 $ & $ -0.00 \pm  0.04 $ & $ -0.00 \pm  0.04$\\
  & \GG & ${-0.005}\pm0.001 $ & ${-0.052}\pm 0.001 $ & ${-0.211} \pm0.02 $ & ${-0.088} \pm -0.02 $ & ${1.406}\pm 0.007 $ & ${0.003} \pm 0.003 $ & ${0.073} \pm 0.003 $ \\
  \hline
  \rowcolor{Gray}
  \tabspace 
  \multirow{2}{*}{HSCJ020241$-$064611}   & ResNet & $ 0.052 \pm  0.071 $ & $ 0.16 \pm  0.11 $ & $ 0.02 \pm  0.08 $ & $ -0.012 \pm  0.073 $ & $ 1.01 \pm  0.11 $ & $ -0.00 \pm  0.04 $ & $ -0.01 \pm  0.04$\\
  & \GG & $-0.007 \pm 0.001 $ & ${0.061} \pm 0.001 $ & ${0.1} \pm 0.1 $ & ${-0.158}^{+0.08}_{-0.07} $ & ${1.262}^{+0.031}_{-0.02} $ & ${0.06} \pm 0.03 $ & ${-0.10} \pm 0.02 $ \\
\end{longtable}
\end{landscape}

\begin{landscape}
\begin{longtable}{c|c|cccccccc}
  \caption{Lens light parameter values obtained with \GG. We give, respectively for both the first and second \sersic \, profile (column 2), the axis ratio $q_\text{ll}$, its orientation $\phi_\text{ll}$, the effective radius $r_\text{eff,ll}$, the \sersic \, index $n_\text{ll}$, and the amplitudes $A_\text{ll}$ for the four different filters $griz$. The lens light center is assumed to match the lens mass center and is thus only quoted in Tab.~\ref{tab:comparison_complex}\label{tab:lenslightpars}}\\
Name    & component & $q_\text{ll1}$ & $\phi_\text{ll1}$ [rad] & $r_\text{eff,ll1}$ & $ n_\text{ll1}$ & $A_\text{g,ll1}$ & $A_\text{r,ll1}$ & $A_\text{i,ll1}$ & $A_\text{z,ll1}$ \\
      &           & $q_\text{ll2}$ & $\phi_\text{ll2}$ [rad] & $r_\text{eff,ll2}$ & $ n_\text{ll2}$ & $A_\text{g,ll2}$ & $A_\text{r,ll2}$ & $A_\text{i,ll2}$ & $A_\text{z,ll2}$ \\
  \hline
  \endfirsthead
  \caption*{\textbf{Table \ref{tab:lenslightpars} Continued:} Lens light parameter values obtained with \GG.}\\
Name    & component & $q_\text{ll1}$ & $\phi_\text{ll1}$ [rad] & $r_\text{eff,ll1}$ & $ n_\text{ll1}$ & $A_\text{G,ll1}$ & $A_\text{R,ll1}$ & $A_\text{I,ll1}$ & $A_\text{Z,ll1}$ \\
      &           & $q_\text{ll2}$ & $\phi_\text{ll2}$ [rad] & $r_\text{eff,ll2}$ & $ n_\text{ll2}$ & $A_\text{G,ll2}$ & $A_\text{R,ll2}$ & $A_\text{I,ll2}$ & $A_\text{Z,ll2}$ \\
  \hline
\endhead 

  \multirow{2}{*}{HSCJ015618$-$010747}    & 1 &  $ 0.615_{- 0.03}^{+ 0.023} $  & $ 0.420_{- 0.051}^{+ 0.045} $  & $ 2.49_{- 0.11}^{+ 0.12} $  & $ 0.529_{- 0.022}^{+ 0.044} $  & $ 0.088_{- 0.012}^{+ 0.014} $  & $ 0.044 \pm  0.004 $  & $ 0.487_{- 0.053}^{+ 0.06} $  & $ 0.29 \pm  0.05 $\\
  & 2 & $ 0.645 \pm  0.0010 $  & $ 0.69 \pm  0.02 $  & $ 0.63 \pm  0.05 $  & $ 3.30_{- 0.3}^{+ 0.21} $  & $ 1.66_{- 0.2}^{+ 0.22} $  & $ 0.371_{- 0.041}^{+ 0.051} $  & $ 3.87_{- 0.44}^{+ 0.5} $  & $ 6.24_{- 0.7}^{+ 0.82} $ \\
  \hline
  \tabspace
  \multirow{2}{*}{HSCJ020141$-$030946}   & 1 &  $ 0.563_{- 0.004}^{+ 0.003} $  & $ 1.615 \pm  0.004 $  & $ 0.71 \pm  0.03 $  & $ 4.62 \pm  0.12 $  & $ 2.7 \pm  0.2 $  & $ 0.45 \pm  0.03 $  & $ 5.7 \pm  0.4 $  & $ 8.42_{- 0.54}^{+ 0.6} $ \\
  & 2 & $ 0.80 \pm  0.02 $  & $ 2.604_{- 0.054}^{+ 0.05} $  & $ 2.366 \pm  0.034 $  & $ 0.764_{- 0.042}^{+ 0.044} $  & $ 0.167_{- 0.006}^{+ 0.007} $  & $ 0.059 \pm  0.002 $  & $ 0.253_{- 0.011}^{+ 0.013} $  & $ 0.30 \pm  0.02 $ \\
  \hline
  \tabspace
  \multirow{2}{*}{HSCJ020241$-$064611}   & 1 & $ 0.946 \pm  0.006 $  & $ 1.611_{- 0.052}^{+ 0.053} $  & $ 0.541 \pm  0.003 $  & $ 1.788_{- 0.021}^{+ 0.022} $  & $ 2.98_{- 0.06}^{+ 0.05} $  & $ 0.709_{- 0.011}^{+ 0.0010} $  & $ 6.900_{- 0.065}^{+ 0.07} $  & $ 5.93_{- 0.3}^{+ 0.41} $  \\
  & 2 & $ 0.972_{- 0.034}^{+ 0.023} $  & $ -0.5_{- 0.6}^{+ 0.5} $  & $ 0.688_{- 0.021}^{+ 0.02} $  & $ 2.91_{- 0.2}^{+ 0.15} $  & $ 0.026_{- 0.021}^{+ 0.034} $  & $ 0.002_{- 0.002}^{+ 0.004} $  & $ 0.014_{- 0.011}^{+ 0.02} $  & $ 2.26_{- 0.3}^{+ 0.24} $ \\
  \hline
  \tabspace
  \multirow{2}{*}{HSCJ020955$-$024442}   & 1 & $ 0.883 \pm  0.013 $  & $ 1.151_{- 0.06}^{+ 0.054} $  & $ 0.979 \pm  0.013 $  & $ 1.139_{- 0.062}^{+ 0.07} $  & $ 0.38 \pm  0.01 $  & $ 0.121 \pm  0.006 $  & $ 0.778_{- 0.024}^{+ 0.022} $  & $ 1.057_{- 0.04}^{+ 0.034} $ \\
\end{longtable}
\end{landscape}

\begin{landscape}
  \small{
\begin{longtable}{c|c|cccccccccc}
\caption{Reconstructed source-light parameter values obtained with \GG. For each \sersic \, profile as specified in column 2 (component $c$), we list the unlensed source center coordinates $x_\text{s}$ and $y_\text{s}$, the axis ratio $q_\text{s}$, its orientation $\phi_\text{s}$, the effective radius $r_\text{eff,s}$, the \sersic \, index $n_\text{s}$, and the amplitudes $A_\text{s}$ for the four different filters $griz$.\label{tab:sourcelightpars}}\\
  Name  & $c$ &  $x_\text{s,c} [\arcsec]$ & $y_\text{s,c} [\arcsec]$ & $q_\text{s,c}$ & $\phi_\text{s,c}$ [rad] & $r_\text{eff,s,c}$ & $ n_\text{s,c}$ & $A_\text{g,s,c}$ & $A_\text{r,s,c}$ & $A_\text{i,s,c}$ & $A_\text{z,s,c}$\\
    \hline
  \endfirsthead
  \caption*{\textbf{Table \ref{tab:sourcelightpars} Continued:} Reconstructed source light parameter values obtained with \GG.}\\
  Name  &  $c$ & $x_\text{s,c} [\arcsec]$ & $y_\text{s,c} [\arcsec]$ & $q_\text{s,c}$ & $\phi_\text{s,c}$ [rad] & $r_\text{eff,s,c}$ & $ n_\text{s,c}$ & $A_\text{g,s,c}$ & $A_\text{r,s,c}$ & $A_\text{i,s,c}$ & $A_\text{z,s,c}$\\
  \hline
  \endhead
  \multirow{1}{*}{HSCJ015618$-$010747}    & 1 & $ 0.67_{- 0.14}^{+ 0.11} $  & $ 0.85_{- 0.05}^{+ 0.04} $  & $ 0.250_{- 0.04}^{+ 0.041} $  & $ -1.36 \pm  0.04 $  & $ 0.01_{- 0.012}^{+ 0.022} $  & $ 2.6 \pm  0.9 $  & $ 93_{- 30}^{+ 18} $  & $ 116_{- 38}^{+ 23} $  & $ 84_{- 31}^{+ 29} $  & $ 238_{- 76}^{+ 47} $ \\
  \tabspace
  \multirow{1}{*}{HSCJ020141$-$030946}   & 1 & $ -0.360_{- 0.012}^{+ 0.03} $  & $ -0.03 \pm  0.03 $  & $ 0.140 \pm  0.006 $  & $ -2.89 \pm  0.01 $  & $ 0.240_{- 0.006}^{+ 0.007} $  & $ 0.52_{- 0.02}^{+ 0.03} $  & $ 2.205_{- 0.09}^{+ 0.094} $  & $ 3.04_{- 0.12}^{+ 0.13} $  & $ 3.44_{- 0.14}^{+ 0.15} $  & $ 3.6 \pm  0.2 $  \\
  \tabspace
  \multirow{1}{*}{HSCJ020241$-$064611}   & 1 & $ 0.20_{- 0.23}^{+ 0.21} $  & $ 0.70_{- 0.2}^{+ 0.14} $  & $ 0.384_{- 0.084}^{+ 0.08} $  & $ 1.04_{- 0.11}^{+ 0.13} $  & $ 0.133_{- 0.011}^{+ 0.012} $  & $ 1.03_{- 0.21}^{+ 0.3} $  & $ 0.87_{- 0.14}^{+ 0.2} $  & $ 1.14_{- 0.2}^{+ 0.22} $  & $ 0.85_{- 0.14}^{+ 0.2} $  & $ 0.97_{- 0.2}^{+ 0.23} $ \\
  \tabspace
  \multirow{1}{*}{HSCJ020955$-$024442}   & 1 &  $ 0.124 \pm  0.044 $  & $ -0.14 \pm  0.03 $  & $ 0.27 \pm  0.02 $  & $ 0.024_{- 0.024}^{+ 0.022} $  & $ 0.082 \pm  0.004 $  & $ 0.514_{- 0.011}^{+ 0.03} $  & $ 0.709_{- 0.035}^{+ 0.04} $  & $ 2.55_{- 0.11}^{+ 0.12} $  & $ 1.92_{- 0.09}^{+ 0.1} $  & $ 1.707_{- 0.092}^{+ 0.11} $ \\
  \tabspace
  \multirow{1}{*}{HSCJ021737$-$051329}   & 1 &  $ 0.071_{- 0.043}^{+ 0.041} $  & $ -0.26_{- 0.13}^{+ 0.04} $  & $ 0.383_{- 0.041}^{+ 0.04} $  & $ 0.07_{- 0.06}^{+ 0.1} $  & $ 0.059_{- 0.005}^{+ 0.01} $  & $ 0.59_{- 0.06}^{+ 0.2} $  & $ 9.16_{- 3.09}^{+ 1.63} $  & $ 9.60_{- 3.23}^{+ 1.71} $  & $ 10.7_{- 3.6}^{+ 1.9} $  & $ 11.93_{- 4.02}^{+ 2.14} $ \\
\end{longtable}
  }
  \end{landscape}

\iftrue
  
\begin{figure}[h!]
  \begin{center}
    \includegraphics[trim=0 0 0 0, clip, width=\textwidth]{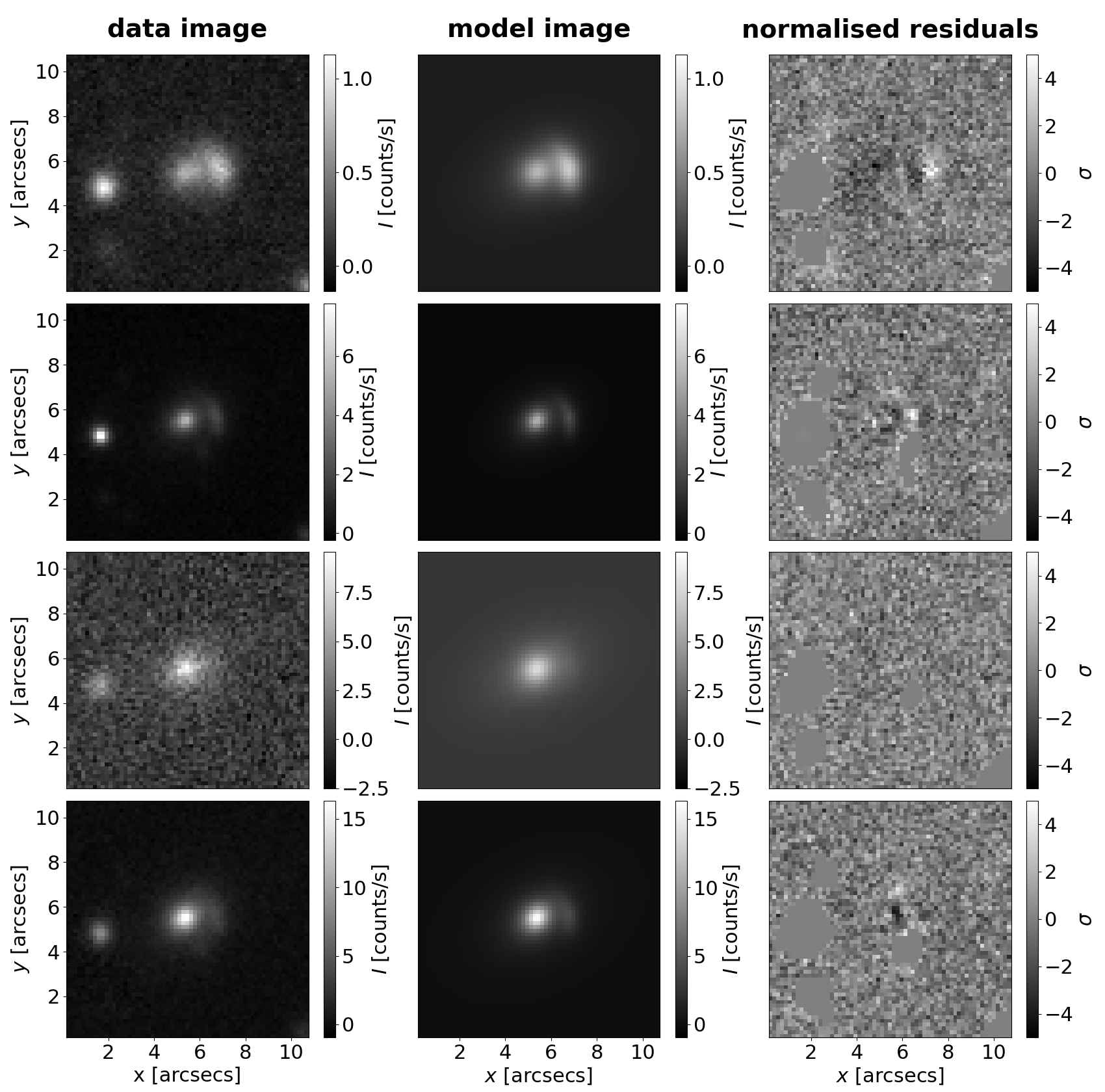}
    \caption{Fit of lens HSCJ015618$-$010747. Top to bottom: $griz$ filters.\label{fig:model_47}}
    \end{center}
  \end{figure}
    \begin{figure}[h!]
      \begin{center}
        \includegraphics[trim=0 0 0 0, clip, width=\textwidth]{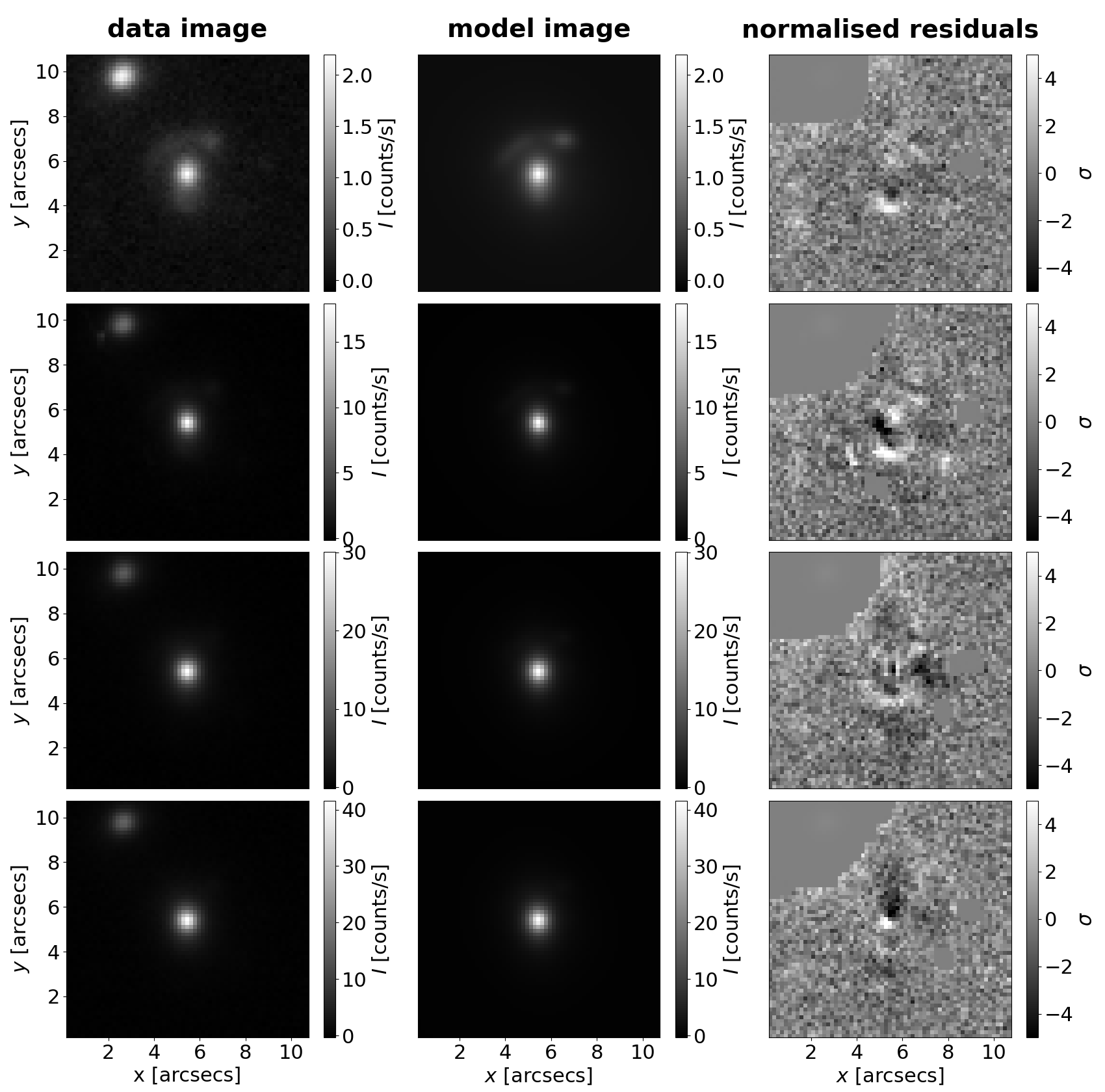}
    \caption{Fit of lens HSCJ020141$-$030946. Top to bottom: $griz$ filters. \label{fig:model_101}}
      \end{center}
\end{figure}

\begin{figure}[h!]
  \begin{center}
    \includegraphics[trim=0 0 0 0, clip, width=\textwidth]{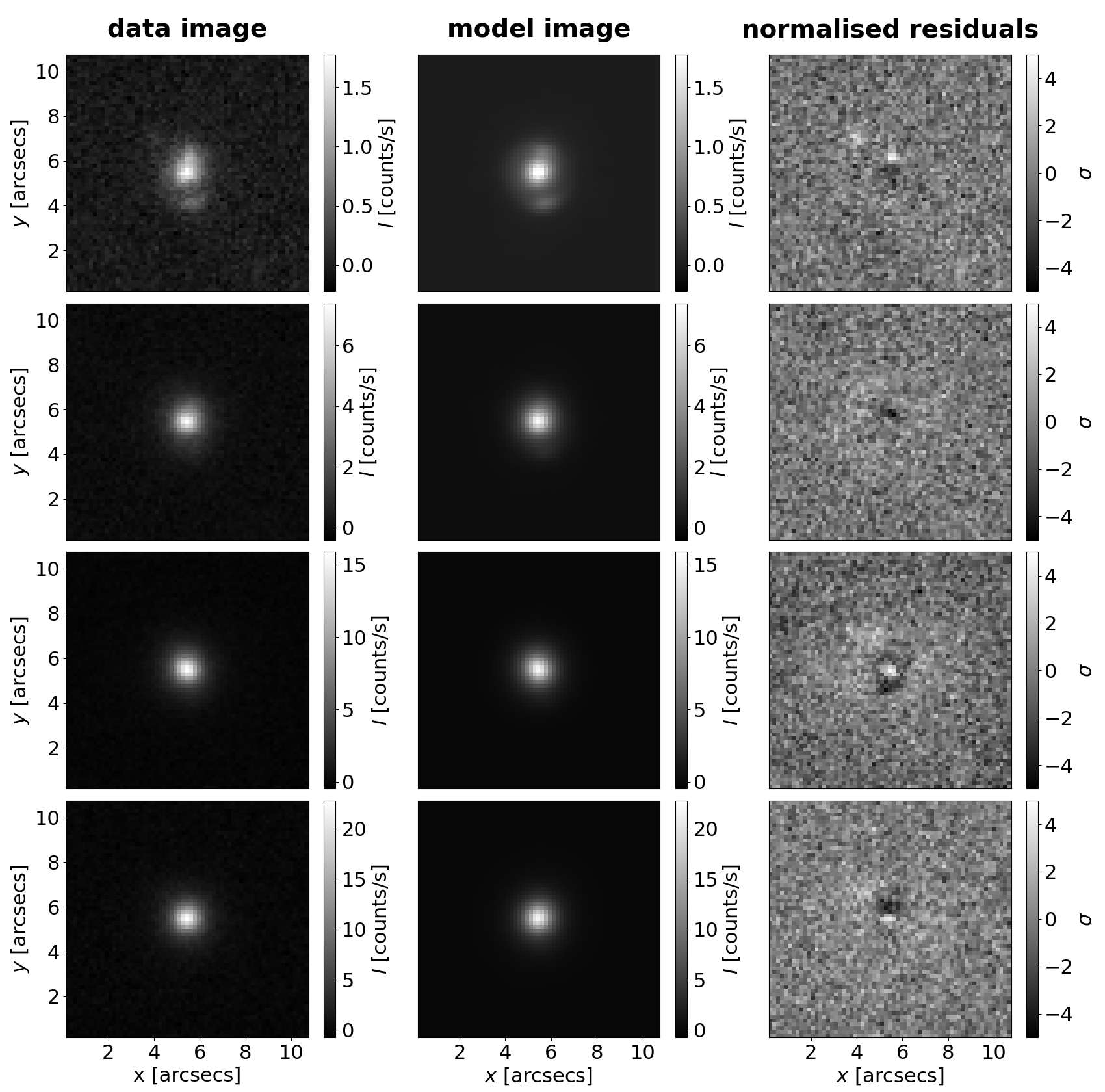}
    \caption{Fit of lens HSCJ020241$-$064611. Top to bottom: $griz$ filters.\label{fig:model_112}}
  \end{center}
  \end{figure}
  \begin{figure}[h!]
    \begin{center}
      \includegraphics[trim=0 0 0 0, clip, width=\textwidth]{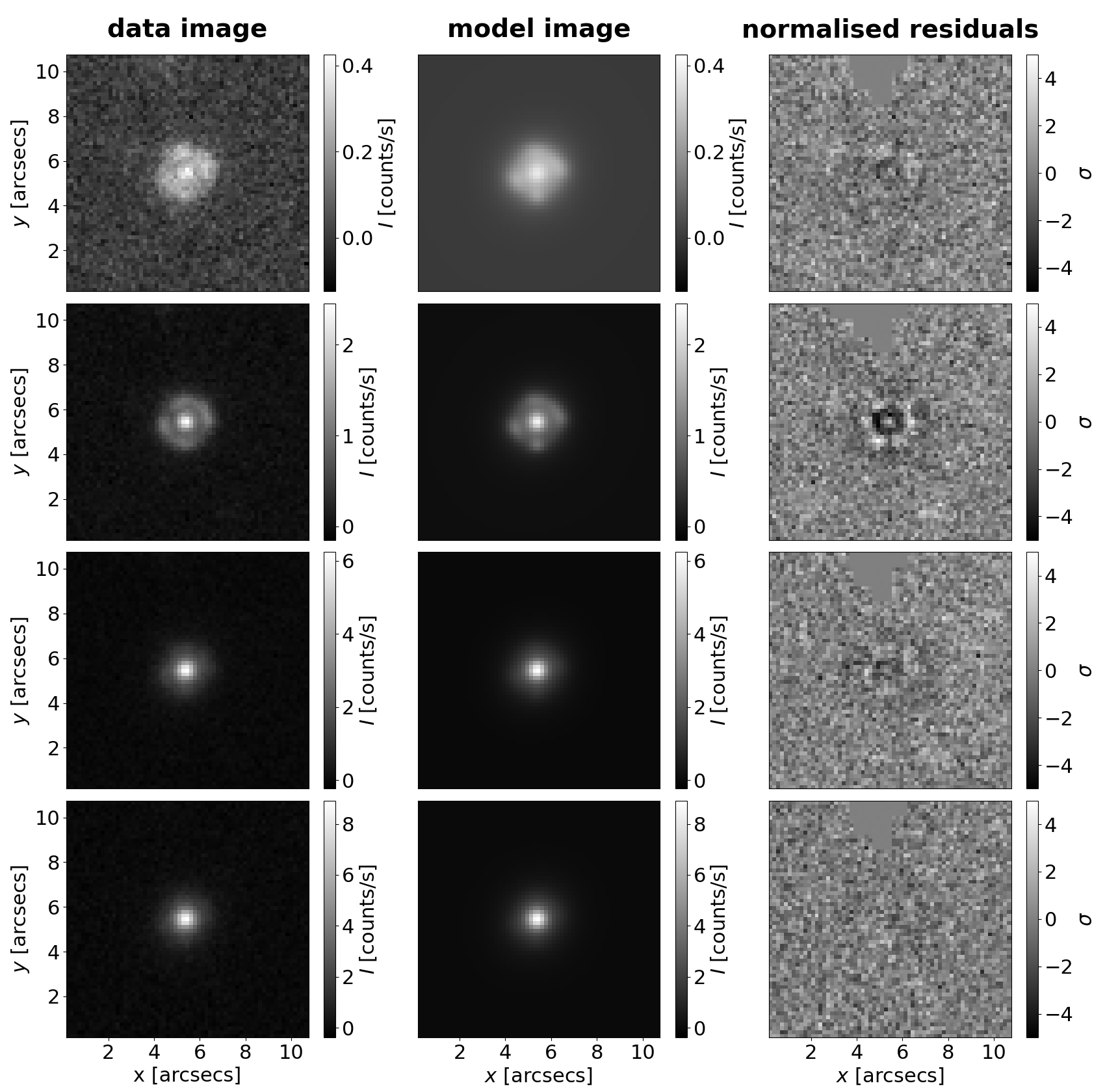}
    \caption{Fit of lens HSCJ020955$-$024442. Top to bottom: $griz$ filters.\label{fig:model_155}}
    \end{center}
\end{figure}

\begin{figure}[h!]
  \begin{center}
    \includegraphics[trim=0 0 0 0, clip, width=\textwidth]{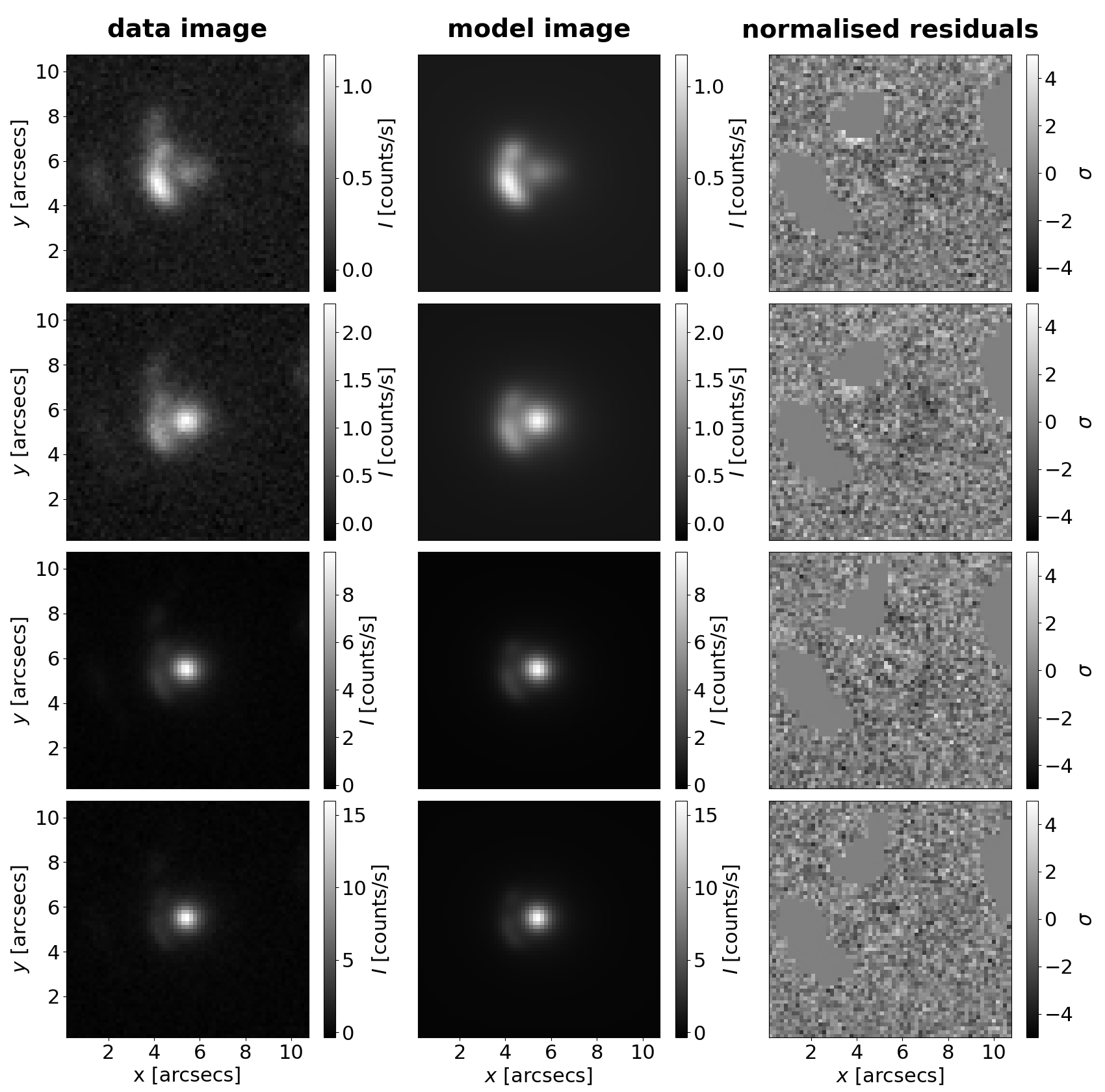}
    \caption{Fit of lens HSCJ021737$-$051329. Top to bottom: $griz$ filters.\label{fig:model_222}}
  \end{center}
\end{figure}
  \begin{figure}[h!]
    \begin{center}
      \includegraphics[trim=0 0 0 0, clip, width=\textwidth]{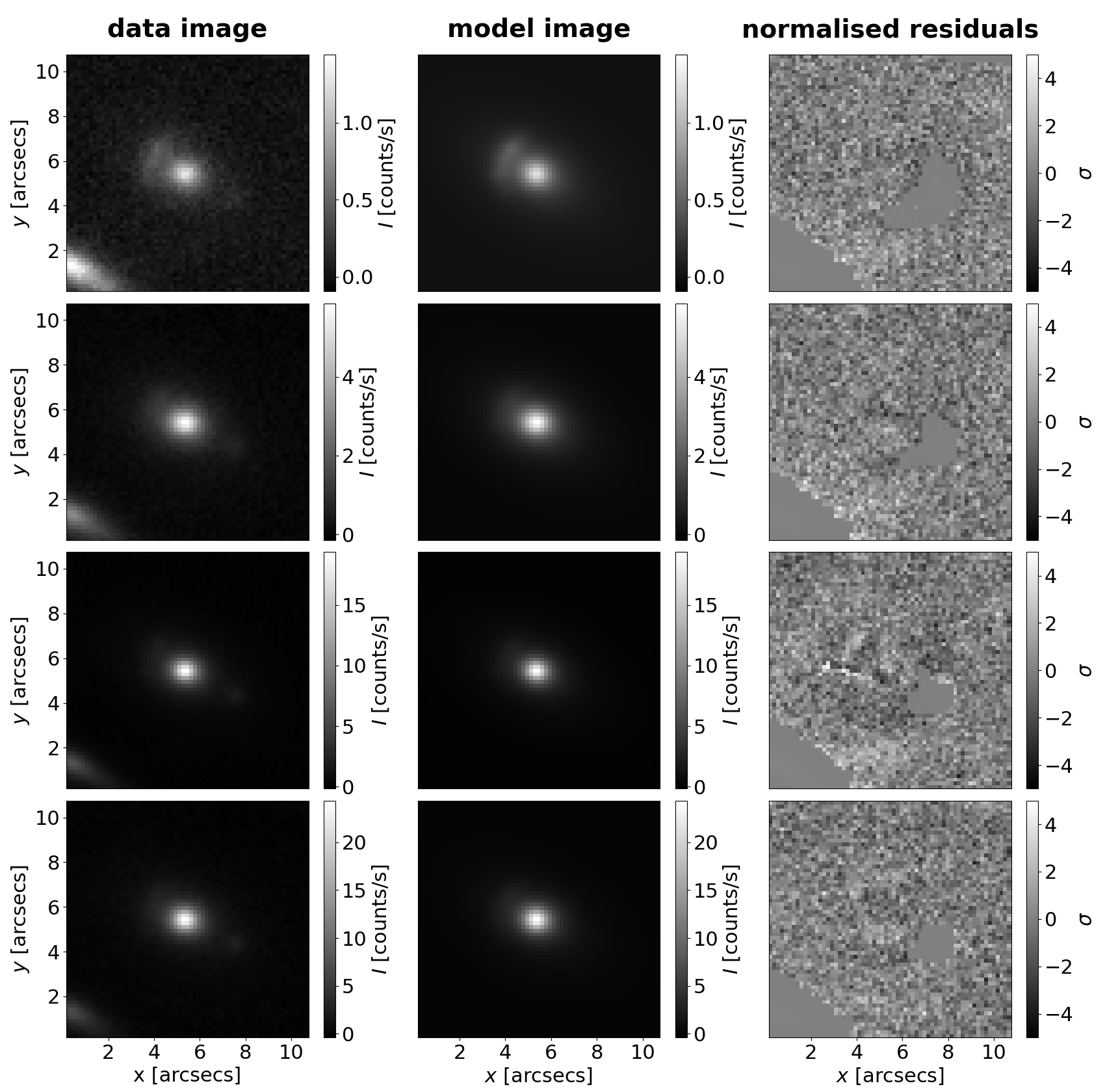}
    \caption{Fit of lens HSCJ022346$-$053418. Top to bottom: $griz$ filters.\label{fig:model_276}}
    \end{center}
\end{figure}

\begin{figure}[h!]
  \begin{center}
    \includegraphics[trim=0 0 0 0, clip, width=\textwidth]{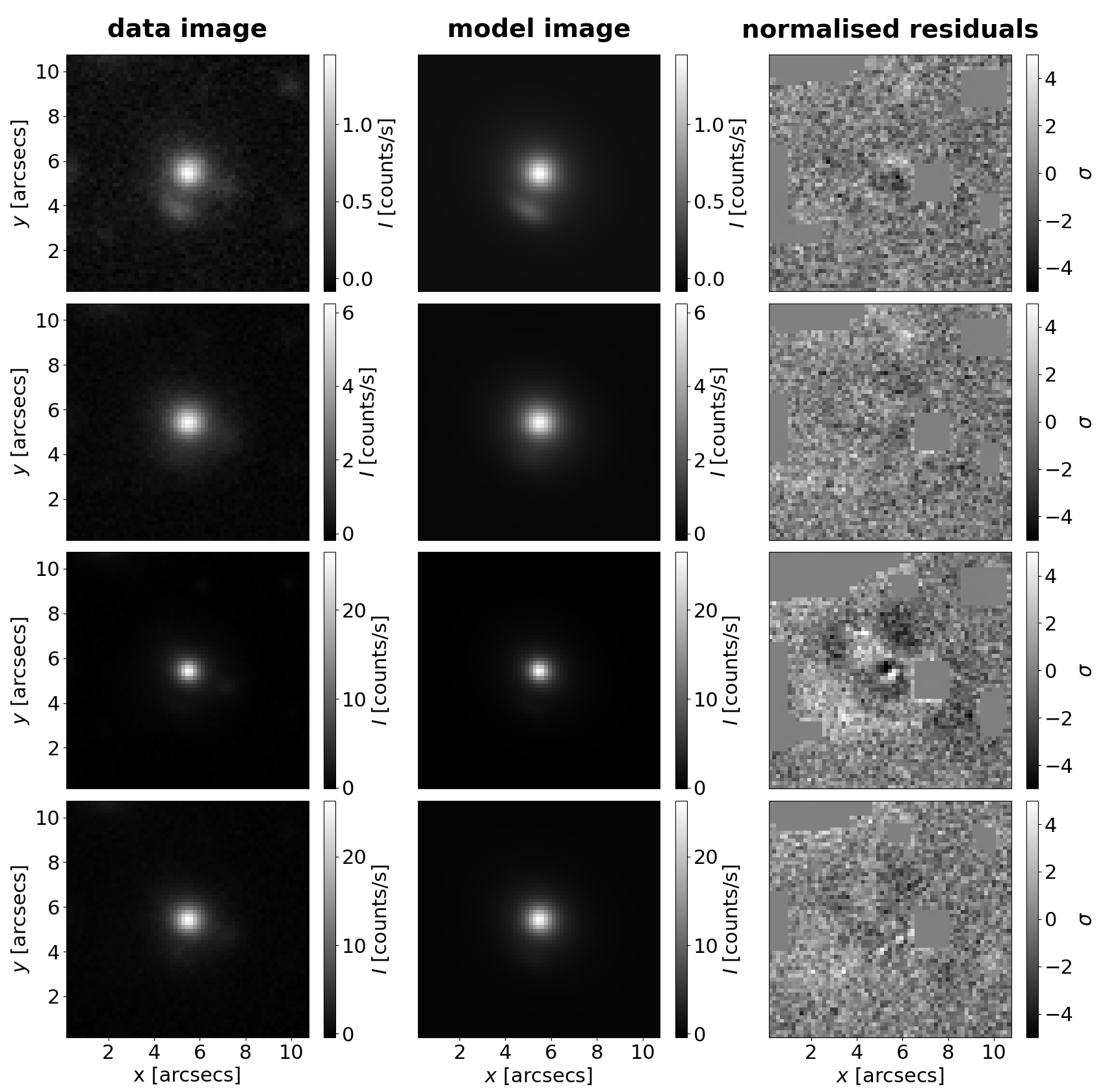}
    \caption{Fit of lens HSCJ022610$-$042011. Top to bottom: $griz$ filters.\label{fig:model_291}}
  \end{center}
\end{figure}
  \begin{figure}[h!]
\begin{center}
  \includegraphics[trim=0 0 0 0, clip, width=\textwidth]{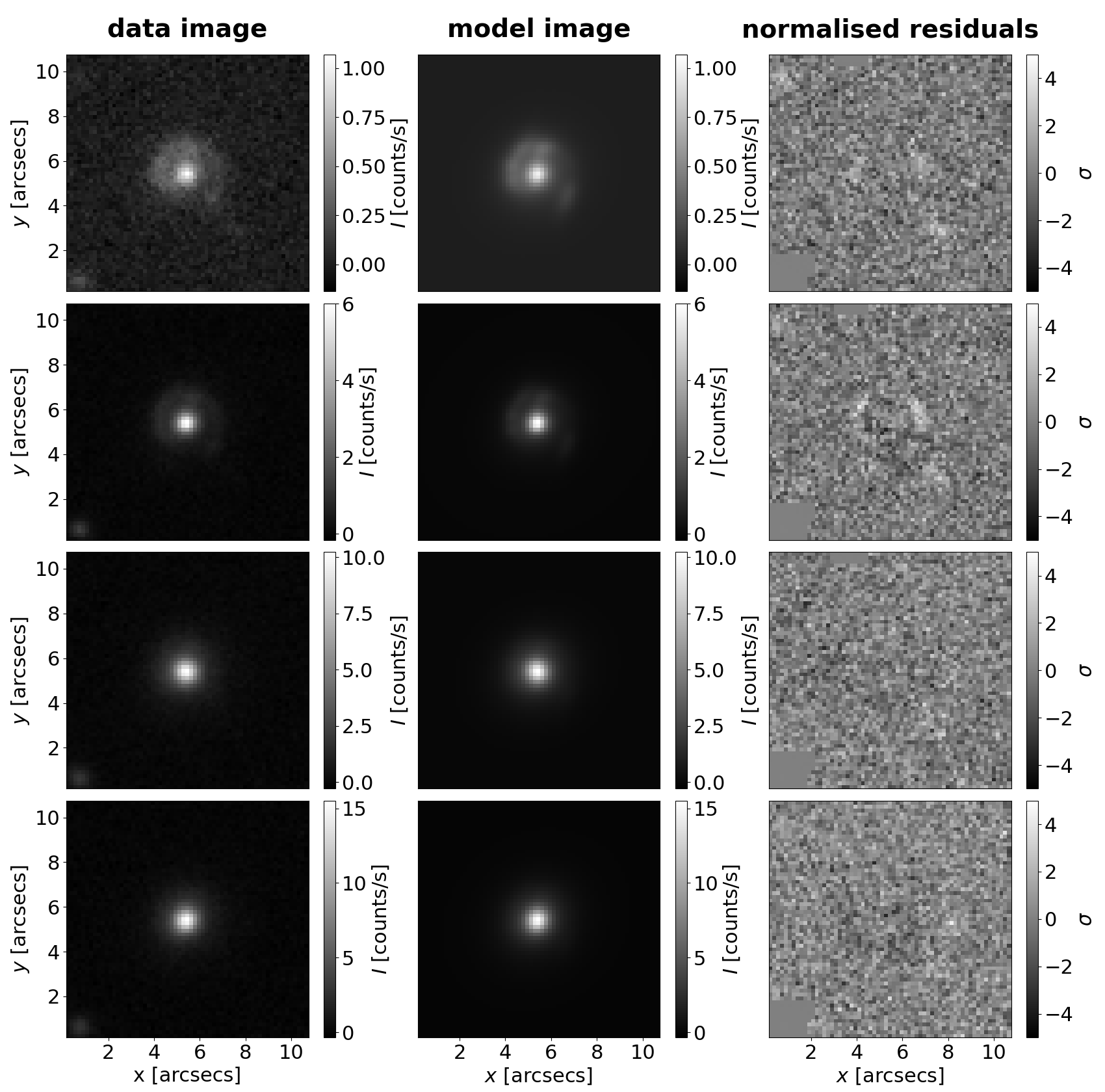}
    \caption{Fit of lens HSCJ023217$-$021703. Top to bottom: $griz$ filters.\label{fig:model_348}}
\end{center}
\end{figure}

\begin{figure}[h!]
  \begin{center}
    \includegraphics[trim=0 0 0 0, clip, width=\textwidth]{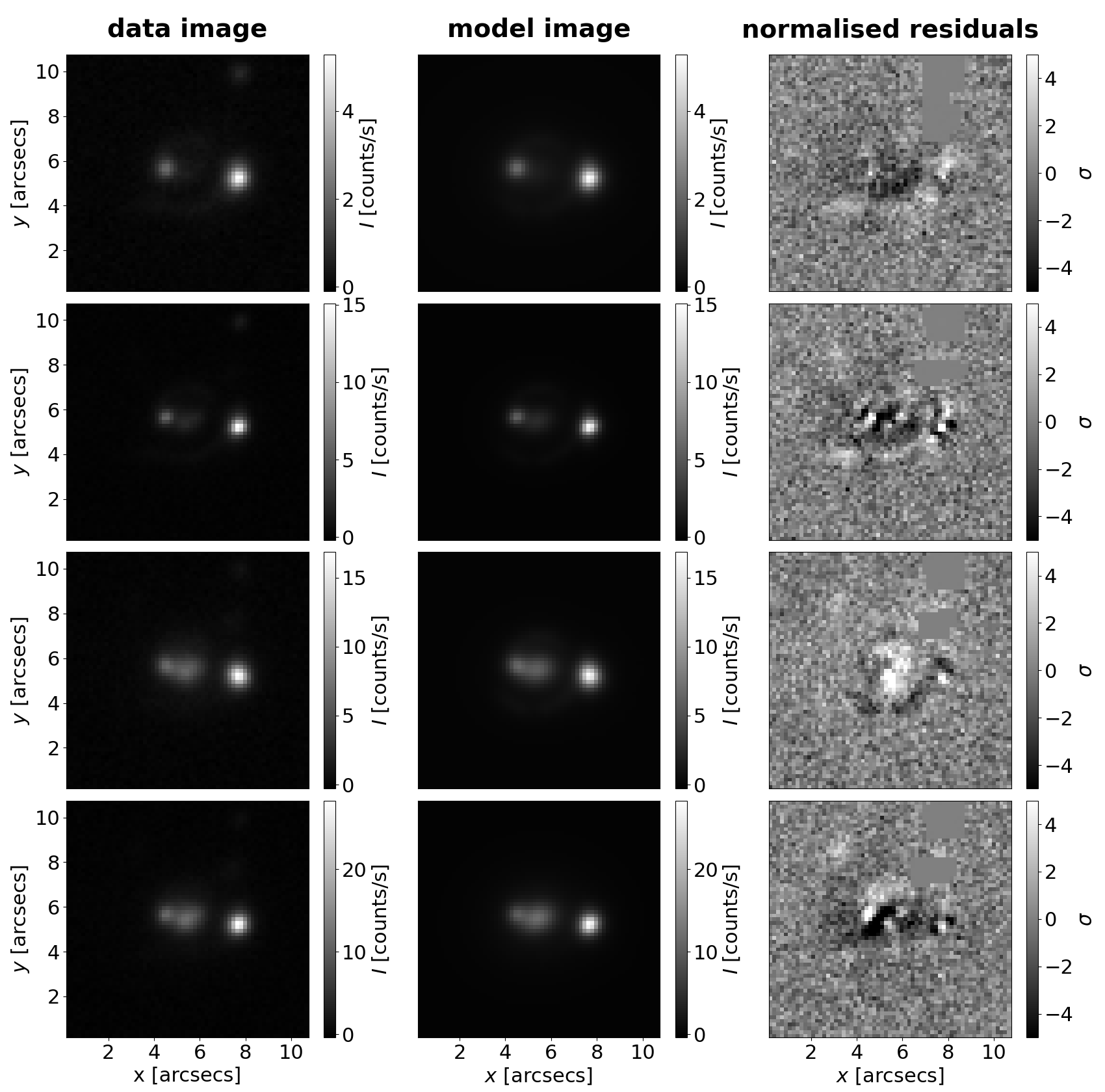}
    \caption{Fit of lens HSCJ023322$-$020530. Top to bottom: $griz$ filters.\label{fig:model_359}}
  \end{center}
\end{figure}
  \begin{figure}[h!]
    \begin{center}
    \includegraphics[trim=0 0 0 0, clip, width=\textwidth]{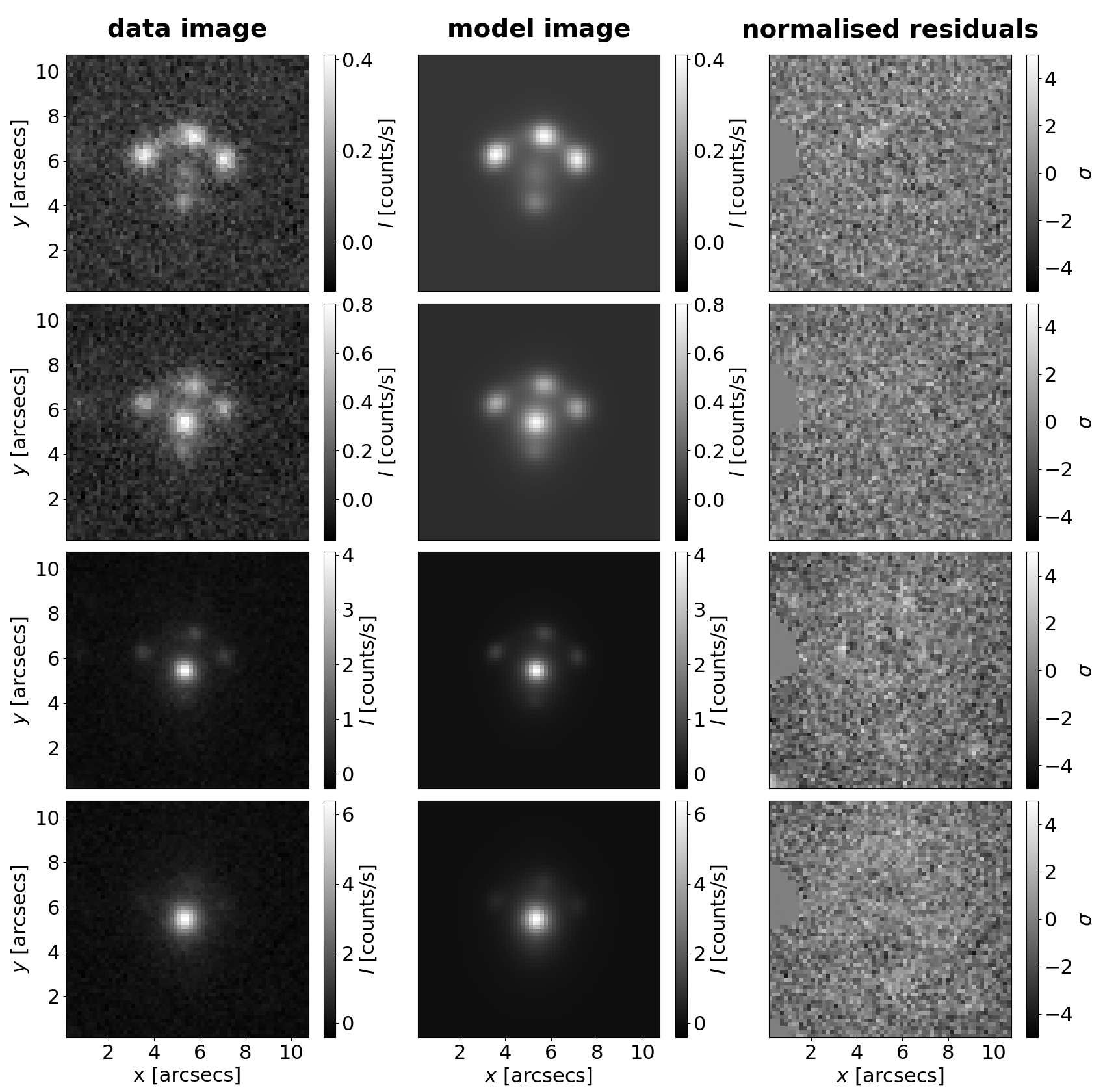}
    \caption{Fit of lens HSCJ085046$+$00390. Top to bottom: $griz$ filters.\label{fig:model_485}}
    \end{center}
\end{figure}

  \begin{figure}[h!]
  \begin{center}
    \includegraphics[trim=0 0 0 0, clip, width=\textwidth]{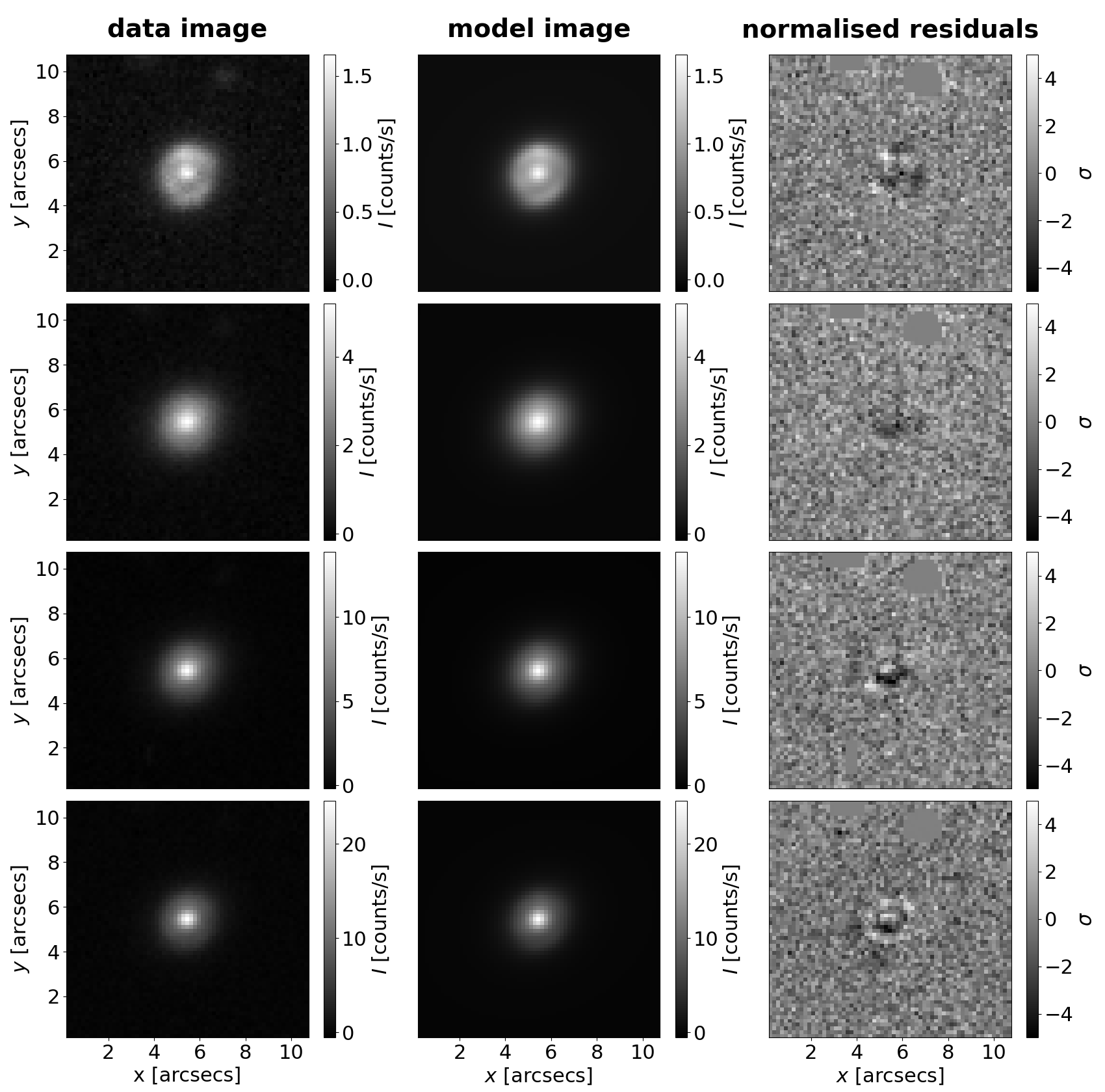}
    \caption{Fit of lens HSCJ085855$-$010208. Top to bottom: $griz$ filters.\label{fig:model_550}}
  \end{center}
  \end{figure}
  \begin{figure}[h!]
  \begin{center}
    \includegraphics[trim=0 0 0 0, clip, width=\textwidth]{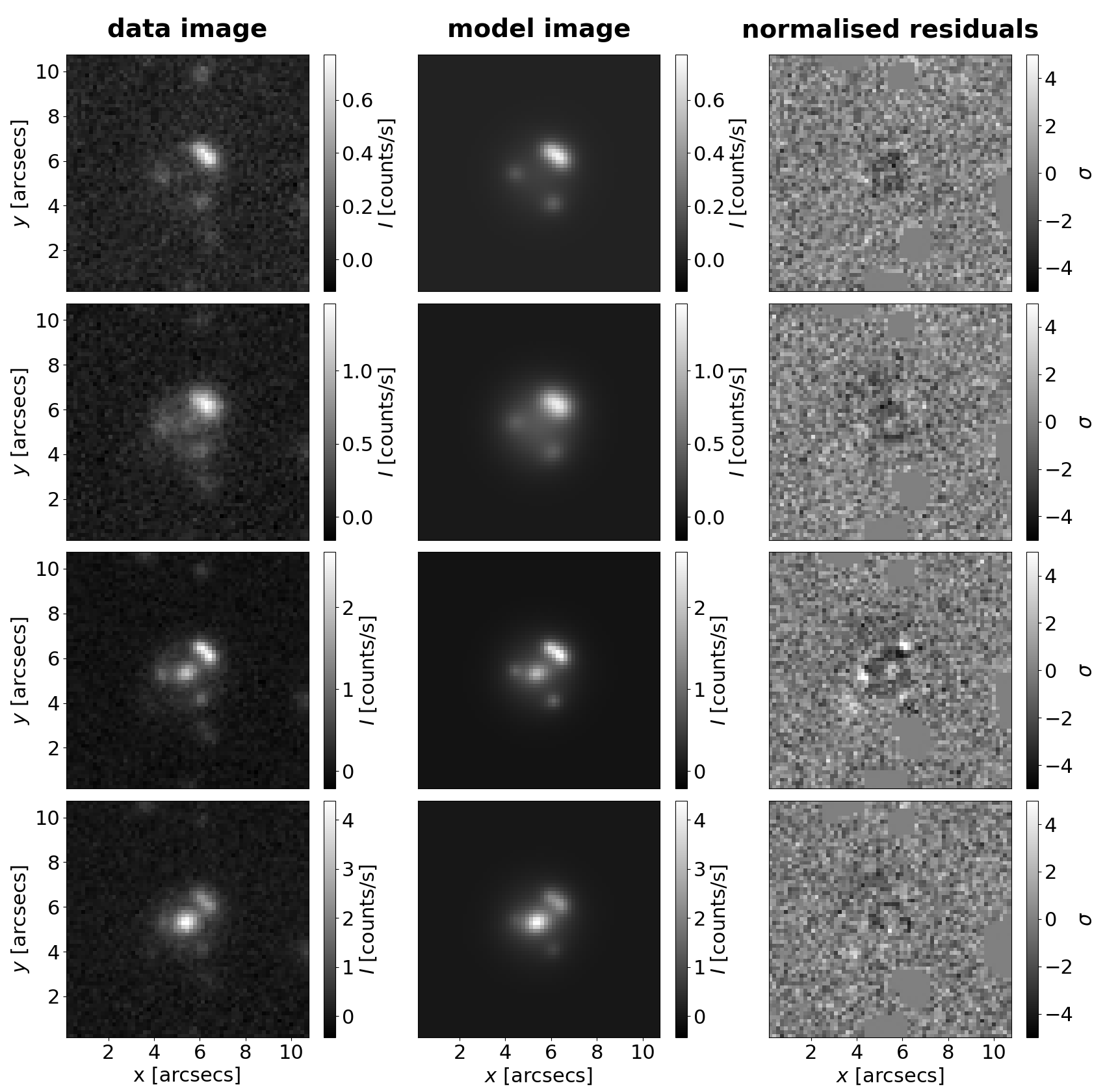}
    \caption{Fit of lens HSCJ090429$-$010228. Top to bottom: $griz$ filters.\label{fig:model_593}}
  \end{center}
\end{figure}

\begin{figure}[h!]
  \begin{center}
    \includegraphics[trim=0 0 0 0, clip, width=\textwidth]{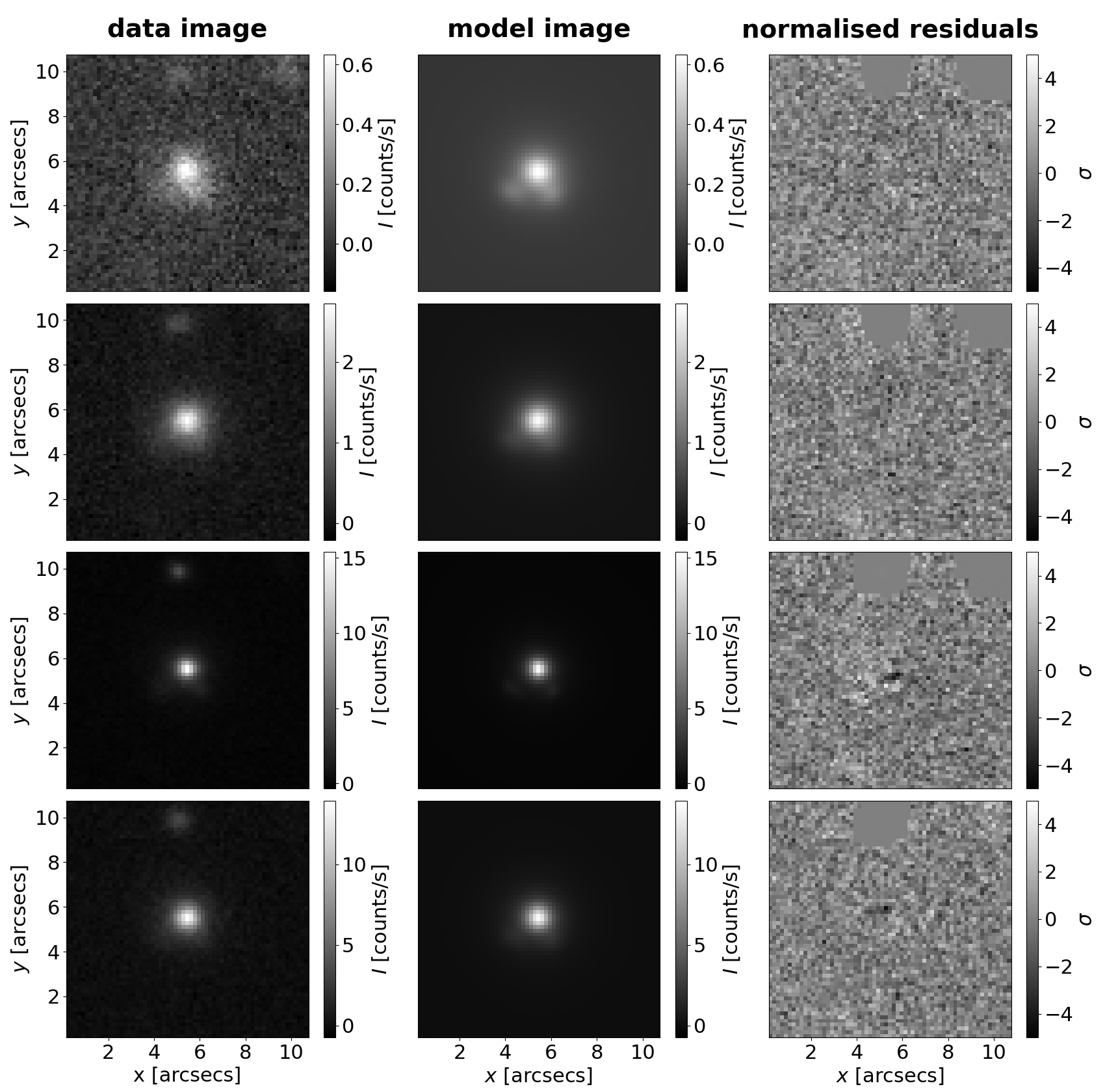}
    \caption{Fit of lens HSCJ094427$-$014742. Top to bottom: $griz$ filters.\label{fig:model_780}}
\end{center}
\end{figure}
  \begin{figure}[h!]
    \begin{center}
      \includegraphics[trim=0 0 0 0, clip, width=\textwidth]{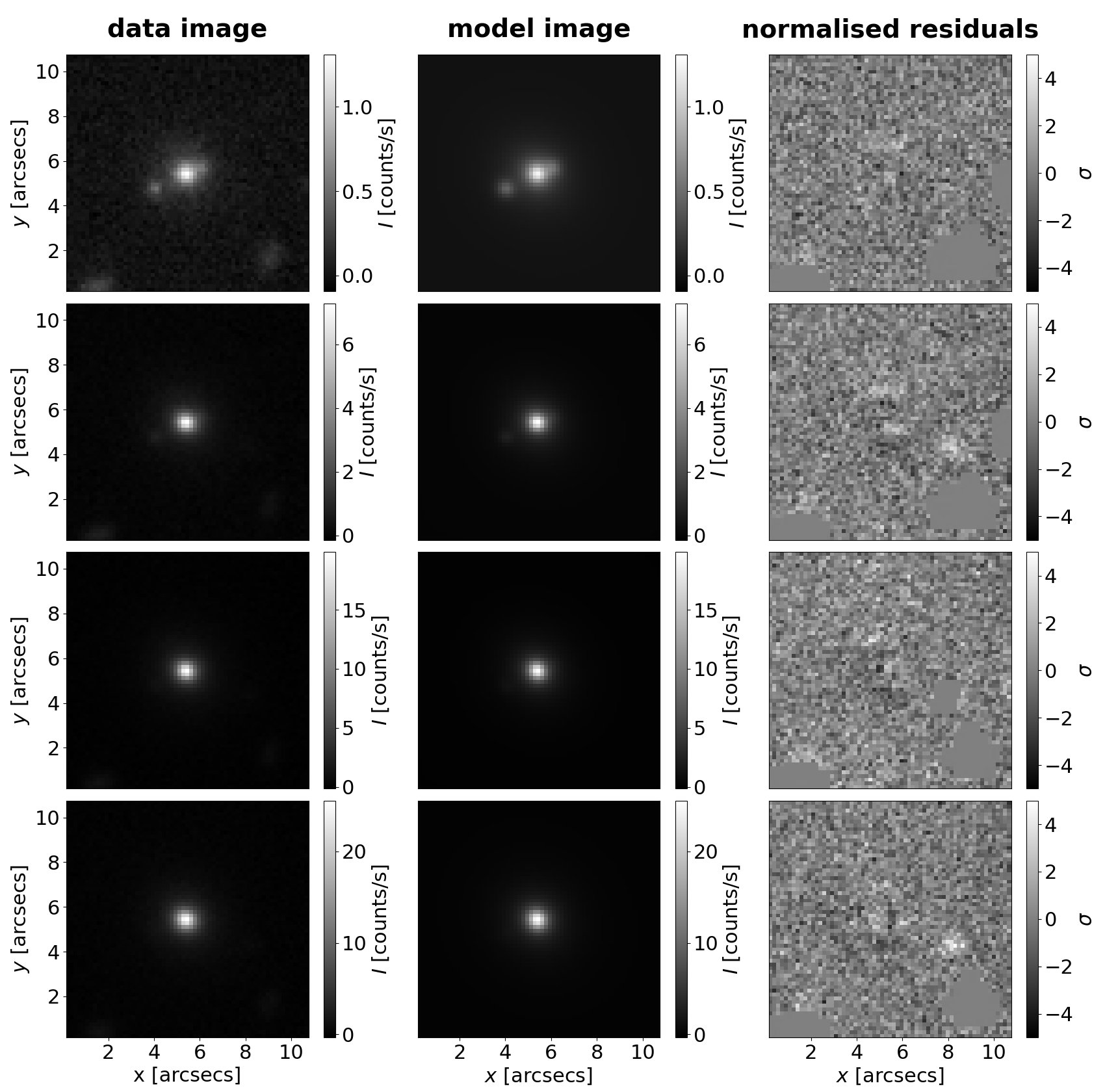}
    \caption{Fit of lens HSCJ120623$+$001507. Top to bottom: $griz$ filters.\label{fig:model_990}}
    \end{center}
\end{figure}

\begin{figure}[h!]
  \begin{center}
    \includegraphics[trim=0 0 0 0, clip, width=\textwidth]{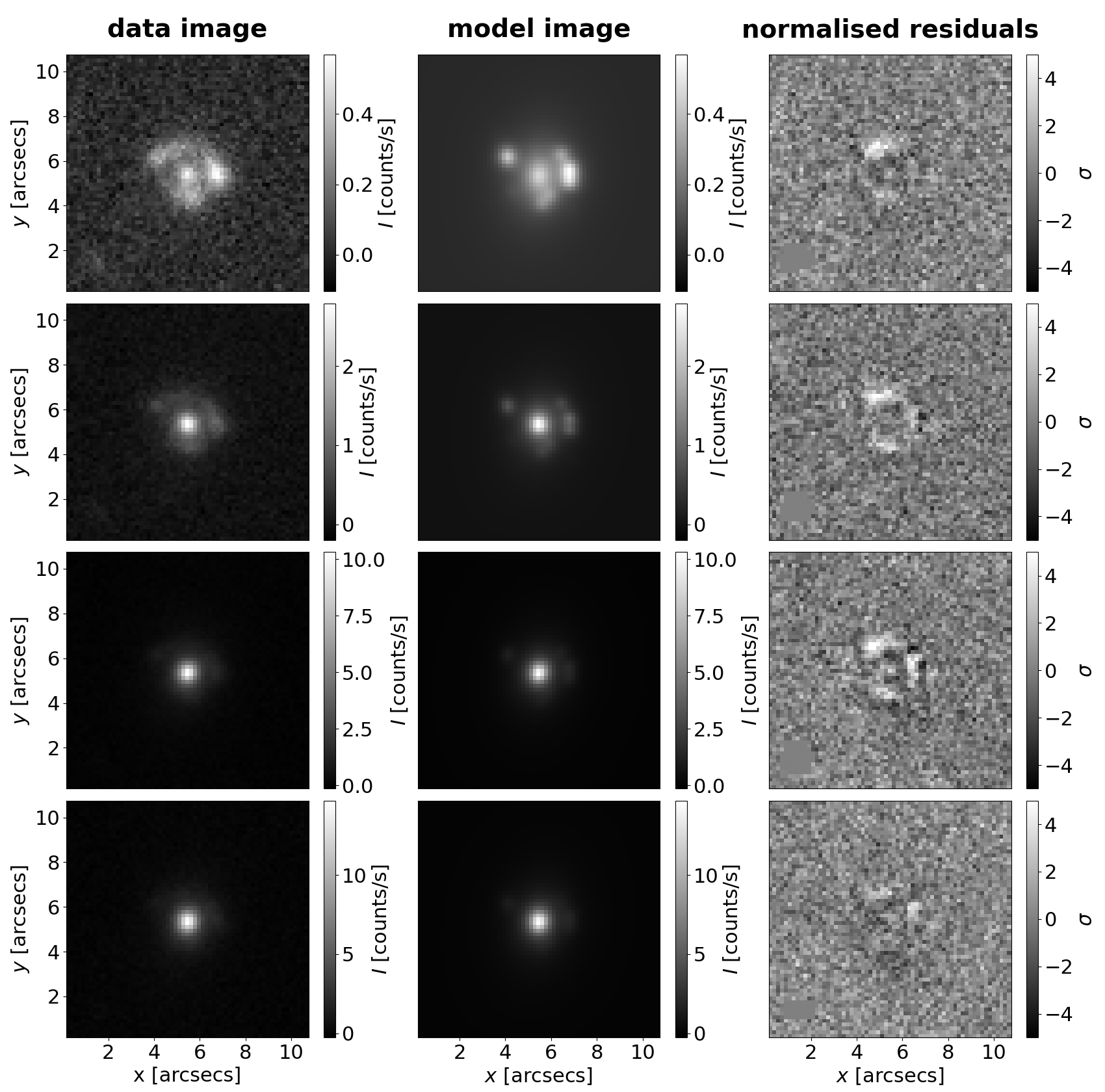}
    \caption{Fit of lens HSCJ121052$-$011905. Top to bottom: $griz$ filters.\label{fig:model_1011}}
  \end{center}
  \end{figure}
  \begin{figure}[h!]
    \begin{center}
      \includegraphics[trim=0 0 0 0, clip, width=\textwidth]{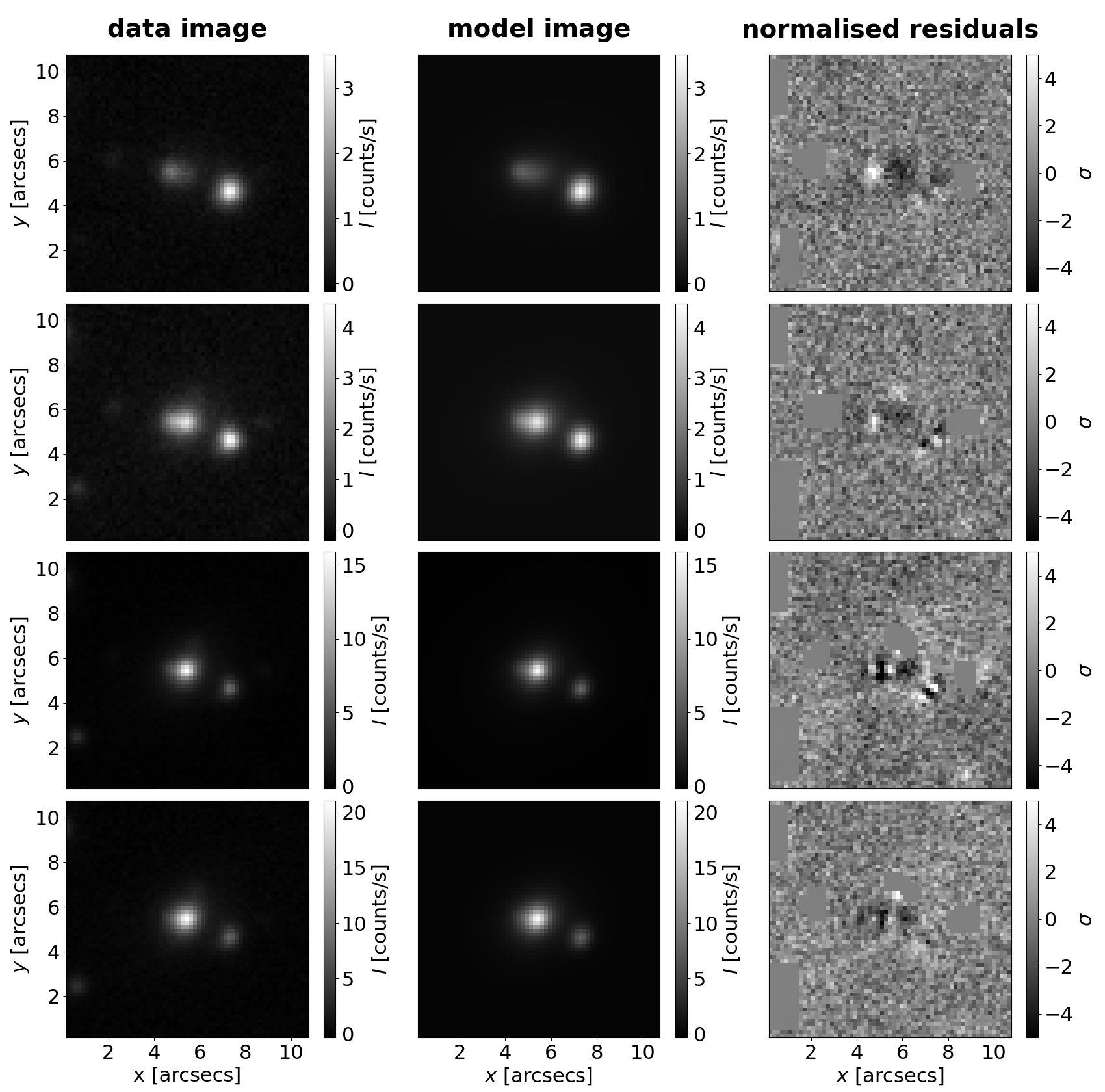}
    \caption{Fit of lens HSCJ121504$+$004726. Top to bottom: $griz$ filters.\label{fig:model_1030}}
    \end{center}
\end{figure}

\begin{figure}[h!]
  \begin{center}
    \includegraphics[trim=0 0 0 0, clip, width=\textwidth]{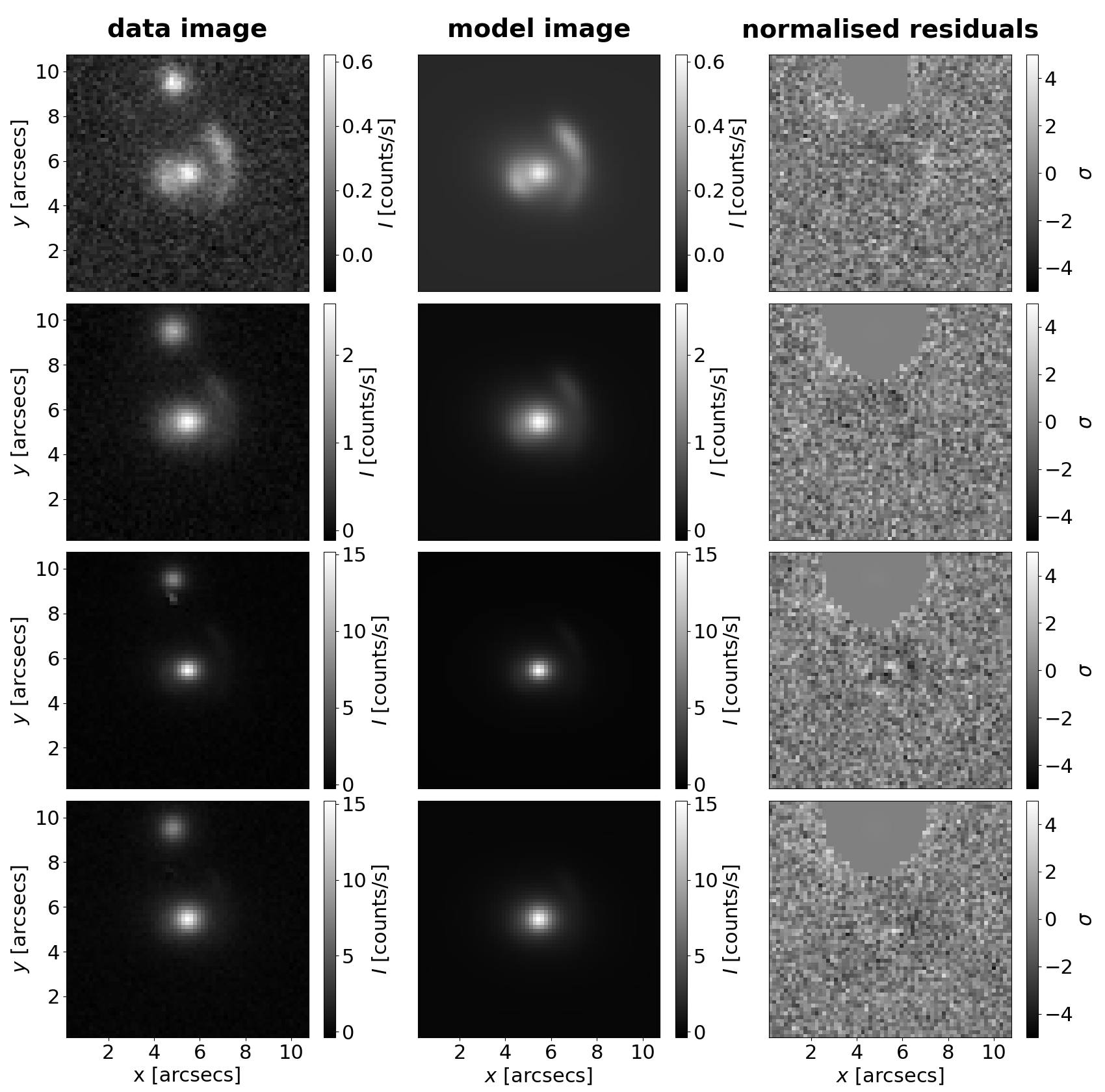}
    \caption{Fit of lens HSCJ124320$-$004517. Top to bottom: $griz$ filters.\label{fig:model_1074}}
  \end{center}
  \end{figure}
  \begin{figure}[h!]
    \begin{center}
      \includegraphics[trim=0 0 0 0, clip, width=\textwidth]{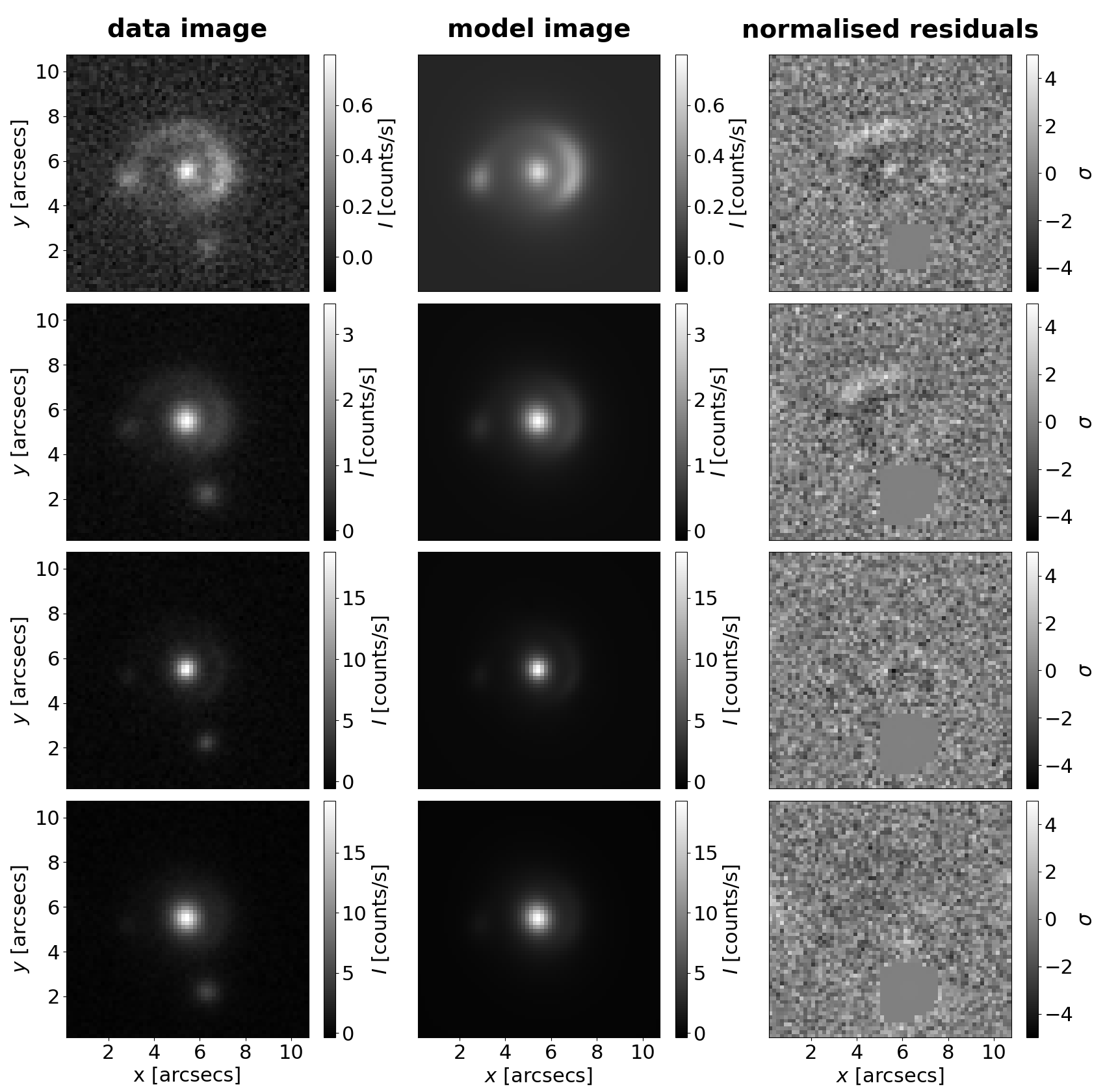}
    \caption{Fit of lens HSCJ125254$+$004356. Top to bottom: $griz$ filters.\label{fig:model_1084}}
    \end{center}
\end{figure}

\begin{figure}[h!]
  \begin{center}
    \includegraphics[trim=0 0 0 0, clip, width=\textwidth]{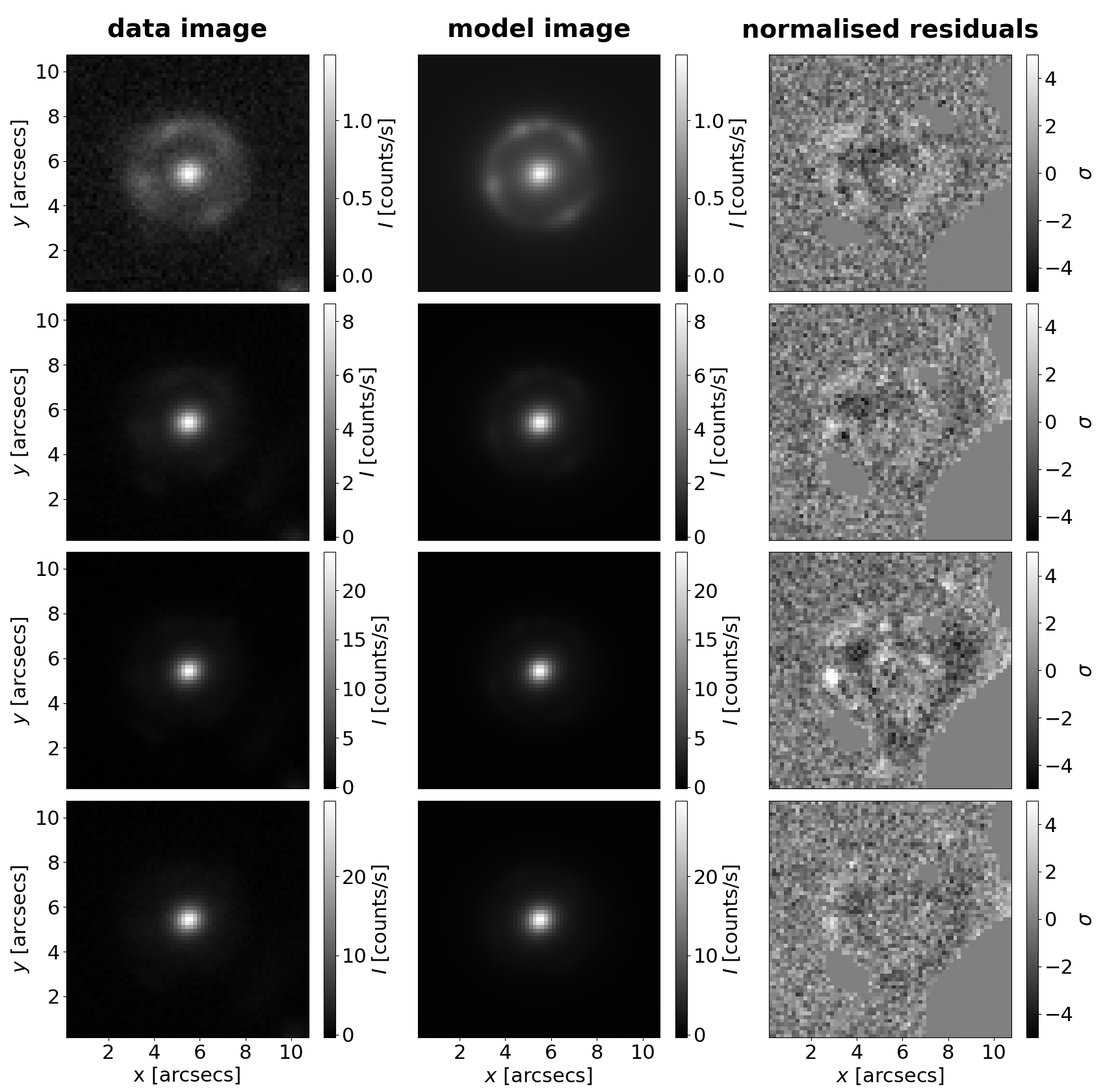}
    \caption{Fit of lens HSCJ135138$+$002839. Top to bottom: $griz$ filters.\label{fig:model_1121}}
  \end{center}
  \end{figure}
  \begin{figure}[h!]
    \begin{center}
      \includegraphics[trim=0 0 0 0, clip, width=\textwidth]{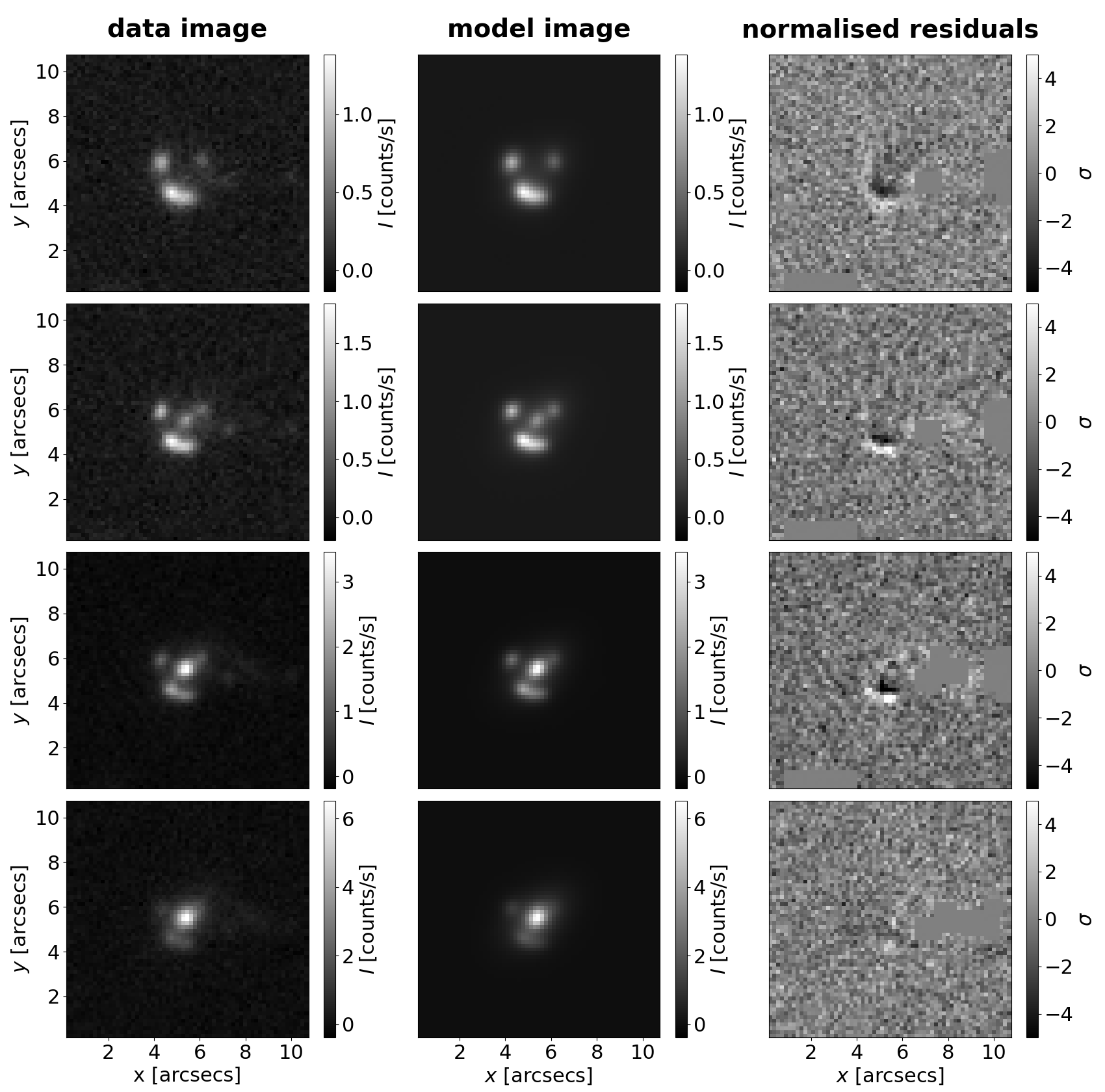}
    \caption{Fit of lens HSCJ141136$-$010215. Top to bottom: $griz$ filters.\label{fig:model_1188}}
    \end{center}
\end{figure}

\begin{figure}[h!]
  \begin{center}
    \includegraphics[trim=0 0 0 0, clip, width=\textwidth]{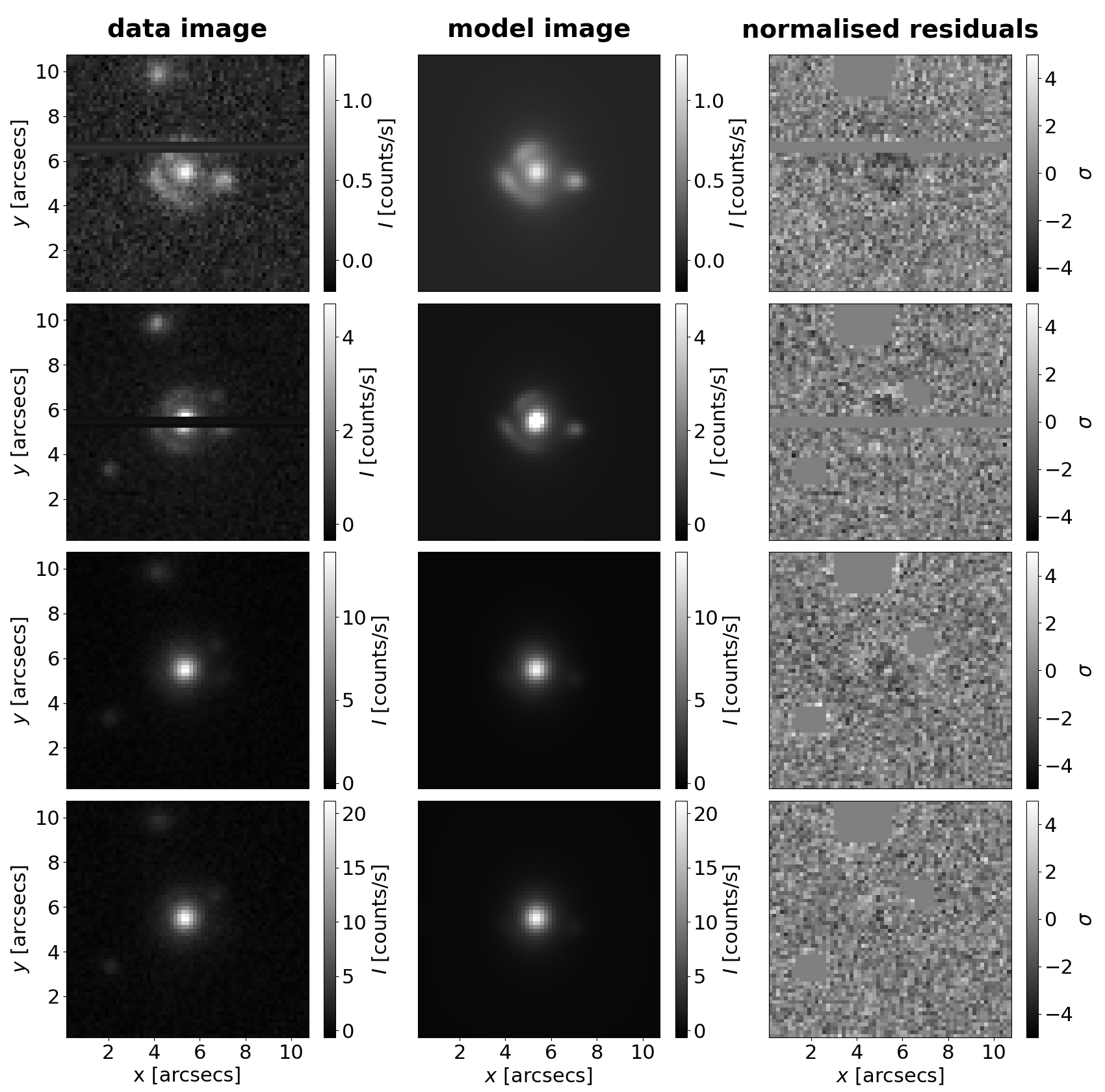}
    \caption{Fit of lens HSCJ141815$+$015832. Top to bottom: $griz$ filters.\label{fig:model_1244}}
  \end{center}
  \end{figure}
  \begin{figure}[h!]
    \begin{center}
      \includegraphics[trim=0 0 0 0, clip, width=\textwidth]{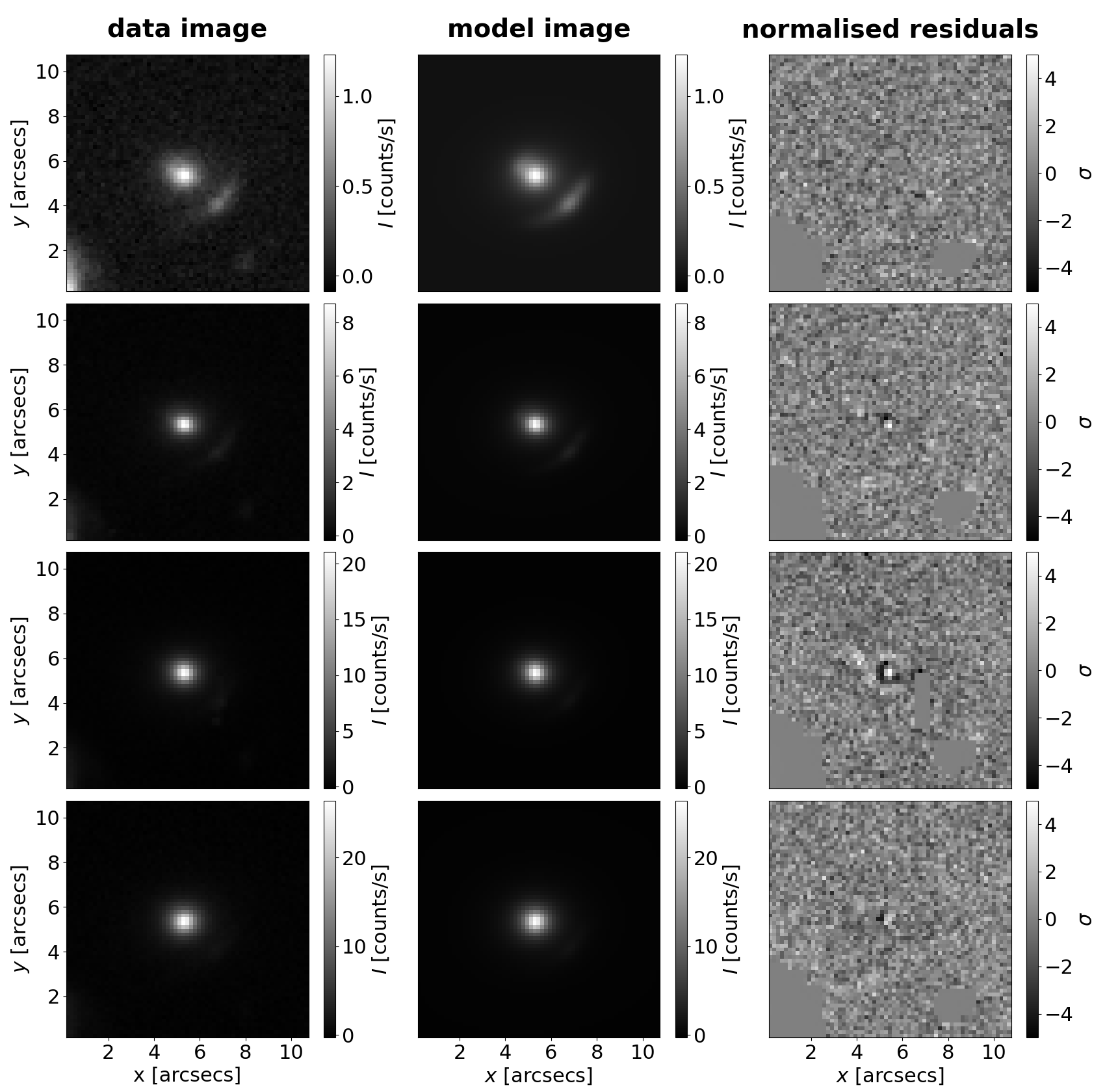}
    \caption{Fit of lens HSCJ142720$+$001916. Top to bottom: $griz$ filters.\label{fig:model_1318}}
    \end{center}
\end{figure}

\begin{figure}[h!]
  \begin{center}
    \includegraphics[trim=0 0 0 0, clip, width=\textwidth]{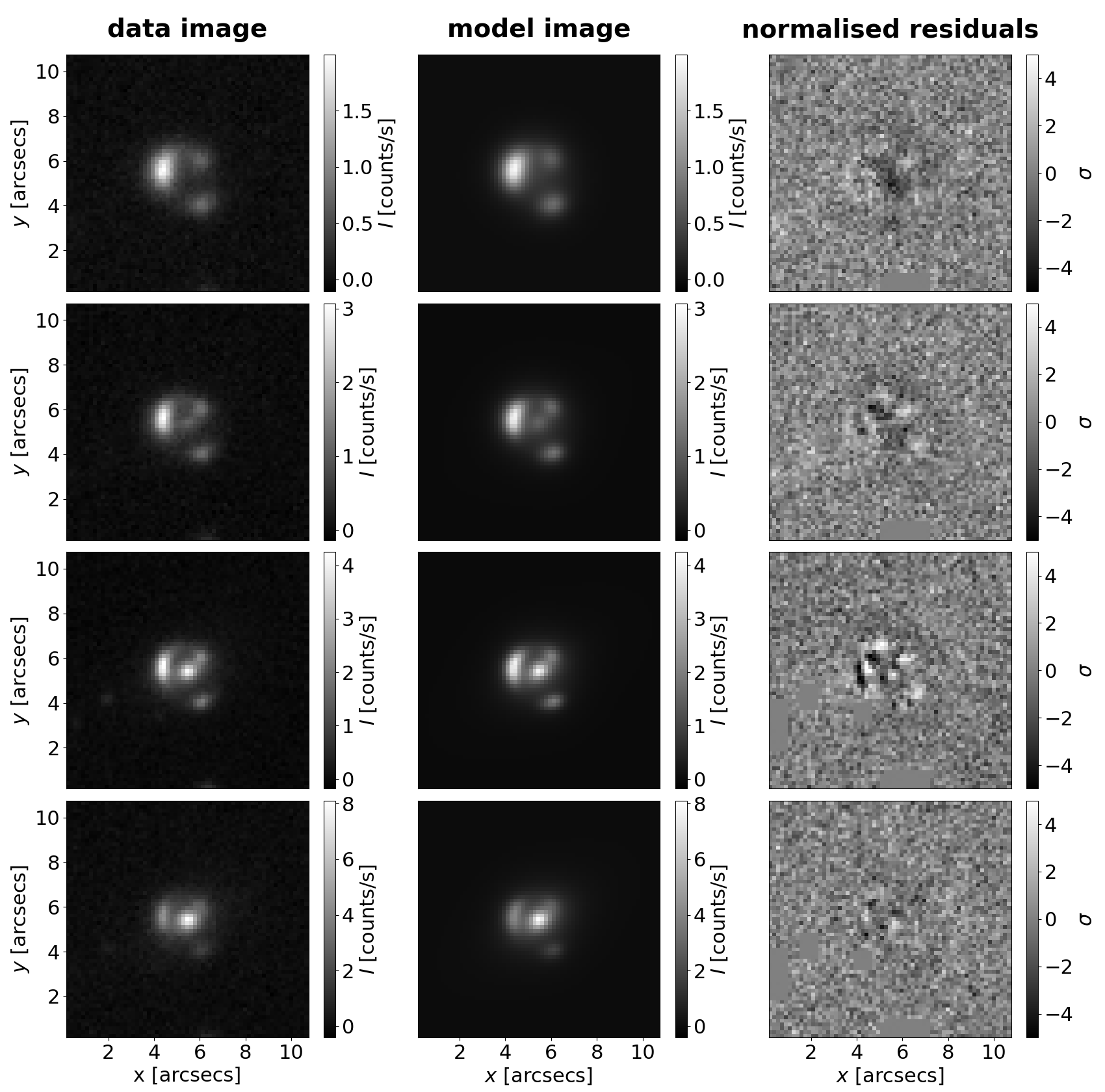}
    \caption{Fit of lens HSCJ144320$-$012537. Top to bottom: $griz$ filters.\label{fig:model_1390}}
  \end{center}
  \end{figure}
\begin{figure}[h!]
  \begin{center}
    \includegraphics[trim=0 0 0 0, clip, width=\textwidth]{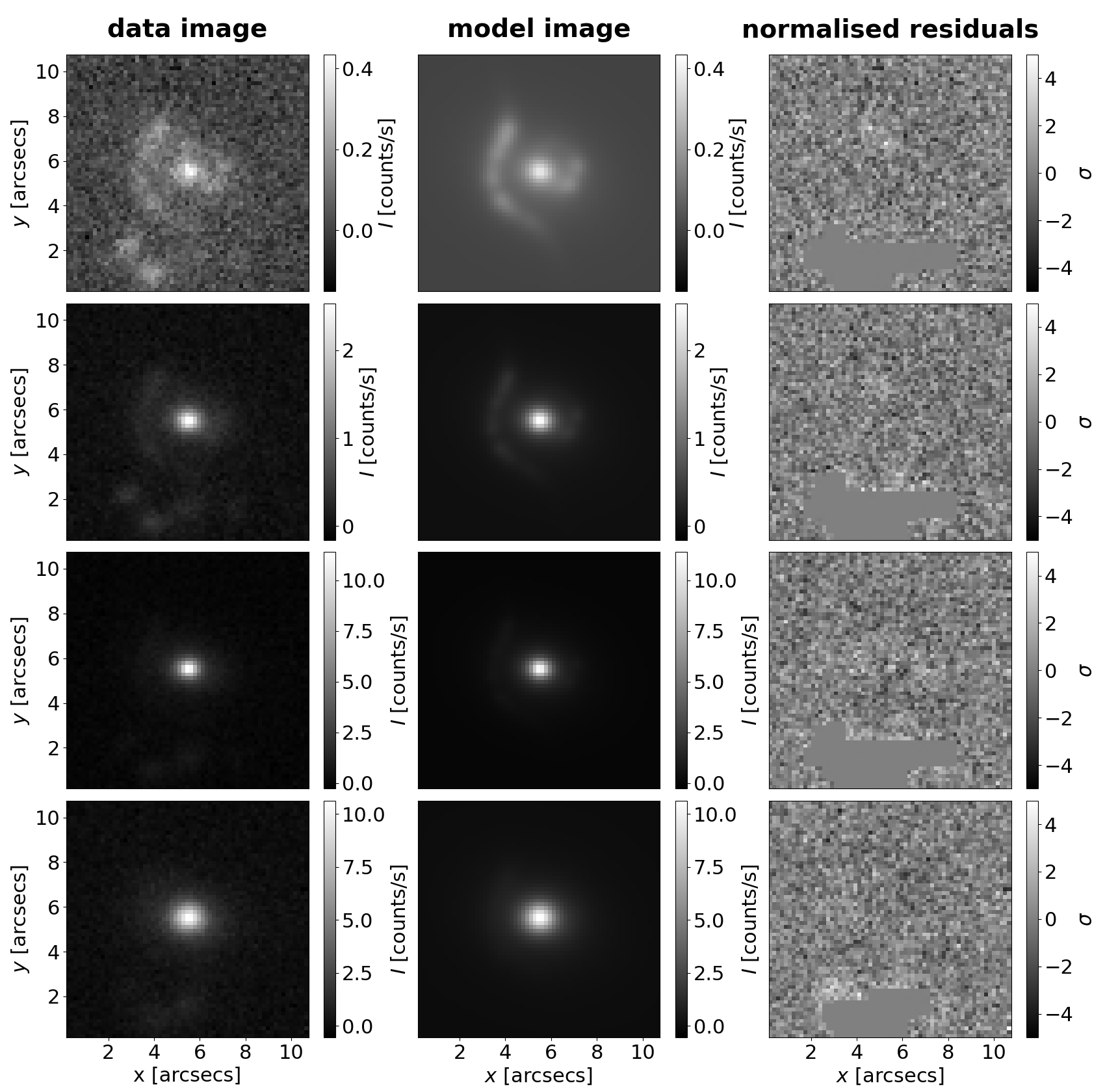}
    \caption{Fit of lens HSCJ145242$+$425731. Top to bottom: $griz$ filters.\label{fig:model_1433}}
  \end{center}
\end{figure}

\begin{figure}[h!]
  \begin{center}
    \includegraphics[trim=0 0 0 0, clip, width=\textwidth]{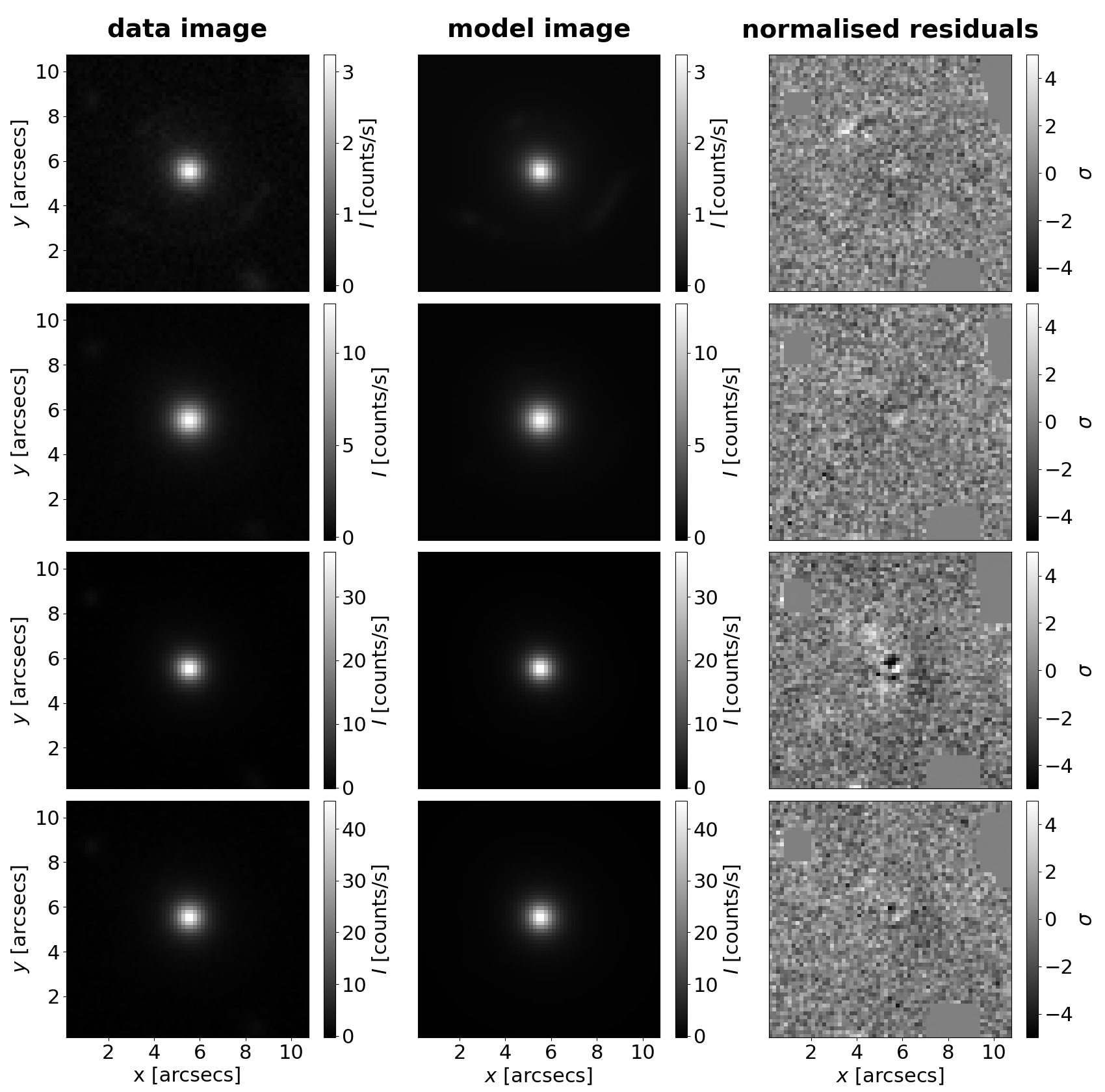}
    \caption{Fit of lens HSCJ150021$-$004936. Top to bottom: $griz$ filters.\label{fig:model_1470}}
    \end{center}
  \end{figure}
\begin{figure}[h!]
      \begin{center}
        \includegraphics[trim=0 0 0 0, clip, width=\textwidth]{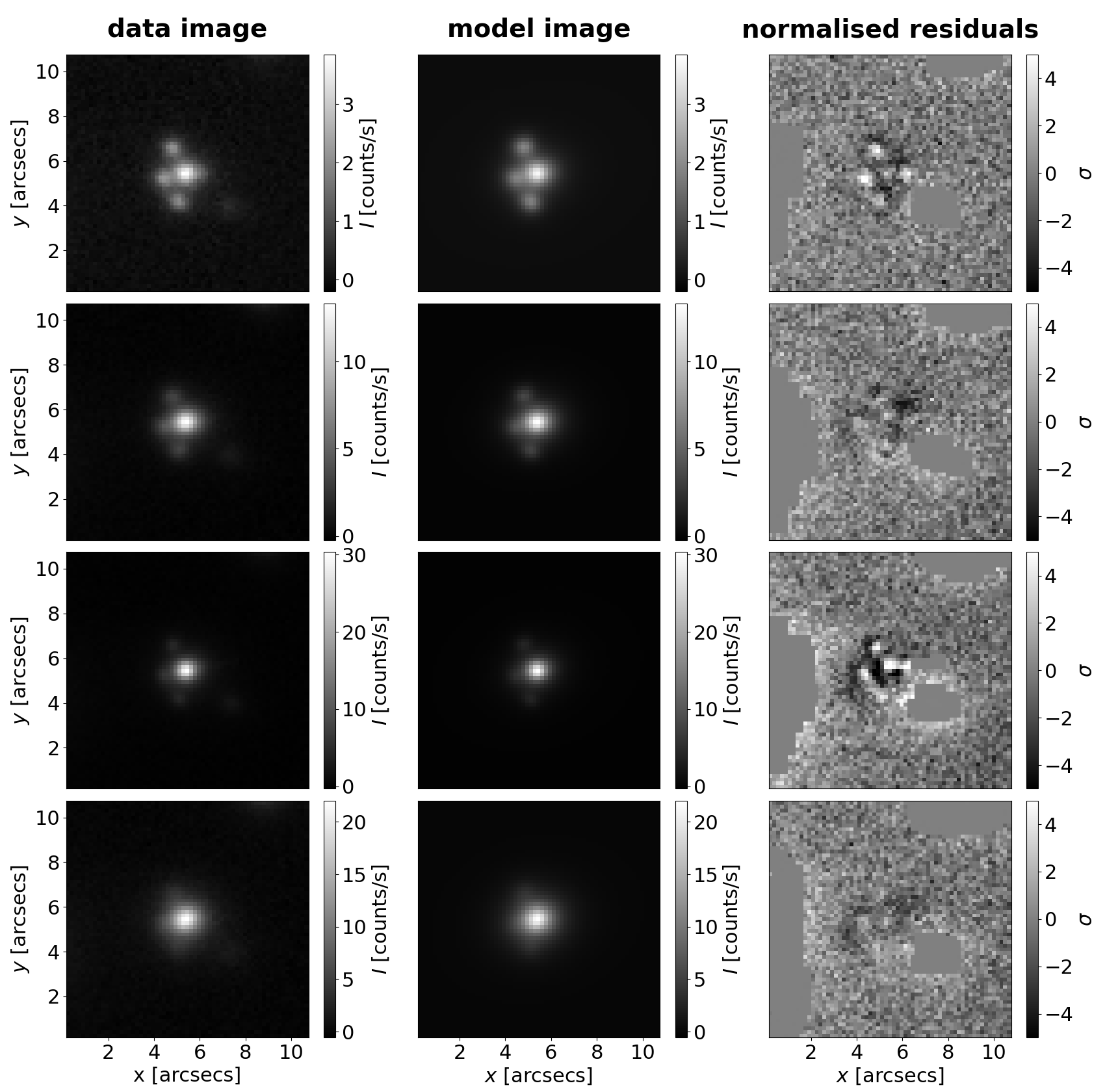}
    \caption{Fit of lens HSCJ150112$+$422113. Top to bottom: $griz$ filters.\label{fig:model_1472}}
      \end{center}
\end{figure}
  
\begin{figure}[h!]
  \begin{center}
    \includegraphics[trim=0 0 0 0, clip, width=\textwidth]{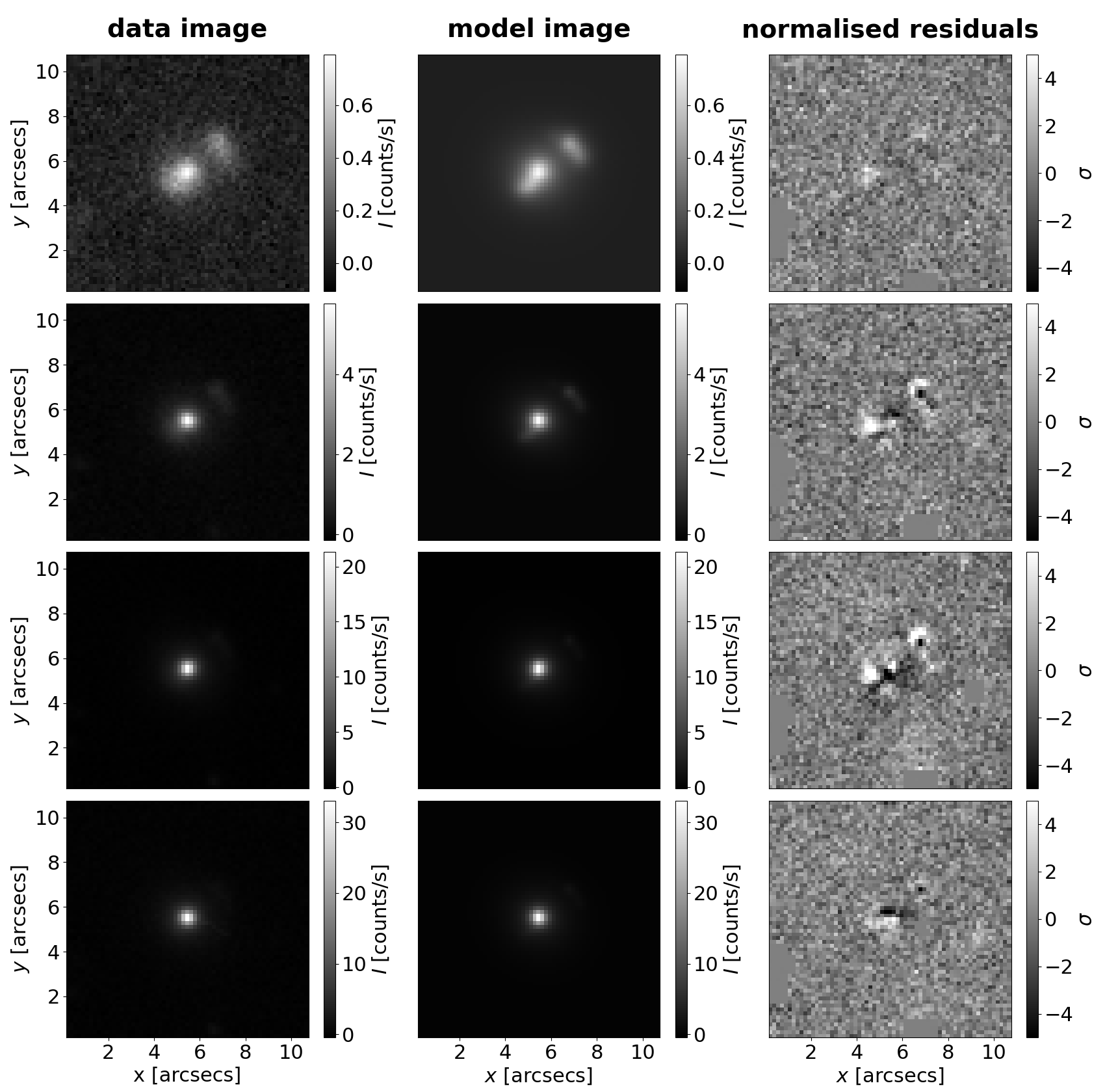}
    \caption{Fit of lens HSCJ223733$+$005015. Top to bottom: $griz$ filters.\label{fig:model_1826}}
  \end{center}
\end{figure}
\begin{figure}[h!]
  \begin{center}
    \includegraphics[trim=0 0 0 0, clip, width=\textwidth]{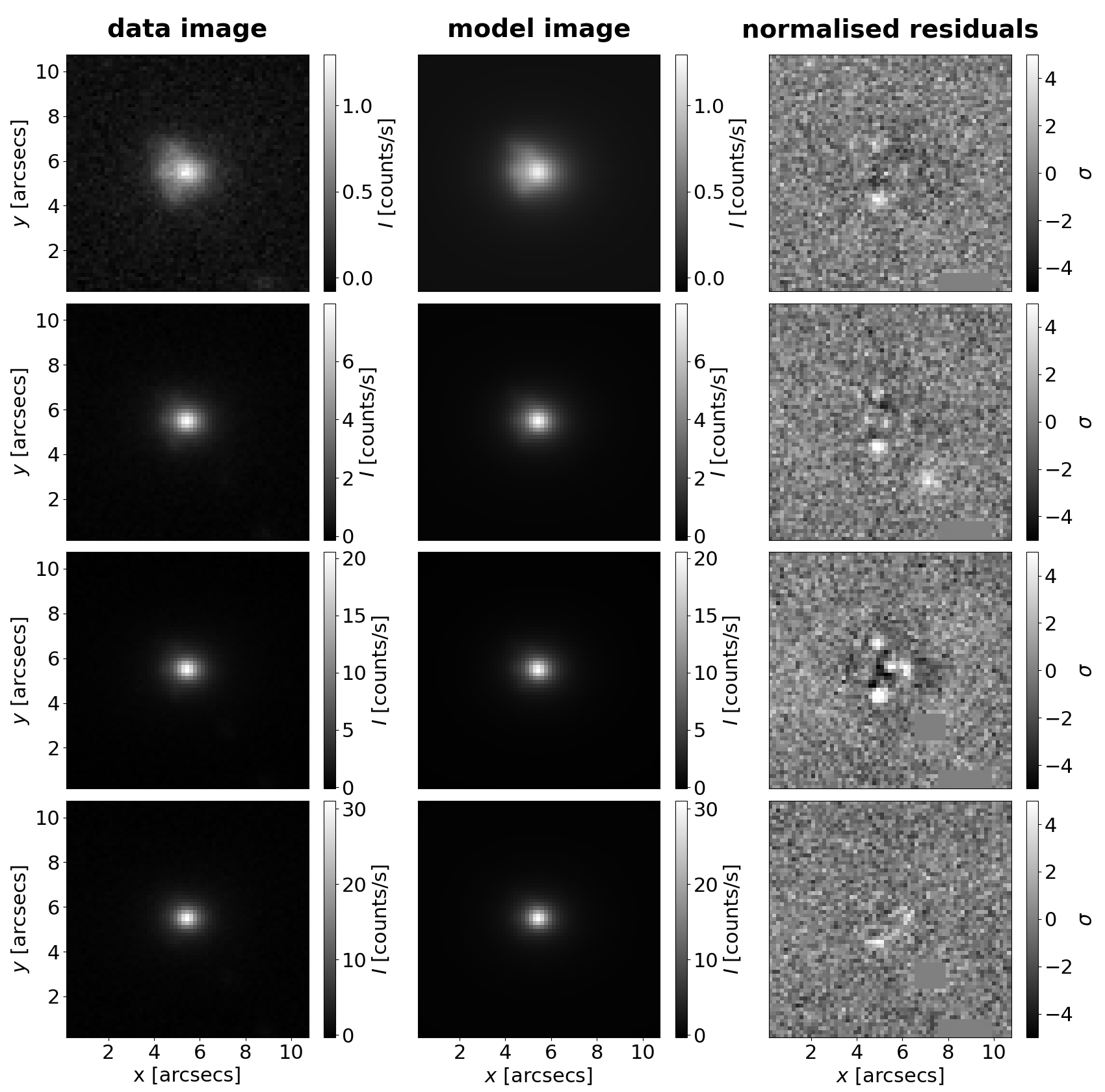}
    \caption{Fit of lens HSCJ230335$+$003703. Top to bottom: $griz$ filters.\label{fig:model_1908}}
  \end{center}
\end{figure}

\begin{figure}[h!]
  \begin{center}
    \includegraphics[trim=0 0 0 0, clip, width=\textwidth]{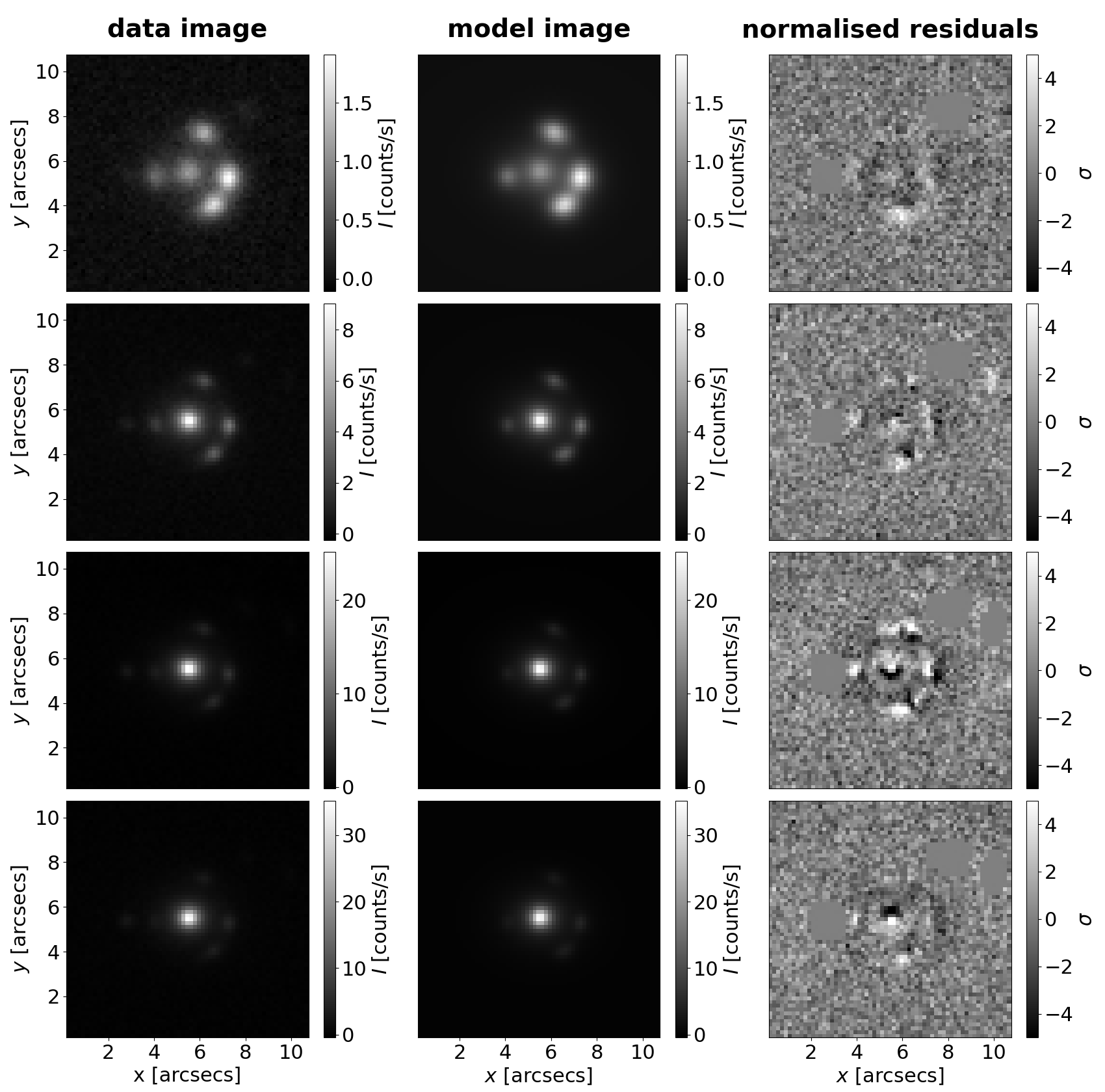}
    \caption{Fit of lens HSCJ230521$-$000211. Top to bottom: $griz$ filters.\label{fig:model_1911}}
  \end{center}
\end{figure}
\begin{figure}[h!]
  \begin{center}
    \includegraphics[trim=0 0 0 0, clip, width=\textwidth]{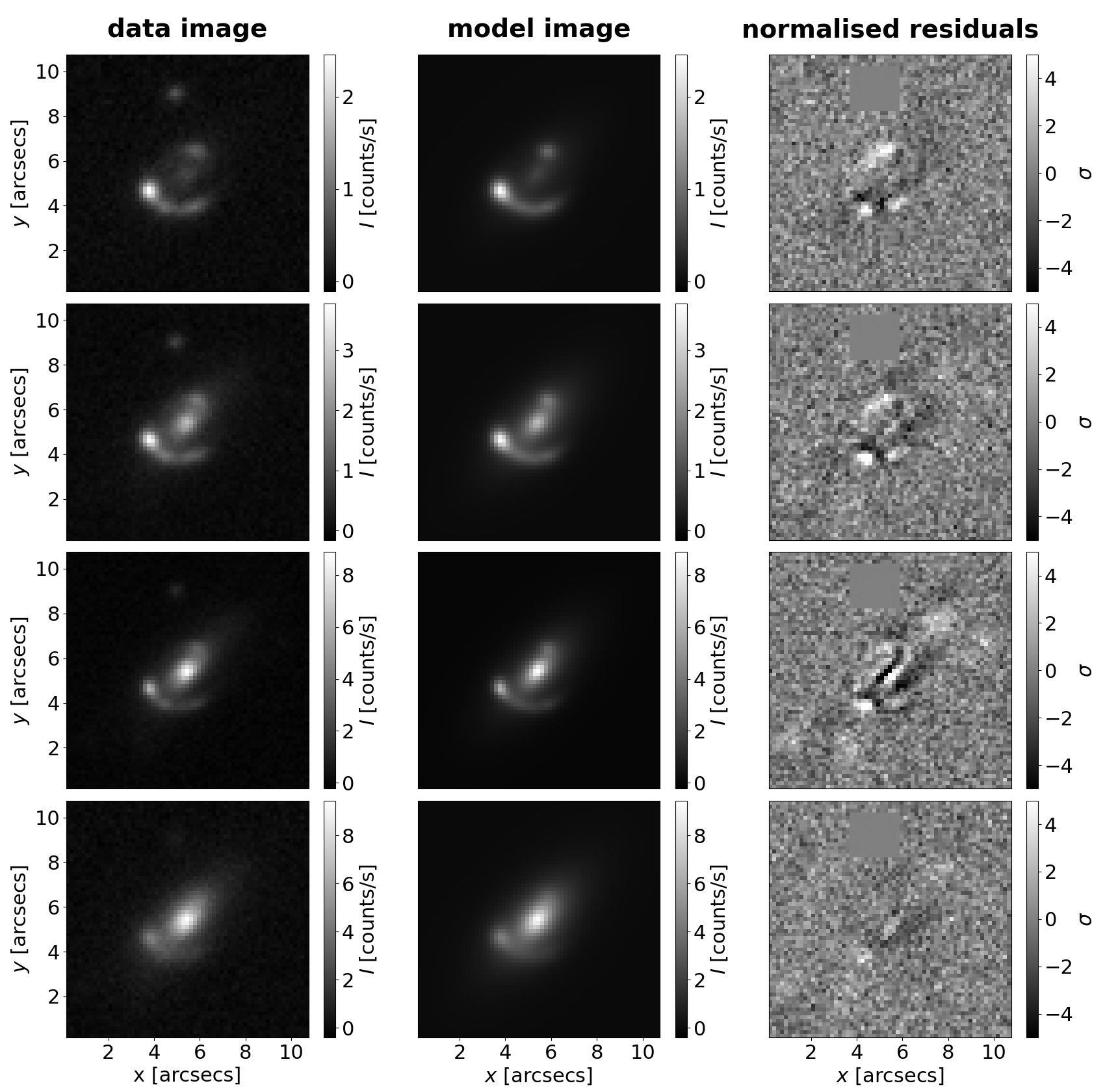}
    \caption{Fit of lens HSCJ233130$+$003733. Top to bottom: $griz$ filters.\label{fig:model_1975}}
  \end{center}
\end{figure}

\fi
\FloatBarrier

\begin{multicols}{2}
\section{Predicted image positions, Fermat potential, and time delays}
\label{app:ImPosdetails}

Tab.~\ref{tab:comparison_ImPos} lists the predicted image positions for all 31 lenses using both the mass model obtained with \GG \, and obtained with the network. We further report the distance $d$ (see Eq.~(\ref{eq:d})) and the source positions. For the \GG \, model, the latter is the magnification-weighted mean position of the final MCMC chain, and for the network the obtained source position from the first appearing image, both with respect to the image center. The corresponding Fermat potential differences and time delays of the predicted images are reported in Tab.~\ref{tab:comparison_Fermat}.
\end{multicols}

\onecolumn
\LTcapwidth=\textwidth
\begin{landscape}
\begin{longtable}{c|c|cccccccc|c|cc}
\caption{Resulting image positions and source position obtained with the \GG \, model and predicted by the network. \label{tab:comparison_ImPos}}\\

  Name   & method &\multicolumn{9}{c|}{image positions} & \multicolumn{2}{c}{source positions} \\
         &        &$\theta_\text{x,A} ~ [\arcsec]$&$\theta_\text{y,A} ~ [\arcsec]$&$\theta_\text{x,B} ~ [\arcsec]$&$\theta_\text{y,B} ~ [\arcsec]$&$\theta_\text{x,C} ~ [\arcsec]$&$\theta_\text{y,C} ~ [\arcsec]$&$\theta_\text{x,D} ~ [\arcsec]$&$\theta_\text{y,D} ~ [\arcsec]$& $d ~ [\arcsec]$ & $x_\text{s} ~ [\arcsec]$ & $y_\text{s} ~ [\arcsec]$ \\
  \hline
  \endfirsthead
  \caption*{\textbf{Table \ref{tab:comparison_ImPos} Continued:} Image positions and source position obtained from our \GG \, and predicted by the network.}\\
  Name   & method &\multicolumn{9}{c|}{image positions} & \multicolumn{2}{c}{predicted source positions} \\
         &        &$\theta_\text{x,A} ~ [\arcsec]$&$\theta_\text{y,A} ~ [\arcsec]$&$\theta_\text{x,B} ~ [\arcsec]$&$\theta_\text{y,B} ~ [\arcsec]$&$\theta_\text{x,C} ~ [\arcsec]$&$\theta_\text{y,C} ~ [\arcsec]$&$\theta_\text{x,D} ~ [\arcsec]$&$\theta_\text{y,D} ~ [\arcsec]$& $d ~ [\arcsec]$ & $x_\text{s} ~ [\arcsec]$ & $y_\text{s} ~ [\arcsec]$ \\
  \hline
  \endhead
  \rowcolor{Gray}
  \multirow{2}{*}{HSCJ015618$-$010747} & ResNet & $ 1.34 $ & $ -0.28 $  & $ -0.41 $ & $ 0.45 $  & $-$ & $-$  & $-$ & $-$ & $ \multirow{2}{*}{0.236}$ & $0.53 $ & $ 0.46$ \\
  & \GG &  $ 1.35 $ & $ -0.28 $  & $ -0.19 $ & $ 0.03 $  & $-$ & $-$  & $-$ & $-$ & & $\equiv 0.62 $ & $ \equiv 0.85$ \\
  \hline
  \rowcolor{Gray}
  \tabspace
  \multirow{2}{*}{HSCJ020141$-$030946} & ResNet & $ 1.02 $ & $ 1.52 $  & $ -0.15 $ & $ -1.59 $  & $-$ & $-$  & $-$ & $-$ & $ \multirow{2}{*}{0.328}$ & $ 0.21 $ & $ 0.08$ \\
  & \GG & $ 1.02 $ & $ 1.52 $  & $ 0.08 $ & $ -0.98 $  & $-$ & $-$  & $-$ & $-$ & & $\equiv -0.36 $ & $ \equiv -0.03$ \\
  \hline
  \rowcolor{Gray}
  \tabspace
  \multirow{2}{*}{HSCJ020241$-$064611} & ResNet & $ 0.22 $ & $ -1.38 $  & $ 0.02 $ & $ 0.67 $  & $-$ & $-$  & $-$ & $-$ & $ \multirow{2}{*}{0.2}$ & $0.26 $ & $ -0.26$ \\
  & \GG & $ 0.22 $ & $ -1.38 $  & $ 0.27 $ & $ 0.98 $  & $-$ & $-$  & $-$ & $-$ & & $\equiv 0.22 $ & $ \equiv 0.7$\\
  \hline
  \rowcolor{Gray}
  \tabspace
  \multirow{2}{*}{HSCJ020955$-$024442} & ResNet & $ 0.94 $ & $ 0.3 $  & $ -1.03 $ & $ 0.21 $  & $ 0.34 $ & $ 0.84 $  & $ 0.09 $ & $ -1.06 $ & $ \multirow{2}{*}{0.311}$ & $-0.2 $ & $ -0.02$ \\
  & \GG & $ 0.93 $ & $ 0.31 $  & $ -1.08 $ & $ -0.22 $  & $ -0.34 $ & $ 1.01 $  & $ 0.13 $ & $ -0.98 $ & & $\equiv 0.12 $ & $ \equiv -0.14$ \\
  \hline
  \rowcolor{Gray}
  \tabspace
  \multirow{2}{*}{HSCJ021737$-$051329} & ResNet & $ -1.03 $ & $ 0.98 $  & $ 1.05 $ & $ -0.06 $  & $-$ & $-$  & $-$ & $-$ & $ \multirow{2}{*}{/} $ & $0.14 $ & $ 0.35$ \\
  & \GG & $ -1.03 $ & $ 0.98 $  & $ -1.01 $ & $ -0.82 $  & $ -1.3 $ & $ -0.18 $  & $ 0.73 $ & $ 0.1 $ & & $\equiv 0.07 $ & $ \equiv -0.27$ \\
  \hline \multicolumn{1}{c}{} & \multicolumn{1}{c}{} & \\
  \hline 
  \rowcolor{Gray}
  \tabspace
  \multirow{2}{*}{HSCJ022346$-$053418} & ResNet & $ -1.6 $ & $ 0.46 $ & $-$ & $-$ & $-$ & $-$ & $-$ & $-$ & \multirow{2}{*}{$/$} & $-0.876 $ & $ 0.314$ \\
  & \GG &  $ -1.6 $ & $ 0.47 $  & $ -1.06 $ & $ 1.23 $  & $ -1.26 $ & $ 1.05 $  & $ 0.72 $ & $ -0.64 $ & & $\equiv -0.296 $ & $ \equiv 0.224$ \\
  \hline\rowcolor{Gray}
  \tabspace
  \multirow{2}{*}{HSCJ022610$-$042011} & ResNet & $ -0.23 $ & $ -1.75 $  & $ -0.08 $ & $ 0.58 $  & $-$ & $-$  & $-$ & $-$ & $ \multirow{2}{*}{0.129}$ & $0.02 $ & $ -0.57$ \\
  & \GG & $ -0.23 $ & $ -1.75 $  & $ 0.08 $ & $ 0.38 $  & $-$ & $-$  & $-$ & $-$ & & $\equiv -0.28 $ & $ \equiv -0.28$ \\
  \hline\rowcolor{Gray}
  \tabspace
  \multirow{2}{*}{HSCJ023217$-$021703} & ResNet & $ 1.23 $ & $ -1.03 $  & $ 0.15 $ & $ 1.31 $  & $-$ & $-$  & $-$ & $-$ & $ \multirow{2}{*}{/} $ & $0.08 $ & $ -0.06$ \\ 
  & \GG & $ 1.25 $ & $ -1.01 $  & $ -1.03 $ & $ 0.72 $  & $ 0.17 $ & $ 1.08 $  & $ -1.03 $ & $ -0.62 $ & & $\equiv 0.37 $ & $ \equiv -0.03$ \\
  \hline\rowcolor{Gray}
  \tabspace
  \multirow{2}{*}{HSCJ023322$-$020530} & ResNet & $ 2.24 $ & $ -0.2 $  & $ -0.39 $ & $ 0.17 $  & $-$ & $-$  & $-$ & $-$ & $ \multirow{2}{*}{0.289}$ & $0.69 $ & $ -0.14$ \\
  & \GG & $ 2.24 $ & $ -0.2 $  & $ -0.96 $ & $ 0.22 $  & $-$ & $-$  & $-$ & $-$ & & $\equiv 0.1 $ & $ \equiv -0.73$ \\
  \hline\rowcolor{Gray}
  \tabspace
  \multirow{2}{*}{HSCJ085046$+$003905} & ResNet & $ -1.86 $ & $ 0.88 $  & $ 1.13 $ & $ -0.71 $  & $-$ & $-$  & $-$ & $-$ & $ \multirow{2}{*}{/} $ & $-0.16 $ & $ 0.62$ \\
  & \GG & $ -1.86 $ & $ 0.88 $  & $ 1.72 $ & $ 0.65 $  & $ 0.17 $ & $ 1.73 $  & $ -0.16 $ & $ -1.29 $ & & $\equiv -0.41 $ & $ \equiv 0.55$ \\
  \hline \multicolumn{1}{c}{} & \multicolumn{1}{c}{} & \\ 
  \hline
  \rowcolor{Gray}
  \tabspace
  \multirow{2}{*}{HSCJ085855$-$010208} & ResNet & $ 0.51 $ & $ -0.99 $  & $ 0.61 $ & $ 0.84 $  & $-$ & $-$  & $-$ & $-$ & $ \multirow{2}{*}{0.423}$ & $0.35 $ & $ -0.22$ \\
  & \GG & $ 0.51 $ & $ -0.99 $  & $ -0.23 $ & $ 0.98 $  & $-$ & $-$  & $-$ & $-$ & & $\equiv -0.04 $ & $ \equiv -0.62$ \\
  \hline
  \rowcolor{Gray}
  \tabspace
  \multirow{2}{*}{HSCJ090429$-$010228} & ResNet & $ 0.54 $ & $ -1.41 $  & $ 0.99 $ & $ 1.07 $  & $ -0.82 $ & $ -1.03 $  & $ -1.32 $ & $ 0.19 $ & $ \multirow{2}{*}{0.776}$ & $0.32 $ & $ -0.02$ \\
  & \GG & $ 0.55 $ & $ -1.41 $  & $ 0.79 $ & $ 0.85 $  & $ 0.98 $ & $ 0.62 $  & $ -1.02 $ & $ -0.0 $ & & $\equiv 0.41 $ & $ \equiv -0.42$ \\
  \hline
  \rowcolor{Gray}
  \tabspace
  \multirow{2}{*}{HSCJ094427$-$014742} & ResNet & $ -1.15 $ & $ -0.8 $  & $ 0.28 $ & $ 0.26 $  & $-$ & $-$  & $-$ & $-$ & $ \multirow{2}{*}{0.209}$ & $-0.41 $ & $ -0.13$ \\
  & \GG & $ -1.15 $ & $ -0.8 $  & $ 0.03 $ & $ 0.59 $  & $-$ & $-$  & $-$ & $-$ & & $\equiv -0.68 $ & $ \equiv -0.6$ \\
  \hline
  \rowcolor{Gray}
  \tabspace
  \multirow{2}{*}{HSCJ120623$+$001507} & ResNet & $ -1.36 $ & $ -0.71 $  & $ 0.7 $ & $ 0.34 $  & $-$ & $-$  & $-$ & $-$ & $ \multirow{2}{*}{0.027}$ & $-0.26 $ & $ -0.22$ \\
  & \GG & $ -1.36 $ & $ -0.71 $  & $ 0.65 $ & $ 0.31 $  & $-$ & $-$  & $-$ & $-$ & & $\equiv -0.72 $ & $ \equiv -0.97$ \\
  \hline
  \rowcolor{Gray}
  \tabspace
  \multirow{2}{*}{HSCJ121052$-$011905} & ResNet & $ -1.43 $ & $ 1.17 $  & $ 0.69 $ & $ -0.29 $  & $-$ & $-$  & $-$ & $-$ & $ \multirow{2}{*}{0.612}$ & $-0.43 $ & $ 0.58$ \\
  & \GG &  $ -1.43 $ & $ 1.17 $  & $ 0.11 $ & $ -1.36 $  & $-$ & $-$  & $-$ & $-$ & & $\equiv 0.08 $ & $ \equiv 0.26$ \\
  \hline  \multicolumn{1}{c}{} & \multicolumn{1}{c}{} & \\
  \hline
  \rowcolor{Gray}
  \tabspace
  \multirow{2}{*}{HSCJ121504$+$004726} & ResNet & $ 1.86 $ & $ -0.82 $  & $ -0.9 $ & $ 0.37 $  & $-$ & $-$  & $-$ & $-$ & $ \multirow{2}{*}{0.167}$ & $0.49 $ & $ 0.34$ \\
  & \GG & $ 1.87 $ & $ -0.82 $  & $ -0.73 $ & $ 0.09 $  & $-$ & $-$  & $-$ & $-$ & & $\equiv 1.24 $ & $ \equiv -0.11$ \\
  \hline
  \rowcolor{Gray}
  \tabspace 
  \multirow{2}{*}{HSCJ124320$-$004517} & ResNet & $ 1.4 $ & $ 1.46 $  & $ -1.09 $ & $ 0.03 $  & $-$ & $-$  & $-$ & $-$ & $ \multirow{2}{*}{0.259}$ & $0.55 $ & $ 0.31$ \\
  & \GG & $ 1.4 $ & $ 1.46 $  & $ -0.96 $ & $ -0.47 $  & $-$ & $-$  & $-$ & $-$ & & $\equiv 0.28 $ & $ \equiv -0.42$ \\
  \hline
  \rowcolor{Gray}
  \tabspace
  \multirow{2}{*}{HSCJ125254$+$004356} & ResNet & $ -2.57 $ & $ -0.23 $  & $ 0.7 $ & $ 0.28 $  & $-$ & $-$  & $-$ & $-$ & $ \multirow{2}{*}{0.568}$ & $-1.1 $ & $ 0.13$ \\
  & \GG & $ -2.57 $ & $ -0.23 $  & $ 1.4 $ & $ -0.61 $  & $-$ & $-$  & $-$ & $-$ & & $\equiv -1.04 $ & $ \equiv 0.35$ \\
  \hline
  \rowcolor{Gray}
  \tabspace
  \multirow{2}{*}{HSCJ135138$+$002839} & ResNet & $ 1.3 $ & $ -1.89 $  & $ -1.13 $ & $ 0.69 $  & $-$ & $-$  & $-$ & $-$ & $ \multirow{2}{*}{/} $ & $0.57 $ & $ -0.24$ \\
  & \GG & $ 1.31 $ & $ -1.89 $  & $ -0.57 $ & $ 2.05 $  & $ 1.54 $ & $ 1.63 $  & $ -1.98 $ & $ -0.62 $ & & $\equiv -0.1 $ & $ \equiv -0.36$ \\
  \hline
  \rowcolor{Gray}
  \tabspace
  \multirow{2}{*}{HSCJ141136$-$010215} & ResNet & $ -1.24 $ & $ 0.38 $  & $ 1.13 $ & $ 0.51 $  & $-$ & $-$  & $-$ & $-$ & $ \multirow{2}{*}{0.214}$ & $-0.1 $ & $ 0.21$ \\
  & \GG & $ -1.24 $ & $ 0.39 $  & $ 0.71 $ & $ 0.51 $  & $-$ & $-$  & $-$ & $-$ & & $\equiv -0.33 $ & $ \equiv 0.22$ \\
  \hline \multicolumn{1}{c}{} & \multicolumn{1}{c}{} & \\
  \hline
  \rowcolor{Gray}
  \tabspace
  \multirow{2}{*}{HSCJ141815$+$015832} & ResNet & $ 1.43 $ & $ -0.61 $  & $ 1.57 $ & $ -0.41 $  & $ 1.49 $ & $ 0.95 $  & $ -1.19 $ & $ 0.48 $ & $ \multirow{2}{*}{/} $ & $0.23 $ & $ 0.46$ \\
  & \GG & $ 1.62 $ & $ -0.31 $  & $ -0.63 $ & $ 1.02 $  & $-$ & $-$  & $-$ & $-$ & & $\equiv 0.0 $ & $ \equiv 0.63$ \\
  \hline
  \rowcolor{Gray}
  \tabspace
  \multirow{2}{*}{HSCJ142720$+$001916} & ResNet & $ 1.5 $ & $ -1.16 $  & $ -0.63 $ & $ 0.72 $  & $-$ & $-$  & $-$ & $-$ & $ \multirow{2}{*}{0.12}$ & $0.33 $ & $ -0.01$ \\
  & \GG & $ 1.5 $ & $ -1.15 $  & $ -0.74 $ & $ 0.51 $  & $-$ & $-$  & $-$ & $-$ & & $\equiv 0.18 $ & $ \equiv -1.1$ \\
  \hline
  \rowcolor{Gray}
  \tabspace
  \multirow{2}{*}{HSCJ144320$-$012537} & ResNet & $ 0.61 $ & $ -1.42 $  & $ -0.61 $ & $ 0.86 $ & $-$ & $-$  & $-$ & $-$ & $ \multirow{2}{*}{/} $ & $-0.1 $ & $ -0.29$ \\ 
  & \GG & $ 0.61 $ & $ -1.41 $  & $ -1.08 $ & $ 0.33 $  & $ -1.11 $ & $ -0.05 $  & $ 0.59 $ & $ 0.65 $ & & $\equiv 0.37 $ & $ \equiv 0.01$ \\
  \rowcolor{Gray}
  \tabspace
  \multirow{2}{*}{HSCJ145242$+$425731} & ResNet & $ -1.43 $ & $ 1.71 $  & $ 1.77 $ & $ -0.07 $  & $-$ & $-$  & $-$ & $-$ & $ \multirow{2}{*}{0.085}$ & $0.0 $ & $ 0.12$ \\
  & \GG & $ -1.42 $ & $ 1.72 $  & $ 1.63 $ & $ -0.02 $  & $-$ & $-$  & $-$ & $-$ & & $\equiv -0.66 $ & $ \equiv 0.04$ \\
  \hline
  \rowcolor{Gray}
  \tabspace
  \multirow{2}{*}{HSCJ150021$-$004936} & ResNet & $ -2.53 $ & $ -2.23 $ & $-$ & $-$ & $-$ & $-$ & $-$ & $-$ & \multirow{2}{*}{$/$} & $-1.216 $ & $ -0.956$ \\
  & \GG & $ -2.53 $ & $ -2.23 $  & $ -0.25 $ & $ 2.49 $  & $-$ & $-$  & $-$ & $-$ & & $\equiv -0.506 $ & $ \equiv 0.344$ \\
  \hline \multicolumn{1}{c}{} & \multicolumn{1}{c}{} & \\
  \hline 
  \rowcolor{Gray}
  \tabspace
  \multirow{2}{*}{HSCJ150112$+$422113} & ResNet & $ -0.63 $ & $ 1.22 $  & $ -0.61 $ & $ -0.82 $  & $ -1.18 $ & $ 0.47 $  & $ 0.72 $ & $ 0.25 $ & $ \multirow{2}{*}{0.367}$ & $-0.07 $ & $ 0.13$  \\
  & \GG & $ -0.62 $ & $ 1.22 $  & $ -0.33 $ & $ -1.28 $  & $ -1.06 $ & $ -0.22 $  & $ 0.56 $ & $ 0.1 $ & & $\equiv 0.4 $ & $ \equiv -0.74$ \\
  \hline
  \rowcolor{Gray}
  \tabspace
  \multirow{2}{*}{HSCJ223733$+$005015} & ResNet & $ 0.28 $ & $ 1.89 $  & $ 0.13 $ & $ -0.7 $  & $-$ & $-$  & $-$ & $-$ & $ \multirow{2}{*}{0.328}$ & $0.12 $ & $ 0.79$ \\
  & \GG &  $ 0.28 $ & $ 1.89 $  & $ -0.51 $ & $ -0.62 $  & $-$ & $-$  & $-$ & $-$ & & $\equiv -0.34 $ & $ \equiv 1.18$ \\
  \hline
  \rowcolor{Gray}
  \tabspace
  \multirow{2}{*}{HSCJ230335$+$003703} & ResNet & $ -0.91 $ & $ 1.07 $  & $ 0.47 $ & $ 0.05 $  & $-$ & $-$  & $-$ & $-$ & $ \multirow{2}{*}{0.296}$ & $-0.1 $ & $ 0.39$ \\
  & \GG & $ -0.91 $ & $ 1.07 $  & $ 0.43 $ & $ -0.53 $  & $-$ & $-$  & $-$ & $-$ & & $\equiv -0.12 $ & $ \equiv 0.48$ \\
  \hline
  \rowcolor{Gray}
  \tabspace
  \multirow{2}{*}{HSCJ230521$-$000211} & ResNet & $ 0.67 $ & $ 1.86 $  & $ 0.98 $ & $ -1.55 $  & $ 1.53 $ & $ 1.25 $  & $ -1.29 $ & $ 0.34 $ & $ \multirow{2}{*}{0.526}$ & $0.51 $ & $ 0.2$ \\
  & \GG & $ 0.67 $ & $ 1.86 $  & $ 1.16 $ & $ -1.41 $  & $ 1.86 $ & $ -0.17 $  & $ -1.3 $ & $ -0.08 $ & & $\equiv 0.41 $ & $ \equiv 0.51$ \\
  \hline
  \rowcolor{Gray}
  \tabspace
  \multirow{2}{*}{HSCJ233130$+$003733} & ResNet & $ -1.66 $ & $ -0.8 $  & $ 0.3 $ & $ 0.55 $  & $-$ & $-$  & $-$ & $-$ & $ \multirow{2}{*}{0.225}$ & $-0.72 $ & $ -0.31$ \\
  & \GG & $ -1.66 $ & $ -0.8 $  & $ 0.43 $ & $ 0.98 $  & $-$ & $-$  & $-$ & $-$ & & $\equiv -0.75 $ & $ \equiv -0.34$ \\
  \hline \multicolumn{1}{c}{} & \multicolumn{1}{c}{} & \\ 
  \hline
  \rowcolor{Gray}
  \tabspace
  \multirow{2}{*}{HSCJ233146$+$013845} & ResNet & $ -0.7 $ & $ 1.5 $  & $ -0.71 $ & $ -1.29 $  & $ -0.77 $ & $ 1.45 $  & $ 1.44 $ & $ 0.25 $ & $ \multirow{2}{*}{0.664}$ & $0.01 $ & $ -0.03$ \\
  & \GG & $ -0.63 $ & $ 1.54 $  & $ -0.81 $ & $ -1.35 $  & $ -1.42 $ & $ -0.41 $  & $ 1.23 $ & $ -0.19 $ & & $\equiv 0.46 $ & $ \equiv -0.22$ \\
\end{longtable}
\end{landscape}
\twocolumn

\onecolumn
\begin{longtable}{c|c|ccc|ccc}
\caption{Resulting time delays and Fermat potential differences. \label{tab:comparison_Fermat}}\\

  Name   & method &\multicolumn{3}{c|}{Fermat potential difference} & \multicolumn{3}{c}{Time delays} \\
         &        & $\Delta \tau_\text{AB}$ & $\Delta \tau_\text{AC}$ & $\Delta \tau_\text{AD}$ & $\Delta t_\text{AB}$  [days]& $\Delta t_\text{AC}$ [days] & $\Delta t_\text{AD}$ [days]\\
  \hline
  \endfirsthead
  \caption*{\textbf{Table \ref{tab:comparison_Fermat} Continued:} Resulting time delays and Fermat potential differences.}\\
  Name   & method &\multicolumn{3}{c|}{Fermat potential difference} & \multicolumn{3}{c}{Time delays} \\
         &        & $\Delta \tau_\text{AB}$ & $\Delta \tau_\text{AC}$ & $\Delta \tau_\text{AD}$ & $\Delta t_\text{AB}$ [days] & $\Delta t_\text{AC}$ [days] & $\Delta t_\text{AD}$ [days] \\
  \hline
  \endhead
  \rowcolor{Gray}
  \multirow{2}{*}{HSCJ015618$-$010747} & ResNet & $ 0.01723 $  & $-$  & $-$   & $ 75.053 $  & $-$  & $-$ \\
  & \GG & $ 0.02752 $  & $-$  & $-$   & $ 119.87 $  & $-$  & $-$\\
  \hline
  \rowcolor{Gray}
  \tabspace
  \multirow{2}{*}{HSCJ020141$-$030946} & ResNet & $ 0.00426 $  & $-$  & $-$   & $ 10.293 $  & $-$  & $-$ \\
  & \GG & $ 0.03385 $  & $-$  & $-$   & $ 81.758 $  & $-$  & $-$ \\
  \hline
  \rowcolor{Gray}
  \tabspace
  \multirow{2}{*}{HSCJ020241$-$064611} & ResNet & $ 0.02721 $  & $-$  & $-$   & $ 72.623 $  & $-$  & $-$ \\
  & \GG & $ 0.01612 $  & $-$  & $-$   & $ 43.023 $  & $-$  & $-$ \\
  \hline
  \rowcolor{Gray}
  \tabspace
  \multirow{2}{*}{HSCJ020955$-$024442} & ResNet & $ -0.00196 $  & $ -0.00097 $  & $ 0.00038 $   & $ -10.741 $  & $ -5.329 $  & $ 2.104 $ \\
  & \GG & $ 0.00023 $  & $ 0.00158 $  & $ 0.0017 $   & $ 1.258 $  & $ 8.63 $  & $ 9.312 $ \\
  \hline
  \rowcolor{Gray}
  \tabspace
  \multirow{2}{*}{HSCJ021737$-$051329} & ResNet & $ 0.03203 $  & $-$  & $-$   & $ 138.68 $  & $-$  & $-$ \\
  & \GG & $ 0.00067 $  & $ 0.00088 $  & $ 0.01561 $   & $ 2.904 $  & $ 3.792 $  & $ 67.607 $ \\
  \hline \multicolumn{1}{c}{} & \multicolumn{1}{c}{} & \\ 
  \hline
  \rowcolor{Gray}
  \tabspace
  \multirow{2}{*}{HSCJ022346$-$053418} & ResNet & $-$ & $-$ & $-$ & $-$ & $-$ & $-$ \\
  & \GG & $ 5e-05 $  & $ 5e-05 $  & $ 0.0206 $   & $ 0.149 $  & $ 0.163 $  & $ 67.709 $ \\
  \hline
  \rowcolor{Gray}
  \tabspace
  \multirow{2}{*}{HSCJ022610$-$042011} & ResNet & $ 0.03564 $  & $-$  & $-$   & $ 150.257 $  & $-$  & $-$ \\
  & \GG & $ 0.04364 $  & $-$  & $-$   & $ 183.97 $  & $-$  & $-$ \\
  \hline
  \rowcolor{Gray}
  \tabspace
  \multirow{2}{*}{HSCJ023217$-$021703} & ResNet & $ 0.0096 $  & $-$  & $-$   & $ 42.48 $  & $-$  & $-$ \\ 
  & \GG & $ 0.01533 $  & $ 0.01744 $  & $ 0.01765 $   & $ 67.842 $  & $ 77.163 $  & $ 78.116 $ \\
  \hline
  \rowcolor{Gray}
  \tabspace
  \multirow{2}{*}{HSCJ023322$-$020530} & ResNet & $ 0.07056 $  & $-$  & $-$   & $ 290.363 $  & $-$  & $-$ \\
  & \GG & $ 0.06284 $  & $-$  & $-$   & $ 258.57 $  & $-$  & $-$ \\
  \hline
  \rowcolor{Gray}
  \tabspace
  \multirow{2}{*}{HSCJ085046$+$003905} & ResNet & $ 0.04762 $  & $-$  & $-$   & $ 1084.504 $  & $-$  & $-$ \\
  & \GG & $ 0.00266 $  & $ 0.00763 $  & $ 0.02735 $   & $ 60.66 $  & $ 173.82 $  & $ 622.87 $ \\
  \hline \multicolumn{1}{c}{} & \multicolumn{1}{c}{} & \\
  \hline 
  \rowcolor{Gray}
  \tabspace
  \multirow{2}{*}{HSCJ085855$-$010208} & ResNet & $ 0.00433 $  & $-$  & $-$   & $ 13.006 $  & $-$  & $-$ \\
  & \GG & $ 0.0048 $  & $-$  & $-$   & $ 14.422 $  & $-$  & $-$ \\
  \hline
  \rowcolor{Gray}
  \tabspace
  \multirow{2}{*}{HSCJ090429$-$010228} & ResNet & $ 0.00689 $  & $ -0.00095 $  & $ -0.00152 $   & $ 40.746 $  & $ -5.639 $  & $ -8.962 $ \\
  & \GG & $ 0.00501 $  & $ 0.00503 $  & $ 0.01339 $   & $ 29.649 $  & $ 29.746 $  & $ 79.18 $ \\
  \hline
  \rowcolor{Gray}
  \tabspace
  \multirow{2}{*}{HSCJ094427$-$014742} & ResNet & $ 0.02533 $  & $-$  & $-$   & $ 108.259 $  & $-$  & $-$ \\
  & \GG & $ 0.0229 $  & $-$  & $-$   & $ 97.888 $  & $-$  & $-$ \\
  \hline
  \rowcolor{Gray}
  \tabspace
  \multirow{2}{*}{HSCJ120623$+$001507} & ResNet & $ 0.03848 $  & $-$  & $-$   & $ 115.37 $  & $-$  & $-$ \\
  & \GG & $ 0.02033 $  & $-$  & $-$   & $ 60.961 $  & $-$  & $-$ \\
  \hline
  \rowcolor{Gray}
  \tabspace
  \multirow{2}{*}{HSCJ121052$-$011905} & ResNet &  $ 0.04452 $  & $-$  & $-$   & $ 196.516 $  & $-$  & $-$ \\
  & \GG & $ 0.02828 $  & $-$  & $-$   & $ 124.82 $  & $-$  & $-$ \\
  \hline \multicolumn{1}{c}{} & \multicolumn{1}{c}{} & \\ 
  \hline
  \rowcolor{Gray}
  \tabspace
  \multirow{2}{*}{HSCJ121504$+$004726} & ResNet & $ 0.03499 $  & $-$  & $-$   & $ 192.668 $  & $-$  & $-$  \\
  & \GG & $ 0.05442 $  & $-$  & $-$   & $ 299.66 $  & $-$  & $-$ \\
  \hline
  \rowcolor{Gray}
  \tabspace 
  \multirow{2}{*}{HSCJ124320$-$004517} & ResNet & $ 0.02291 $  & $-$  & $-$   & $ 115.424 $  & $-$  & $-$ \\
  & \GG & $ 0.03383 $  & $-$  & $-$   & $ 170.42 $  & $-$  & $-$ \\
  \hline
  \rowcolor{Gray}
  \tabspace
  \multirow{2}{*}{HSCJ125254$+$004356} & ResNet &$ 0.05793 $  & $-$  & $-$   & $ 287.874 $  & $-$  & $-$ \\
  & \GG & $ 0.04887 $  & $-$  & $-$   & $ 242.83 $  & $-$  & $-$ \\
  \hline
  \rowcolor{Gray}
  \tabspace
  \multirow{2}{*}{HSCJ135138$+$002839} & ResNet & $ 0.02231 $  & $-$  & $-$   & $ 81.631 $  & $-$  & $-$\\
  & \GG & $ 0.00671 $  & $ 0.00837 $  & $ 0.01065 $   & $ 24.542 $  & $ 30.617 $  & $ 38.966 $ \\
  \hline
  \rowcolor{Gray}
  \tabspace
  \multirow{2}{*}{HSCJ141136$-$010215} & ResNet & $ 0.01143 $  & $-$  & $-$   & $ 70.164 $  & $-$  & $-$ \\
  & \GG & $ 0.00938 $  & $-$  & $-$   & $ 57.564 $  & $-$  & $-$ \\
  \hline \multicolumn{1}{c}{} & \multicolumn{1}{c}{} & \\
  \hline 
  \rowcolor{Gray}
  \tabspace
  \multirow{2}{*}{HSCJ141815$+$015832} & ResNet & $ -0.00304 $  & $ -0.00267 $  & $ -0.01256 $   & $ -9.927 $  & $ -8.716 $  & $ -41.024 $ \\
  & \GG & $ 0.02243 $  & $-$  & $-$   & $ 73.263 $  & $-$  & $-$ \\
  \hline
  \rowcolor{Gray}
  \tabspace
  \multirow{2}{*}{HSCJ142720$+$001916} & ResNet & $ 0.03238 $  & $-$  & $-$   & $ 136.208 $  & $-$  & $-$ \\
  & \GG & $ 0.04613 $  & $-$  & $-$   & $ 194.06 $  & $-$  & $-$ \\
  \hline
  \rowcolor{Gray}
  \tabspace
  \multirow{2}{*}{HSCJ144320$-$012537} & ResNet & $ 0.03096 $  & $-$  & $-$   & $ 379.762 $  & $-$  & $-$ \\ 
  & \GG & $ 0.01136 $  & $ 0.01142 $  & $ 0.01877 $   & $ 139.37 $  & $ 140.13 $  & $ 230.248 $ \\
  \hline
  \rowcolor{Gray}
  \tabspace
  \multirow{2}{*}{HSCJ145242$+$425731} & ResNet & $ 0.01875 $  & $-$  & $-$   & $ 112.52 $  & $-$  & $-$ \\
  & \GG & $ 0.03818 $  & $-$  & $-$   & $ 229.1 $  & $-$  & $-$ \\
  \hline
  \rowcolor{Gray}
  \tabspace
  \multirow{2}{*}{HSCJ150021$-$004936} & ResNet & $-$ & $-$ & $-$ & $-$ & $-$ & $-$ \\
  & \GG & $ 0.08355 $  & $-$  & $-$   & $ 247.76 $  & $-$  & $-$ \\
  \hline \multicolumn{1}{c}{} & \multicolumn{1}{c}{} & \\
  \hline
  \rowcolor{Gray}
  \tabspace
  \multirow{2}{*}{HSCJ150112$+$422113} & ResNet & $ -0.00572 $  & $ -0.00345 $  & $ 0.00179 $   & $ -8.959 $  & $ -5.401 $  & $ 2.81 $ \\
  & \GG & $ 0.00172 $  & $ 0.00661 $  & $ 0.01668 $   & $ 2.688 $  & $ 10.363 $  & $ 26.133 $ \\
  \hline
  \rowcolor{Gray}
  \tabspace
  \multirow{2}{*}{HSCJ223733$+$005015} & ResNet & $ 0.03799 $  & $-$  & $-$   & $ 139.25 $  & $-$  & $-$ \\
  & \GG & $ 0.04041 $  & $-$  & $-$   & $ 148.11 $  & $-$  & $-$ \\
  \hline
  \rowcolor{Gray}
  \tabspace
  \multirow{2}{*}{HSCJ230335$+$003703} & ResNet & $ 0.02729 $  & $-$  & $-$   & $ 104.554 $  & $-$  & $-$ \\
  & \GG & $ 0.01689 $  & $-$  & $-$   & $ 64.698 $  & $-$  & $-$ \\
  \hline
  \rowcolor{Gray}
  \tabspace
  \multirow{2}{*}{HSCJ230521$-$000211} & ResNet & $ -0.01394 $  & $ -0.00262 $  & $ 0.00478 $   & $ -57.831 $  & $ -10.885 $  & $ 19.815 $ \\
  & \GG & $ 0.00542 $  & $ 0.00748 $  & $ 0.0246 $   & $ 22.465 $  & $ 31.03 $  & $ 102.03 $ \\
  \hline
  \rowcolor{Gray}
  \tabspace
  \multirow{2}{*}{HSCJ233130$+$003733} & ResNet &  $ 0.06206 $  & $-$  & $-$   & $ 328.386 $  & $-$  & $-$ \\
  & \GG & $ 0.02388 $  & $-$  & $-$   & $ 126.36 $  & $-$  & $-$ \\
  \hline \multicolumn{1}{c}{} & \multicolumn{1}{c}{} &  \\
  \hline 
  \rowcolor{Gray}
  \tabspace
  \multirow{2}{*}{HSCJ233146$+$013845} & ResNet & $ -0.00984 $  & $ -0.00282 $  & $ 0.00339 $   & $ -38.253 $  & $ -10.977 $  & $ 13.189 $ \\
  & \GG & $ 0.00347 $  & $ 0.00405 $  & $ 0.01425 $   & $ 13.481 $  & $ 15.744 $  & $ 55.417 $ \\
\end{longtable}

\end{document}